# On the Sunspot Group Number Reconstruction: The Backbone Method Revisited

Leif Svalgaard[1] and Kenneth H. Schatten[2]

**Abstract**

We discuss recent papers very critical of our Group Sunspot Number Series (Svalgaard & Schatten [2016]). Unfortunately, we cannot support any of the concerns they raise. We first show that almost always there is simple proportionality between the group counts by different observers and that taking the small, occasional, non-linearities into account makes very little difference. Among other examples: we verify that the RGO group count was drifting the first twenty years of observations. We then show that our group count matches the diurnal variation of the geomagnetic field with high fidelity, and that the heliospheric magnetic field derived from geomagnetic data is consistent with our group number series. We evaluate the 'correction matrix' approach [Usoskin et al. 2016] and show that it fails to reproduce the observational data. We clarify the notion of daisy-chaining and point out that our group number series has no daisy-chaining for the period 1794-1996 and therefore no accumulation of errors over that span. We compare with the cosmic ray record for the last 400+ years and find good agreement. We note that the Active Day Fraction method (of Usoskin et al.) has the fundamental problem that at sunspot maximum, every day is an 'active day' so ADF is nearly always unity and thus does not carry information about the statistics of high solar activity. This 'information shadow' occurs for even moderate group numbers and thus need to be extrapolated to higher activity. The ADF method also fails for 'equivalent observers' who should register the same group counts, but do not. We conclude that the criticism of Svalgaard & Schatten [2016] is invalid and detrimental to progress in the important field of long-term variation of solar activity.

## 1. Introduction

An accurate and agreed upon record of solar activity is important for a space-faring World increasingly dependent on an understanding of and on reliable forecasting of the activity on many time scales. Several workshops have been held by the solar physics community over the past several years [Clette et al., 2014, 2016] with the goal of reconciling the various sunspot series and producing a vetted and agreed upon series that can form the bedrock for studies of solar activity throughout the solar system. But this goal has not been achieved and the field has fragmented into several competing, incompatible series. As Jack Harvey (http://www.leif.org/research/SSN/Harvey.pdf)

[1] Stanford University, Stanford, CA 94305, USA
[2] a.i. solutions, Lanham, MD 20706, USA

pointedly commented at the third Sunspot Number Workshop in Tucson in 2013 "*It's ugly in there!*"

The present article discusses examples of such ugliness as indicators of 'the state of the art' which are providing a disservice to users and are not helpful for their research. Research into understanding long-term solar activity is important because the ground-based solar observations over centuries have yielded results that are not fully understood. In addition, the long-term trends are important for prediction of solar activity and solar-terrestrial relations. Hopefully the situation will improve in the future because progress in a field is based upon the extent to which common goals can be shared among researchers who can agree on methodologies used and build on each others work. Without such direction, fields become fragmented and research can wither on the vine, as seems to be happening currently.

## 2. On Proportionality

In their Section 6, Lockwood et al. [2016b, see also 2016a] state "We find that proportionality of annual means of the results of different sunspot observers is generally invalid and that assuming it causes considerable errors in the long-term." This remarkable statement is simply not true as plotting the annual means of one observer against the annual means of another clearly demonstrates. We show below many examples of such direct proportionality, underscoring that this is not the result of mere assumptions, but can be directly derived from the data themselves; more examples can be found in the spreadsheet data documentation for the Sunspot Group Number reconstruction [Svalgaard & Schatten, 2016; http://www.leif.org/research/gn-data.htm]. Simple direct proportionality accounts for 98-99% of the variation, so is not an assumption, but an observational fact. We concentrate first on the interval ~1870-1905 where the progressive divergence between the Hoyt & Schatten [1998] Group Sunspot Number and the Svalgaard & Schatten [2016] Sunspot Group Number becomes manifest.

We start with the all-important observations by the Zürich observers Johann Rudolf Wolf and Alfred Wolfer who laid the foundations of the sunspot number series, by their own observations supplemented by data from an extensive world-wide network of secondary observers and by research into historical records of centuries past; making the sunspot record the longest running scientific experiment reaching back to the invention of the telescope. We owe to all of them to continue what they began, so here is first, Figure 1, the comparison of Wolf to match Wolfer. The Figures 1 to 10, all have the same format. The left panel shows the regression of the annual group counts by an observer versus the count by Wolfer. Regression lines are fitted both with and without an offset. Usually the two lines are indistinguishable, going through the origin, because the offset is so small. The right panel plots the counts by the observer (blue) and by Wolfer (pink), and the observer's count scaled with the slope of the regression line (orange).

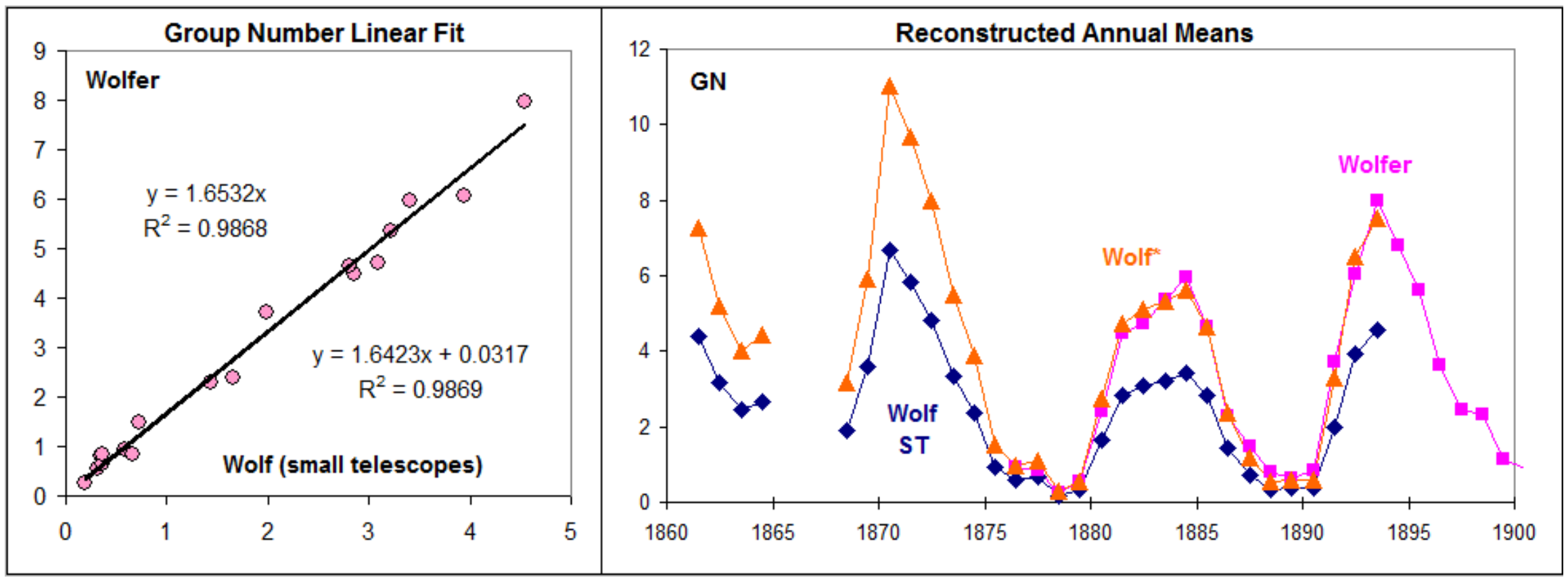

**Figure 1**. Linear fit of Wolf's annual group count (with small telescopes; aperture ~40 mm) to match Wolfer's (with the standard telescope; aperture 82 mm). The offset is insignificantly different from zero, showing that the counts are simply proportional on time scales of a year. The two regression lines with or without an offset are indistinguishable. Note that this is not an assumption, but an observational fact. The right-hand panel shows how well we can reproduce Wolfer's count from Wolf's by simple scaling by a constant factor, the slope of the regression line. The scaled Wolf counts are shown by the orange triangles. Applying the insignificant offset does not make any discernable difference.

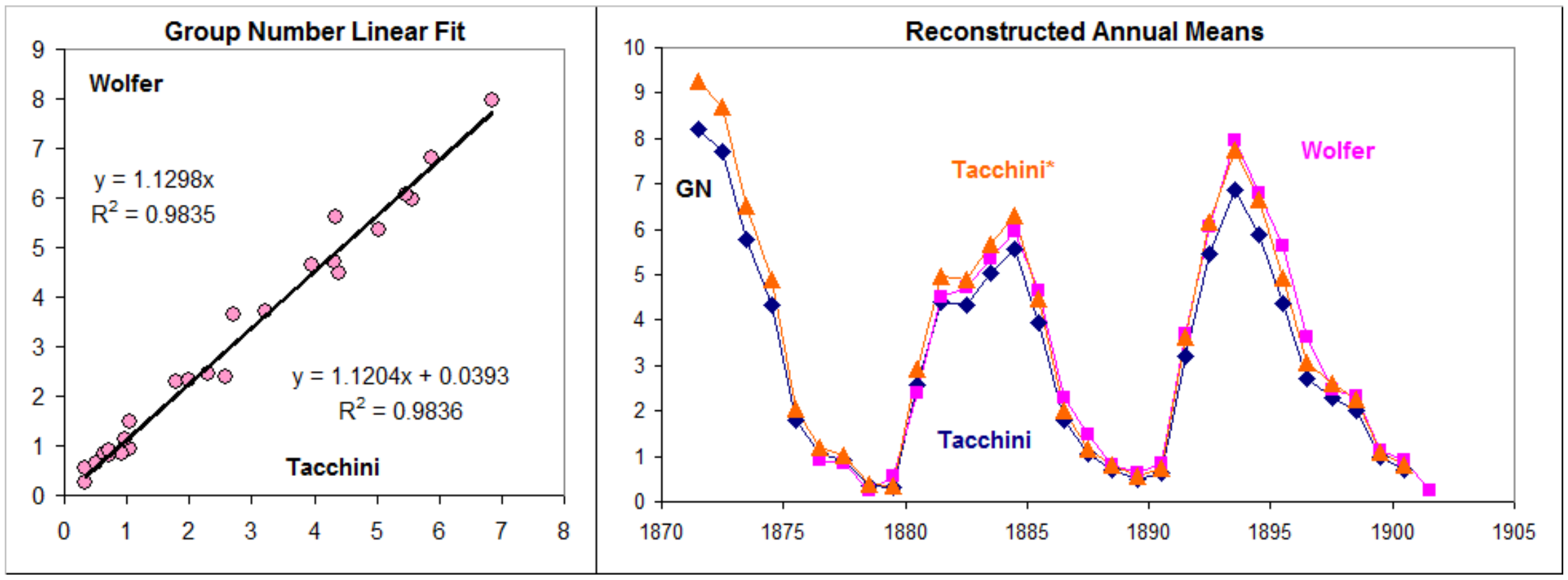

**Figure 2**. As Figure 1, but for the Italian observer Tacchini. Again, simple proportionality is an observational fact.

Pietro Tacchini was an important observer, covering the critical interval 1871-1900 with 7584 daily observations (some made by assistants G. Ferrari and G. de Lisa) obtained with a superb 24-cm Merz refractor (http://www.privatsternwarte.net/250erMerz_HP.jpg). His counts are close to Wolfer's, guarding against any sizable drift over the time interval of most interest. Wolfer's *k*-factor (as published by Wolf) decreased slightly as Wolfer became more experienced, so we would expect a small (but insignificant) increase with time of his group count relative to other observers, but as Figures 1 to 10 show, this is not noticeable so is, indeed, insignificant.

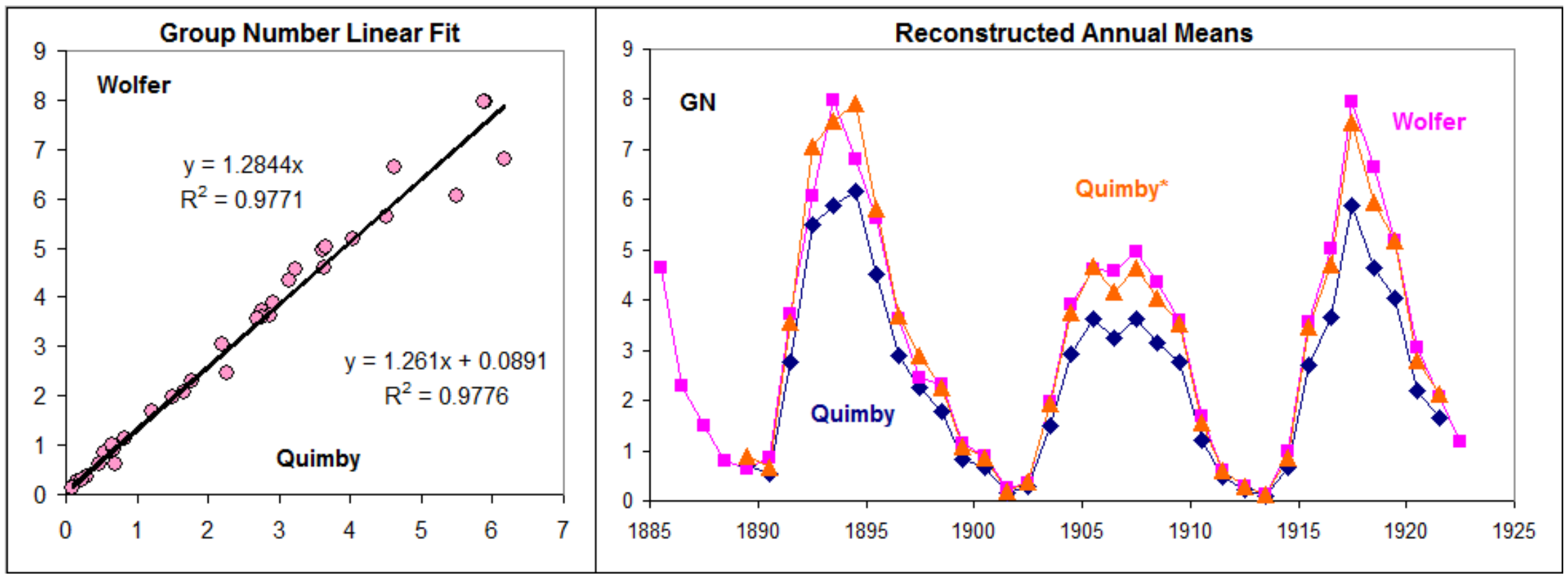

**Figure 3**. As Figure 1, but for the American observer Rev. Quimby. Again, simple proportionality is an observational fact.

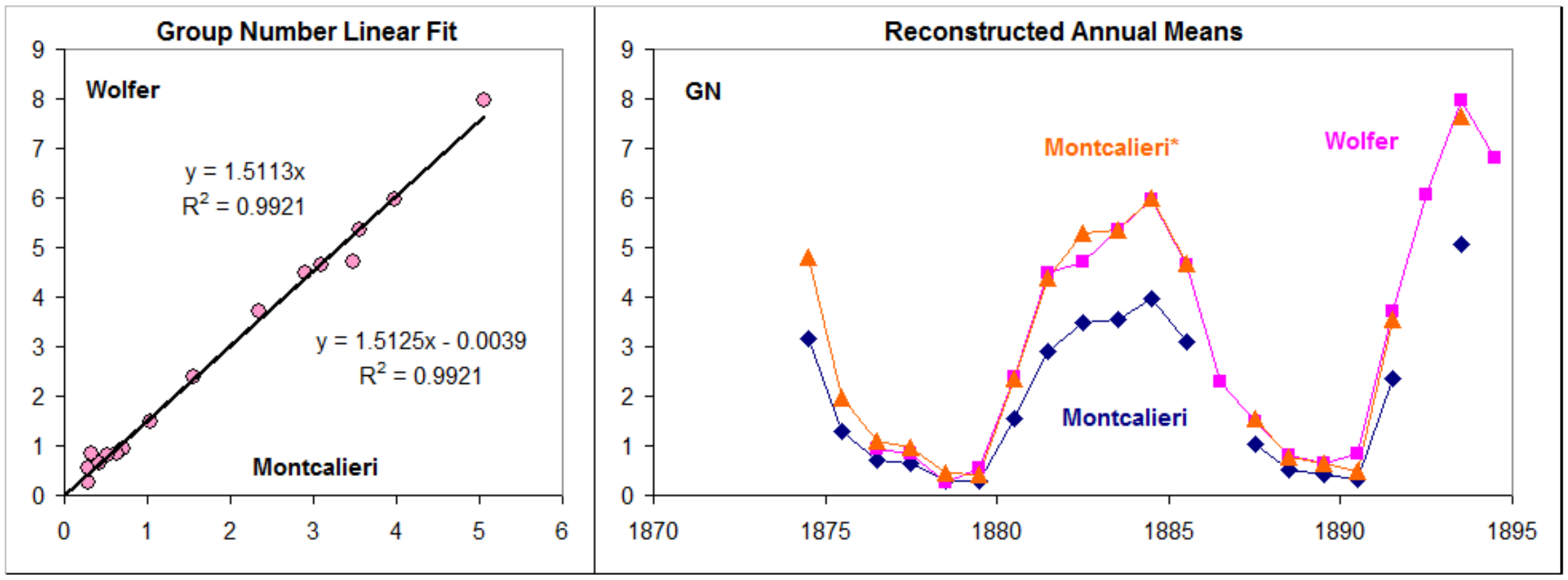

**Figure 4**. As Figure 1, but for the Italian observer at Montcalieri. Again, simple proportionality is an observational fact.

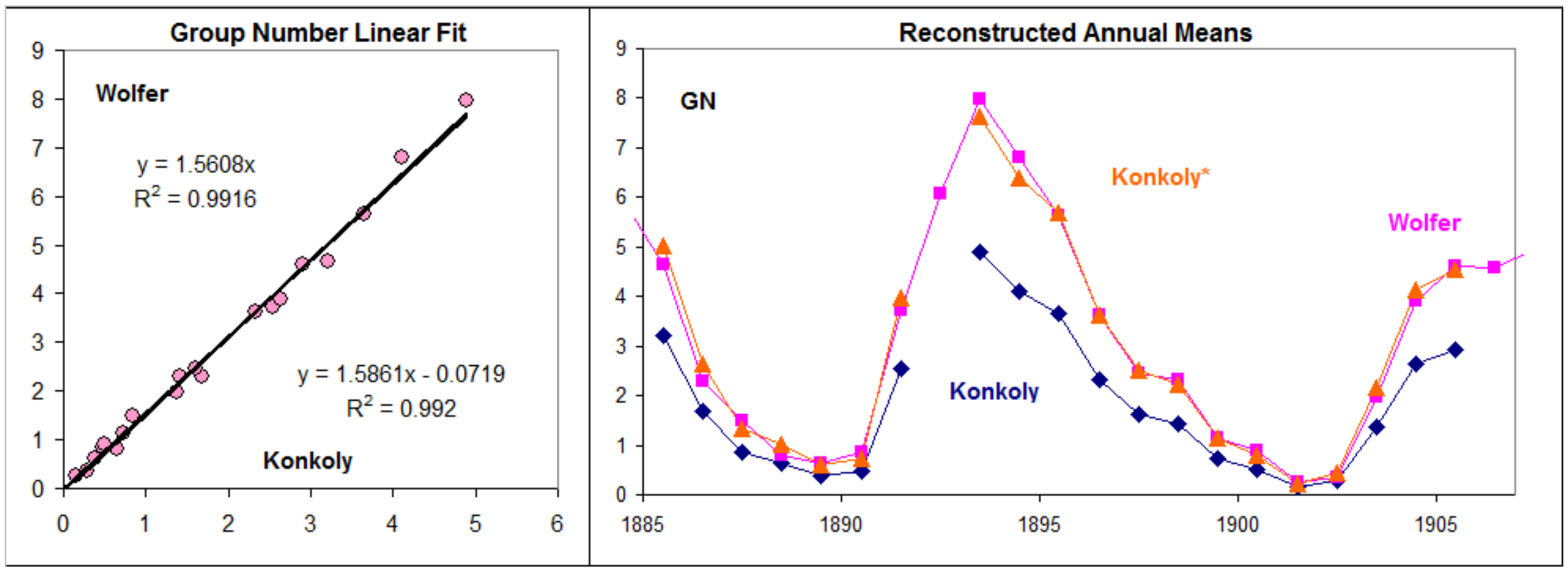

**Figure 5**. As Figure 1, but for the Hungarian observer Miklós Konkoly-Thege. Again, simple proportionality is an observational fact.

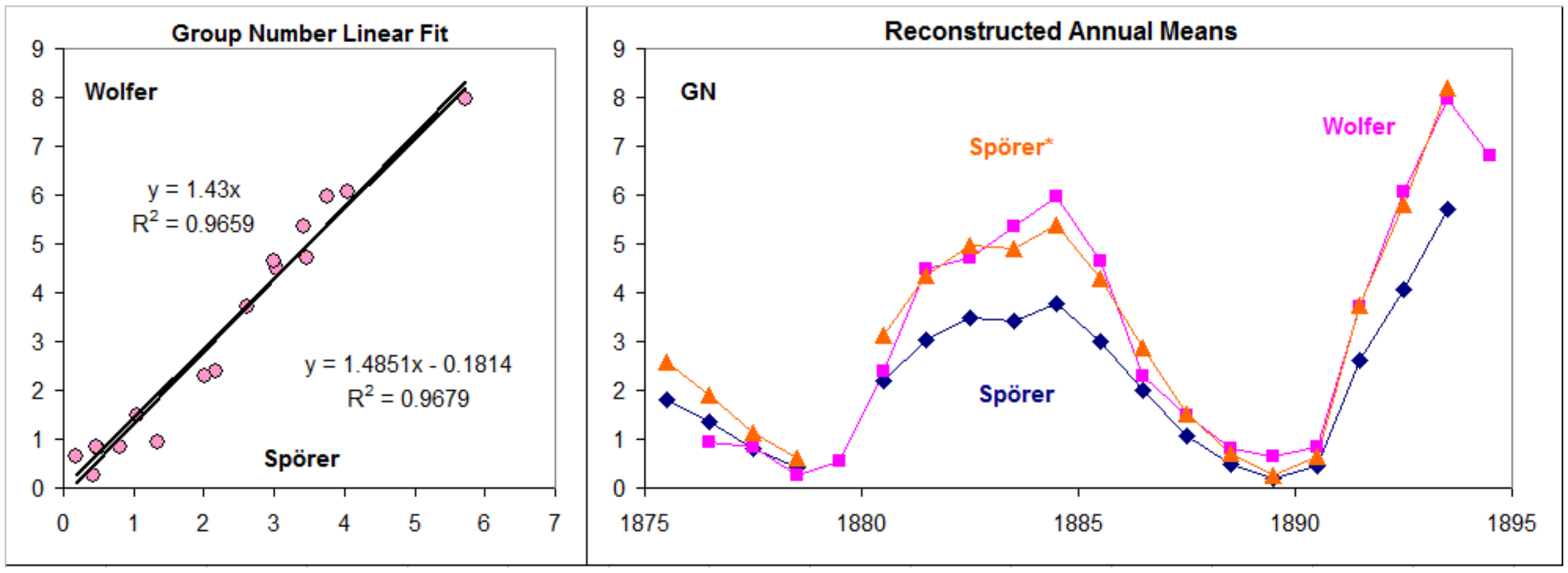

**Figure 6**. As Figure 1, but for the German observer Gustav Spörer. Again, simple proportionality is an observational fact, in spite of the slightly larger scatter.

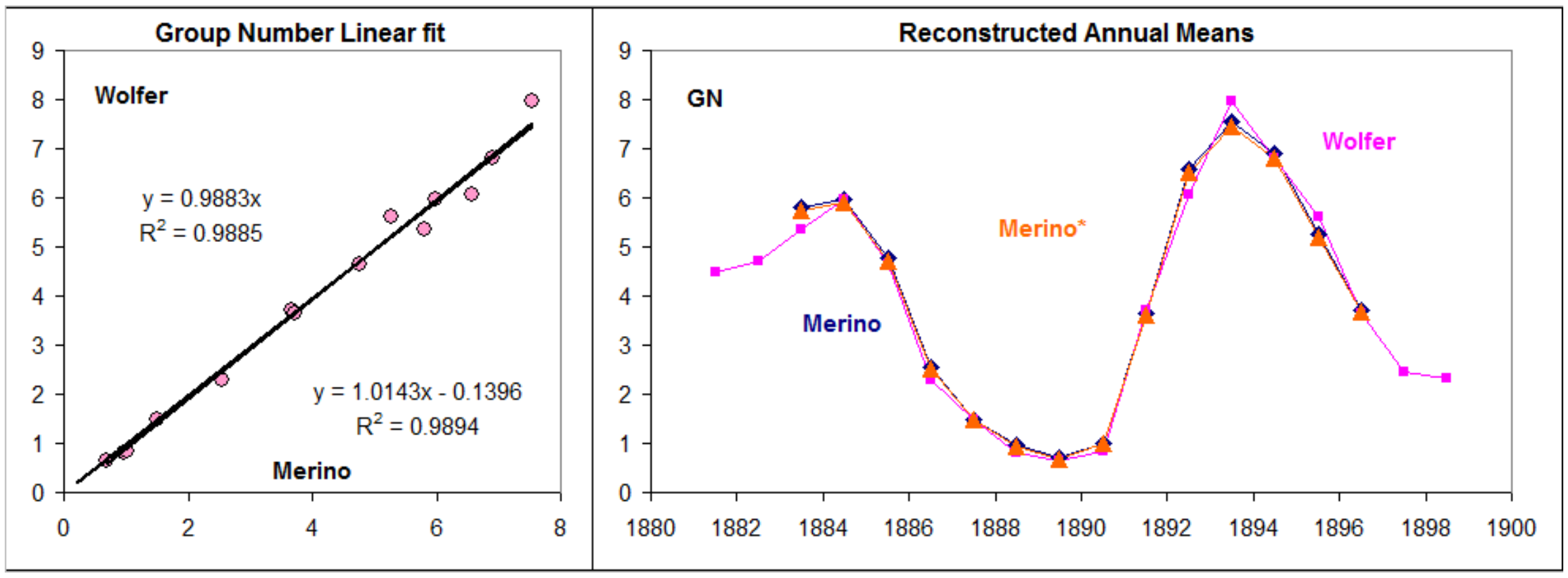

**Figure 7**. As Figure 1, but for the Spanish observer Merino. Again, simple proportionality is an observational fact.

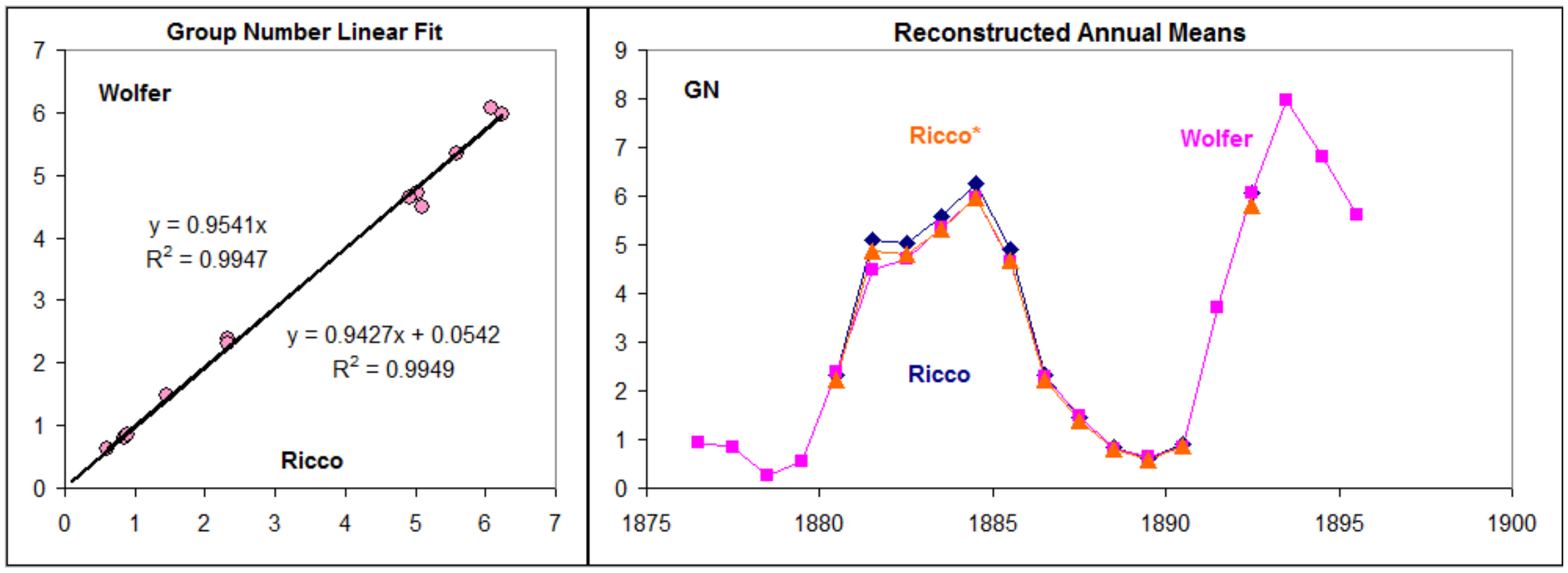

**Figure 8**. As Figure 1, but for the Italian observer Ricco. Again, simple proportionality is an observational fact.

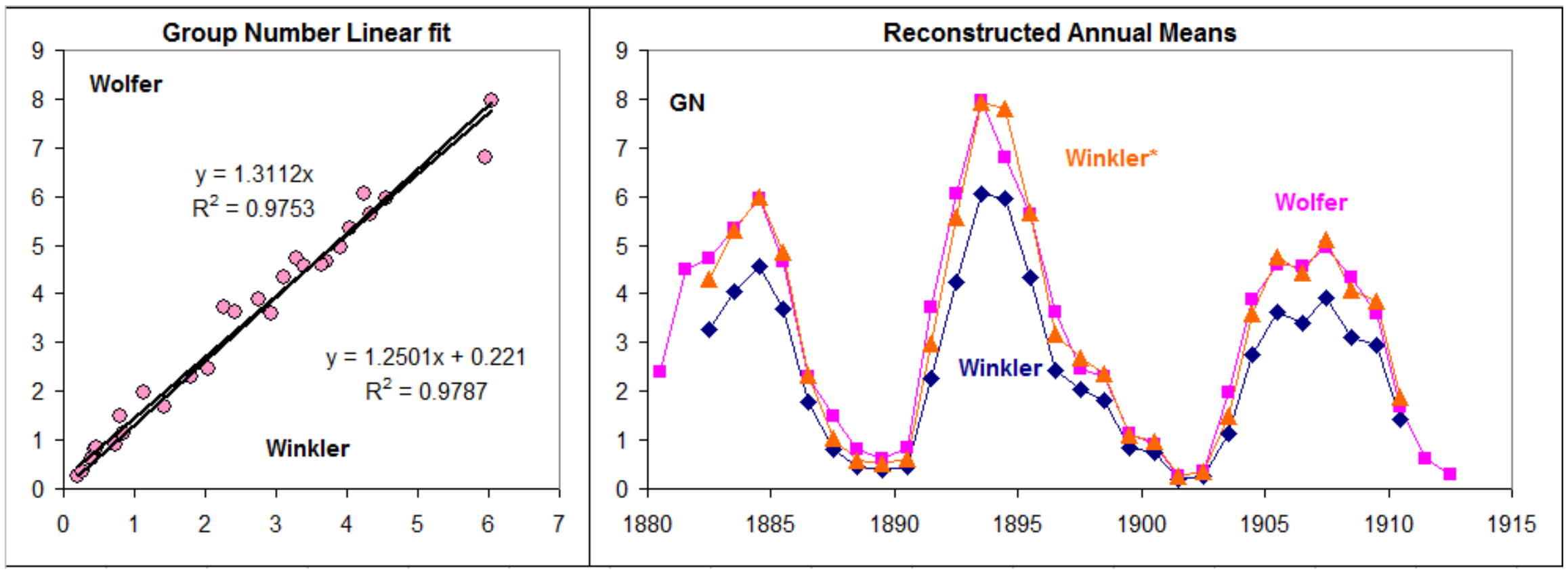

**Figure 9**. As Figure 1, but for the German observer Winkler. Again, simple proportionality is an observational fact, in spite of the slightly larger scatter.

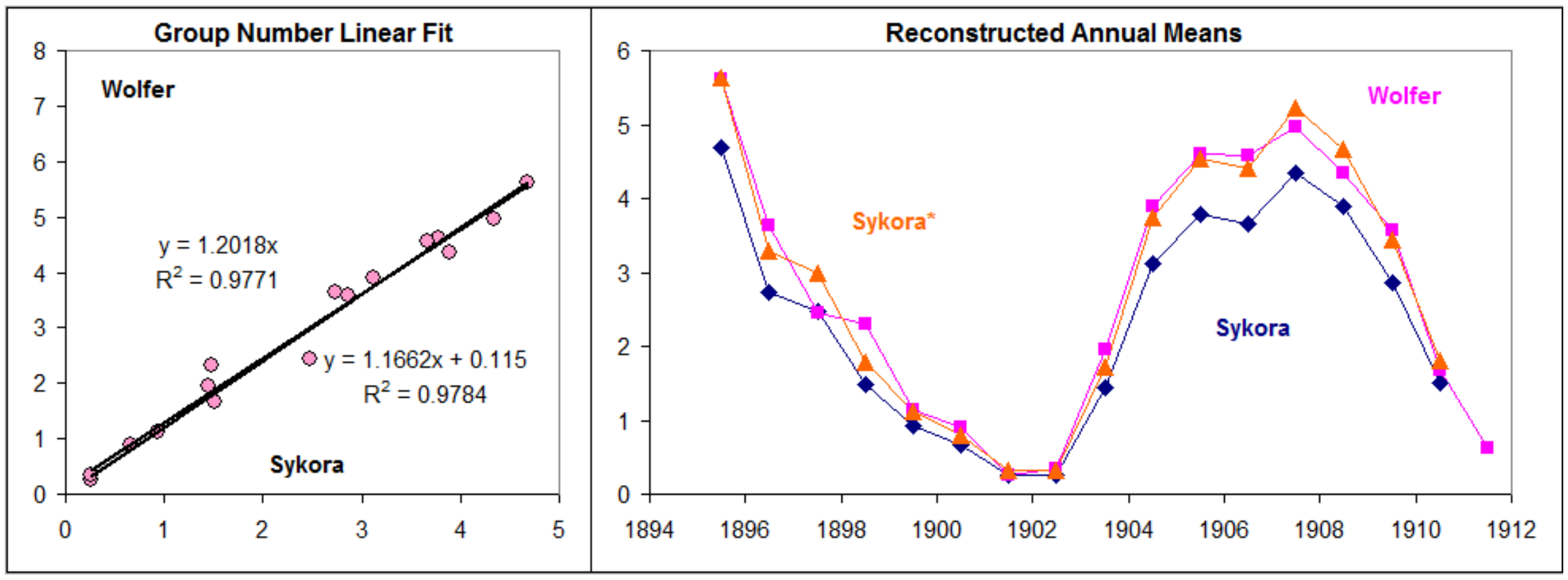

**Figure 10**. As Figure 1, but for the observer Sykora. Again, simple proportionality is an observational fact, not an assumption.

Having shown that linear, proportional scaling works, we can now simply average the reconstructed, scaled, annual means for 1875-1905 (when there are at least four observers each year) for the 11 observers for which we have just demonstrated simple, direct proportionality between their counts, and plot the result, Figure 11. The analysis is straightforward and statistically sound if we assume that the number of groups emerging in a year (and hence the daily average during the year) is a measure of integrated solar activity for that year. This is the only assumption we make, and can even be taken as a *definition* of solar activity for that year when discussing the long-term variation. The rest is derived from the observational data themselves with little freedom to allow different interpretations. As Hoyt et al. [1994] point out "if more than 5% of the days in any one year are randomly observed throughout the year, a reasonable value for the yearly mean can be found", so selection of observers is made with this in mind.

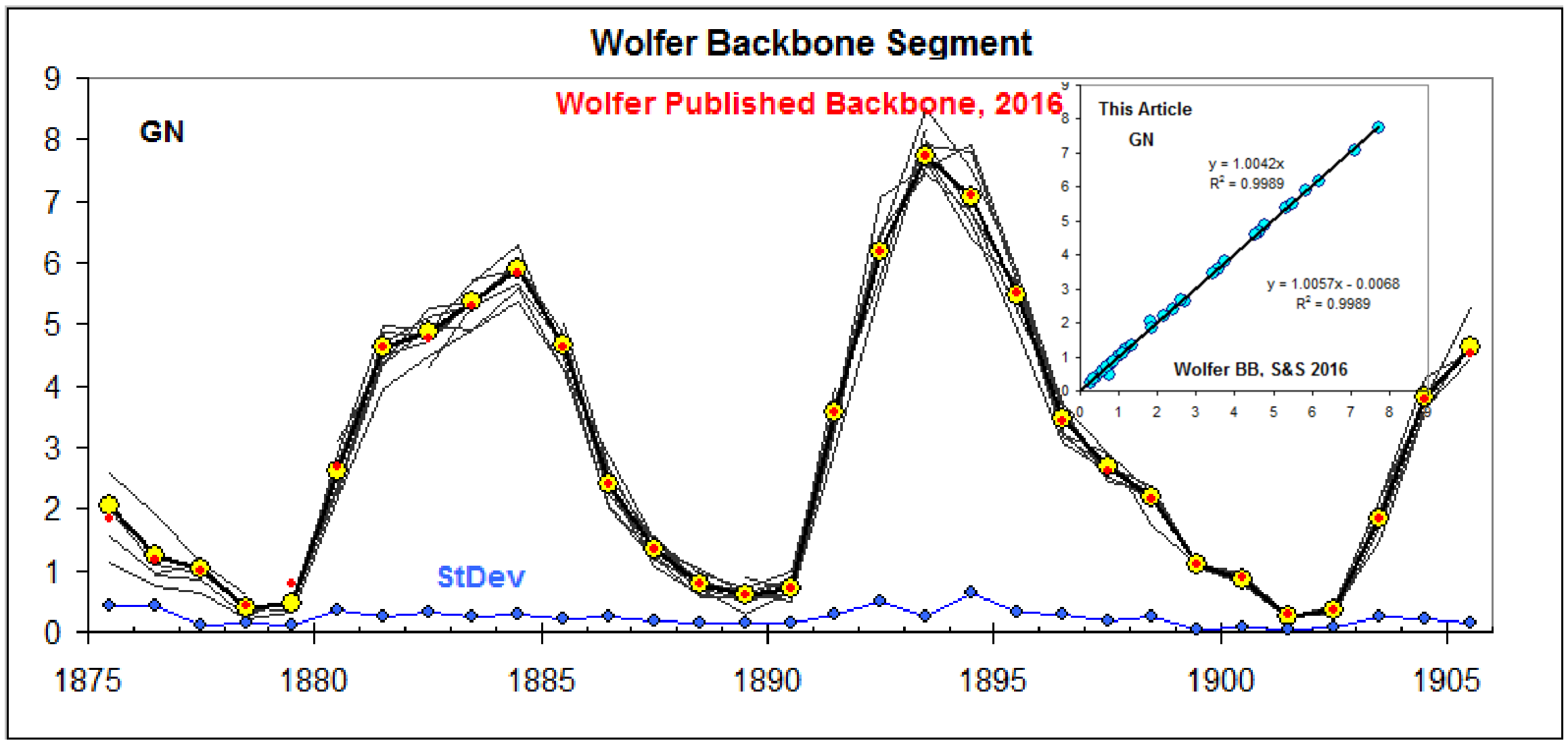

**Figure 11**. The average Wolfer Backbone segment for 1875-1905 constructed by averaging the scaled annual counts from the 11 high-quality observers we are considering. The counts by individual observers are shown as thin gray lines. The average is shown by a heavy black curve with yellow dots. The standard deviation is plotted in blue at the bottom of the graph and is on average 9% of the annual count. The published values from Svalgaard & Schatten [2016] are marked by red dots. The agreement is excellent ($R^2$ = 0.999; see insert) as we would expect from the sound and straightforward analysis.

There are a few cases where the relationship between annual counts for two observers is not quite linear. We then also fitted a power-law to the data if that significantly improved the fit, otherwise the observer was omitted. Figure 12 shows the result for the observer Shea compared to Schwabe for 1847-1864:

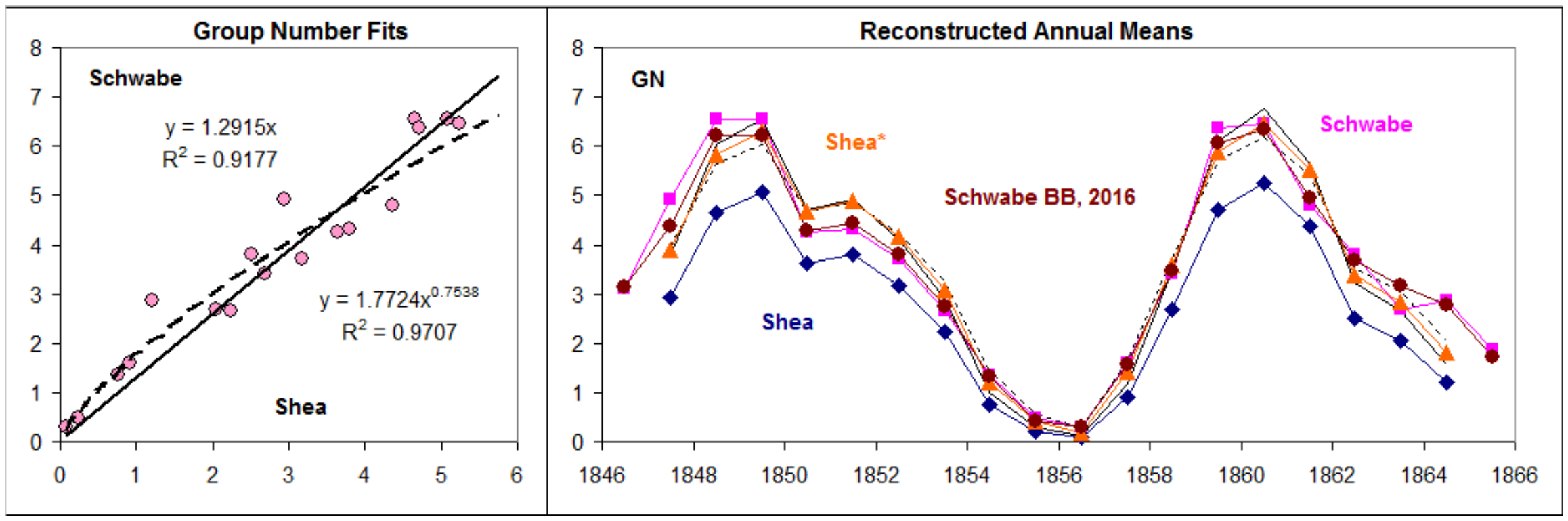

**Figure 12**. Observer Shea scaled to Schwabe, with a linear fit and with a power-law. The latter improving the fit from $R^2$ = 0.92 to $R^2$ = 0.97. We can use both fits for the reconstruction, although the results are not very different. The thin black line is from the linear fit through the origin and the dashed line is for the power-law. The full Schwabe Backbone from Svalgaard & Schatten [2016] is shown for reference (brown dots).

## 4. Non-Linear Backbones with No Daisy-Chaining

We can also construct the backbones using a linear fit with an offset, a power-law, or a 2nd-order fit, taking whichever has the best fit to the primary observer. Figure 13 shows this procedure applied to new Schwabe and Wolfer backbones.

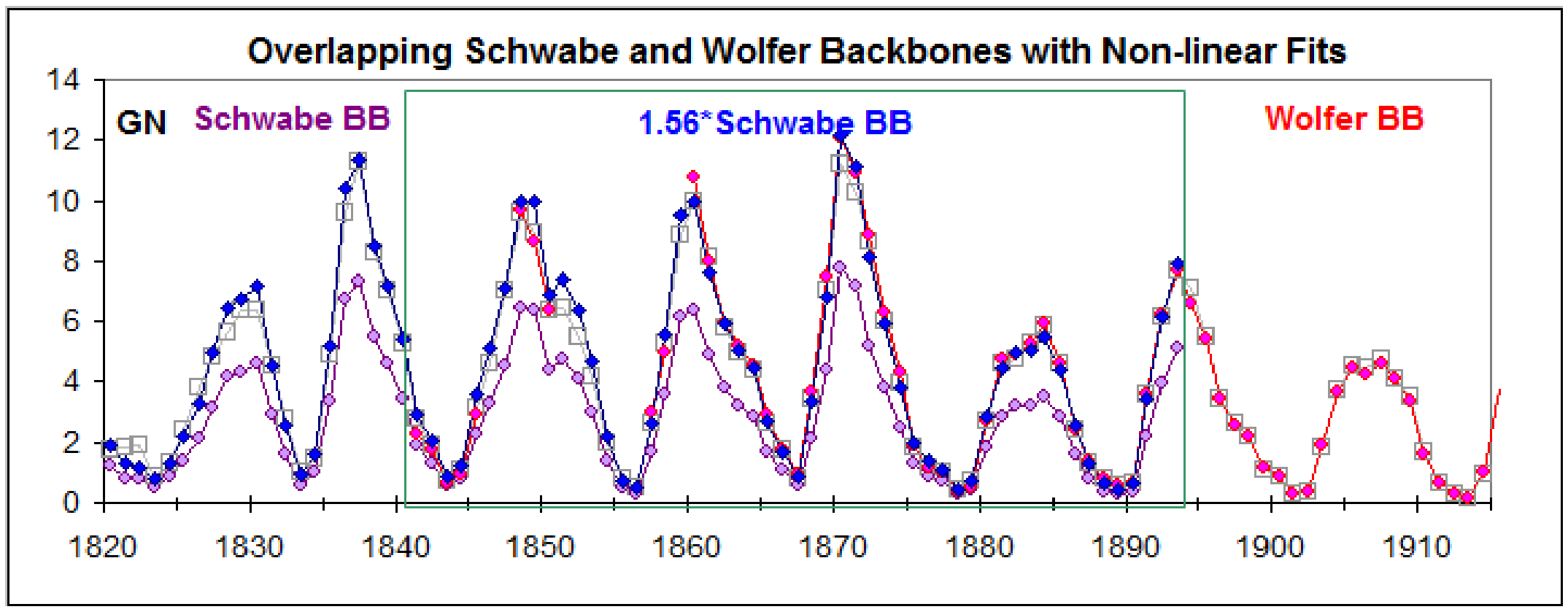

**Figure 13**. First, construct a new Wolfer Backbone (red curve) using linear fits with offsets, power-laws, or 2nd-order polynomial fits, taking whichever has the best fit to the primary observer (average of Wolfer and Tacchini, scaled to Wolfer). Then, construct a new Schwabe Backbone (purple, lower curve) the same way (primary observer Schwabe). The two backbones overlap 1841-1893 (green box) and the scale factor is 1.56±0.03 – (0.09±0.10) which, when applied, yields the (upper) blue curve, matching the red curve with $R^2$ = 0.985. For comparison, the published Svalgaard & Schatten [2016] Backbone is shown by the open grey squares.

Lockwood et al. [2016b] claim "that the factor of 1.48 used by Svalgaard & Schatten (2016) in constructing $R_{BB}$ [the Backbone Series scaling Schwabe to Wolfer] is 20% too large and should be nearer 1.2". As Figure 13 shows, the new scale factor (without invoking proportionality) is not statistically different (at the 95% level) from the 1.48±0.03 found by Svalgaard & Schatten [2016].

Lockwood et al. [2016b] further claim that "our analysis of the join between the Schwabe and Wolfer data sunspot series shows that the uncertainties in daisy-chaining calibrations are large and demonstrates how much the answer depends on which data are used to make such a join." We later in the text (Section 11) demonstrate that the two backbones are not built with daisy-chaining, but at this point we simply construct a single join-less backbone based on Gustav Spörer's observations 1861-1893 spanning the transition from Schwabe to Wolfer without assuming proportionality and also without using any daisy-chaining. That the result depends on which data is used is trivially true, but selecting high-quality observers with long records makes the backbones robust. The new Spörer backbone uses group counts from Spörer (1861-1893), Wolfer (1876-1893), Schwabe (1841-1867, derived by Arlt et al.), Weber (1860-1883), Schmidt (1841-1883), Wolf (1861-1893, small telescope), Wolf (1849-1867, large telescope), Leppig (1867, 1881), Tacchini (1871-1900), Bernaerts (1874-1878), Winkler (1882-1900), and Konkoly (1885-1900), all normalized to Spörer.

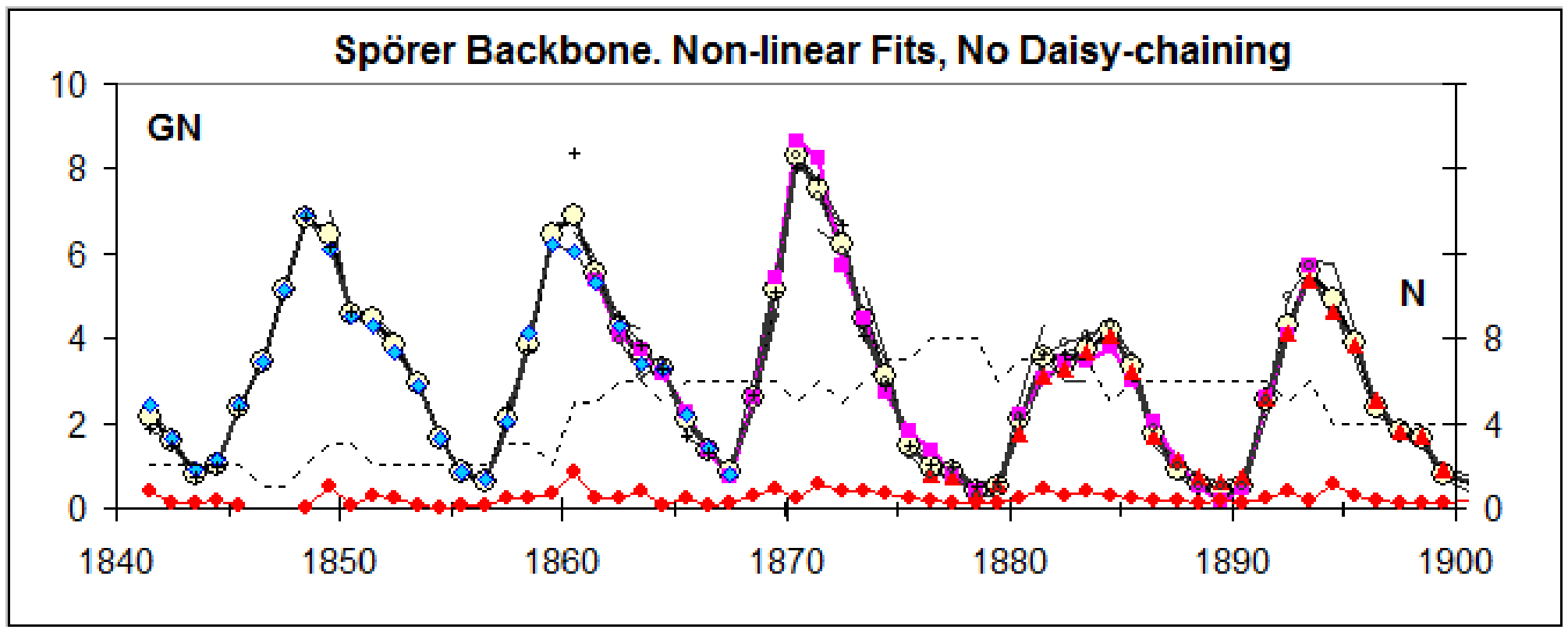

**Figure 14**. Use Spörer's observations (pink squares) as primary observations, then fit to those the group counts from observers who overlap directly with Spörer selecting the functional form of the correlation as either linear with an offset, a power-law, or a 2nd degree polynomial depending on which one provides the best fit. Some scaled data from some observers (e.g. Schwabe, blue diamonds; Wolfer, red triangles) are plotted with distinguishing symbols; the remaining ten observers are shown with thin black curves and the average backbone with large yellow dots. The number, *N*, of observers in each year is shown by the dashed line. The standard deviation is shown by the red symbols at the bottom of the Figure.

As Figure 14 demonstrates, the join-less backbone does not differ from the Wolfer series (red triangles). Formally, the ratio between the Wolfer backbone published by Svalgaard & Schatten [2016] and the new join-less Spörer backbone, shown in Figure 14, is (1.43±0.07)+(0.04±0.21) which within the errors is identical to the ratio (1.42±0.01)-(0.16±0.04) between the observers Wolfer and Spörer. So, the unfounded concern of Lockwood et al. [2016b] on this point can now be put to rest.

It should be clear that there is very little difference between the resulting annual means derived from linear fits through the origin and the non-linear fits, simply because the relationships between observers' counts are so close to simple proportionality in the first place. It is, perhaps, telling that in their invalid criticism of Svalgaard & Schatten [2016], Lockwood et al. [2016a] did not even examine a single case of comparison of two actual observers.

## 5. Group Distributions

Usoskin et al. [2016] marvel at the unlikelihood that "Wolf was missing 40 % of all groups that would have been observed by Wolfer irrespectively of the activity level". We can construct a frequency diagram of daily group counts for simultaneous observations by Wolf and Wolfer. For each bin of group counts (0, 1, 2,…, 13) observed by Wolf, the number of groups observed by Wolfer on the same days defines a series of bins (0, 1, 2,…, 15). The number of observations by Wolfer is then determined for each bin, and a contour plot of the resulting distribution is shown in Figure 15.

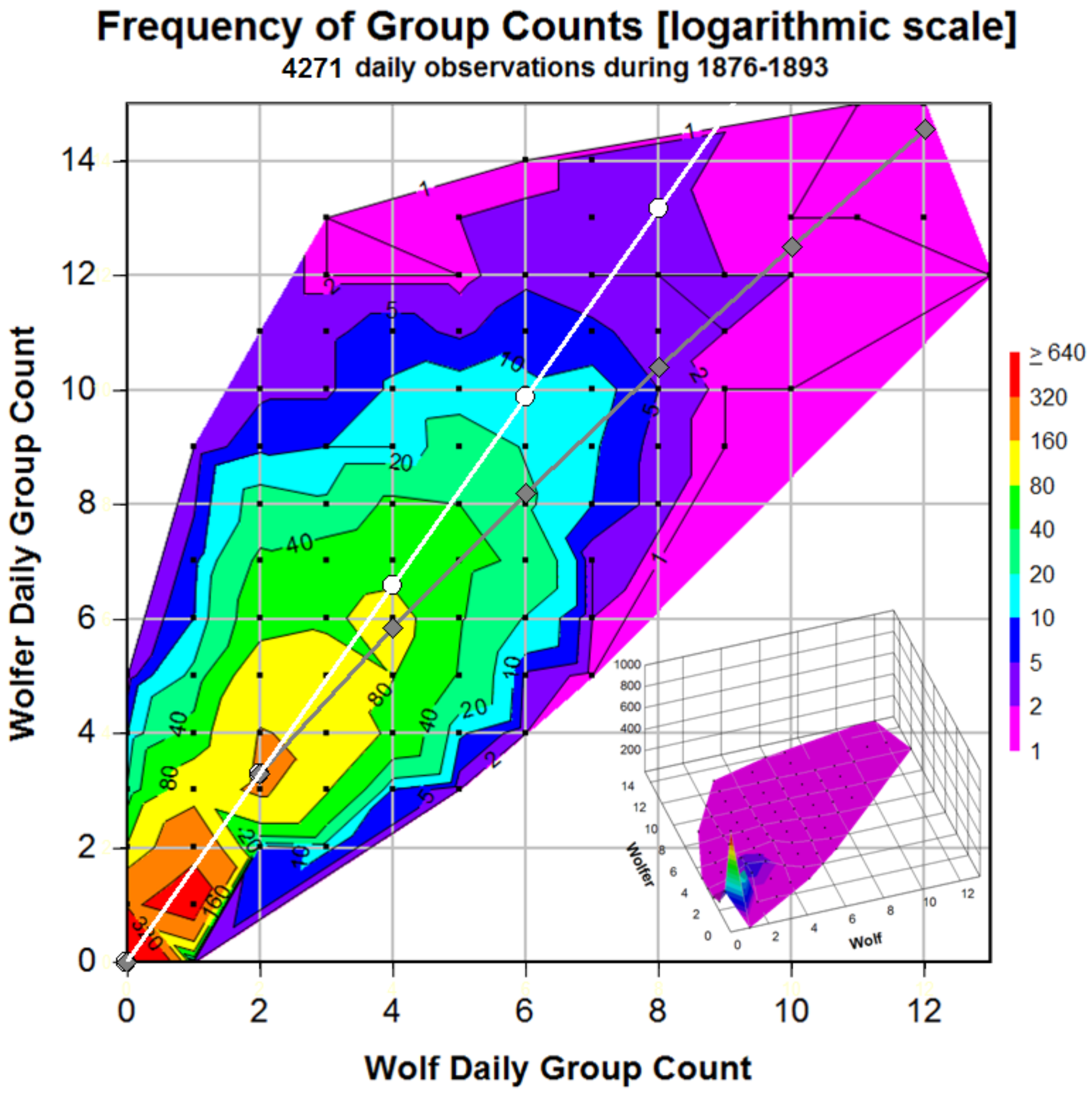

**Figure 15**. Frequency of daily group counts for simultaneous observations by Wolf and Wolfer. For each bin of group counts (0, 1, 2,…, 13) observed by Wolf, the number of groups observed by Wolfer on the same days defines a series of bins (0, 1, 2,…, 15). The number of observations by Wolfer is then determined for each bin, and a contour plot of the resulting distribution is shown in this Figure. Due to the extreme preponderance of the lower group counts (more than 80% of the counts are found in Wolf bins 0 through 3) we use a logarithmic scale (the insert shows a 3D plot of the counts with its sharp peak, 948, at (0, 0)). The white dots on the white line indicate the expected 'ridge' of the distributions corresponding to the value 1.65±0.05 found by Svalgaard & Schatten [2016] to be the ratio between the annual groups counts by Wolfer and Wolf. Gray Diamonds on the grey curve show the Usoskin et al. [2016] 'correction matrix' values (see later text).

There does not seem to be anything unlikely about the 40% mentioned by Usoskin et al. [2016]. The frequency plot is very consistent with their observation.

There are, of course, cases in the early record where there are so few observations made by some observers that the scatter overwhelms the correlation, linear or otherwise. For these observers we have to resort to computing the overall average of all the observations made by the observer, and to compare overall averages covering the years of overlap, much as Hoyt & Schatten [1998] did, exploiting the high autocorrelation (yearly correlation coefficient $R > 0.8$) in the sunspot record.

## 6. RGO Drift of Group Numbers

As to requiring "unlikely drifts in the average of the calibration *k*-factors for historic observers" [Lockwood et al., 2016b] the only requirement is that the group counts reported by the Royal Greenwich Observatory [RGO] were drifting in the early part of the RGO-record compared with the many experienced observers whose records we have used to construct the backbones. Figure 16 shows the progressive drift in evidence before about 1890.

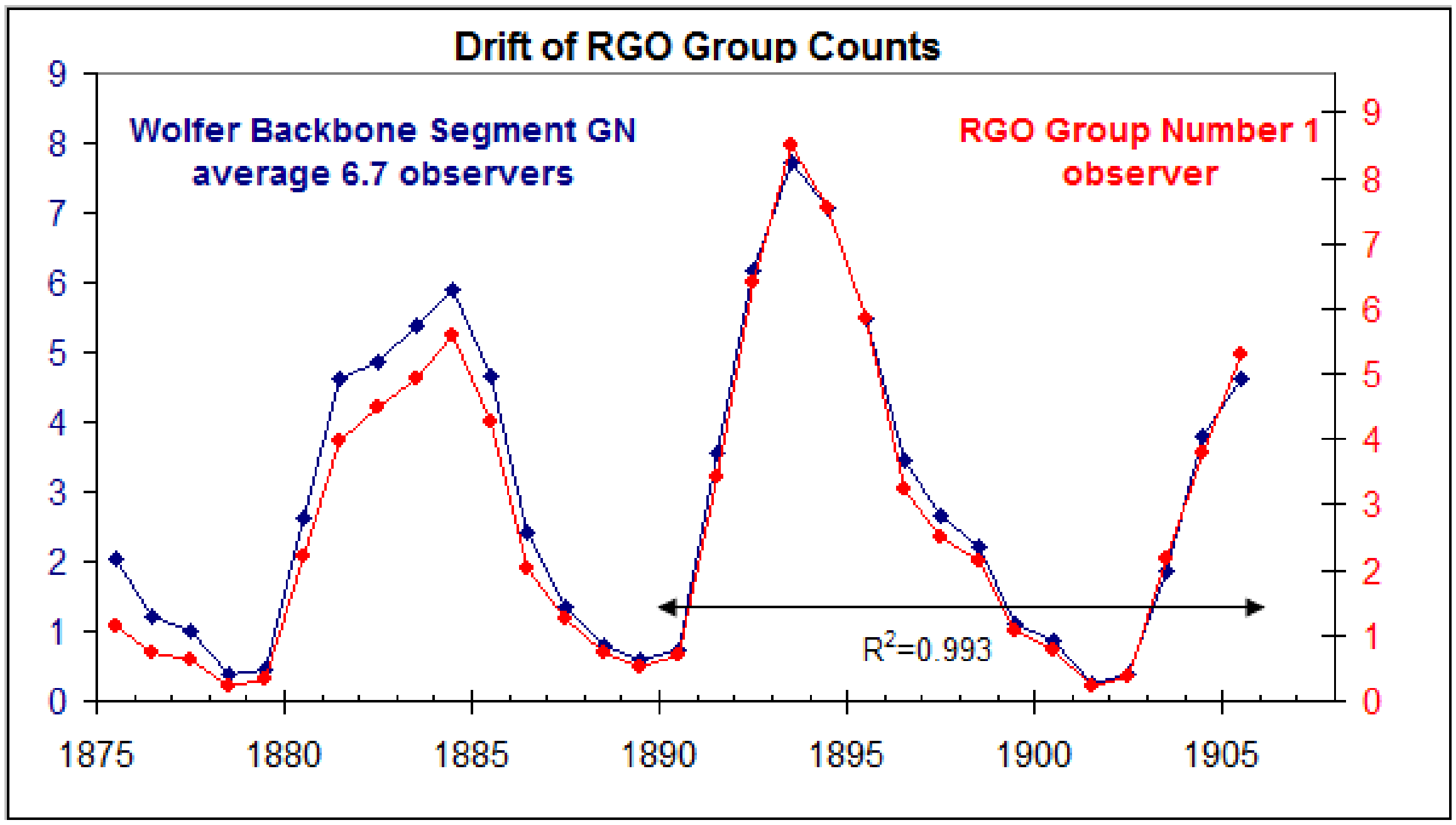

**Figure 16**. Comparing the RGO Group Count with the Wolfer Segment Backbone, after scaling the counts to agree ($R^2 = 0.993$) for 1890-1905.

Determining the areas of sunspots is a straightforward counting of dark 'pixels' on the RGO photographs using a ruled glass plate, while apportioning spots to groups can be very subjective and involves additional difficulties from 'learning curves' and personnel changes. Contrary to popular and often stated belief, counting groups is ***harder***, not easier, than counting spots[3]. We can quantify the drift [or change] in the RGO group counts by comparing the number of groups over, say, each month, with the [daily averaged] areas measured over the same month for the early record before 1890, for the interim record up to 1907, and for the later record. The relationships are weakly non-linear, Figure 17, but it is clear that there is a systematic shift ["the drift"] in the dependence from the earliest observations and forward in time.

[3] Schwabe: "Die schwierigste Aufgabe bei unsern Beobachtungen bleibt die Zählung der Gruppen"

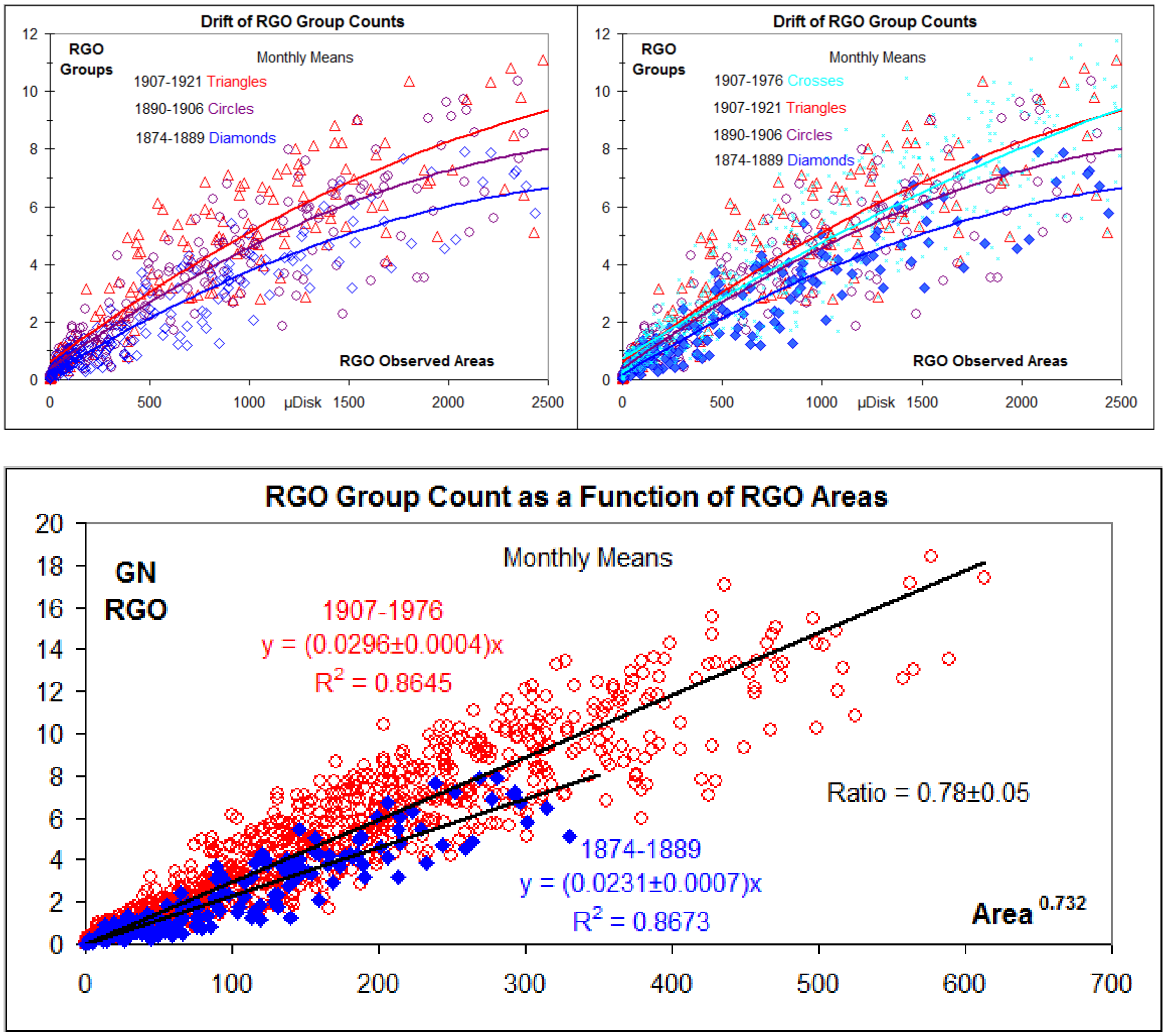

**Figure 17**. The number of groups reported by RGO for [left upper panel] the three intervals 1874-1889, 1890-1906, and 1907-1921. Second order polynomial fits show the progressive increases of the count for equal disk-averaged sunspot areas [observed, foreshortened; Balmaceda et al., 2009]. On the right upper panel we have included the whole interval from 1907 until the end of the RGO data in 1976 shown as small cyan crosses. The difference in level between all that later data and the early data [blue diamonds] is manifest. The lower panel shows the RGO group count as a function of the linearized sunspot areas for the period of the drift [1874-1889, blue diamonds] and since 1907 [red dots] when the drift had abated.

Figure 18 shows over a longer time span the drift in the ratio between counts by RGO and selected high-quality observers with long records. There may be a hint of a slight sunspot cycle variation of the ratio, but both the upper and the lower envelopes show the same drift, strongly suggesting that the drift is not due to a solar cycle variation of the ratio. The ‘drift’ is thus not “unlikely”, but rather an observational fact, likely due to human factors (learning curve; changing definition of what a ‘group’ was) instead of deficiencies in photographs or ‘pixel-counting’. Vaquero independently reached the same conclusion as reported at the second Sunspot Number Workshop in 2012: http://www.leif.org/research/SSN/Vaquero2.pdf.

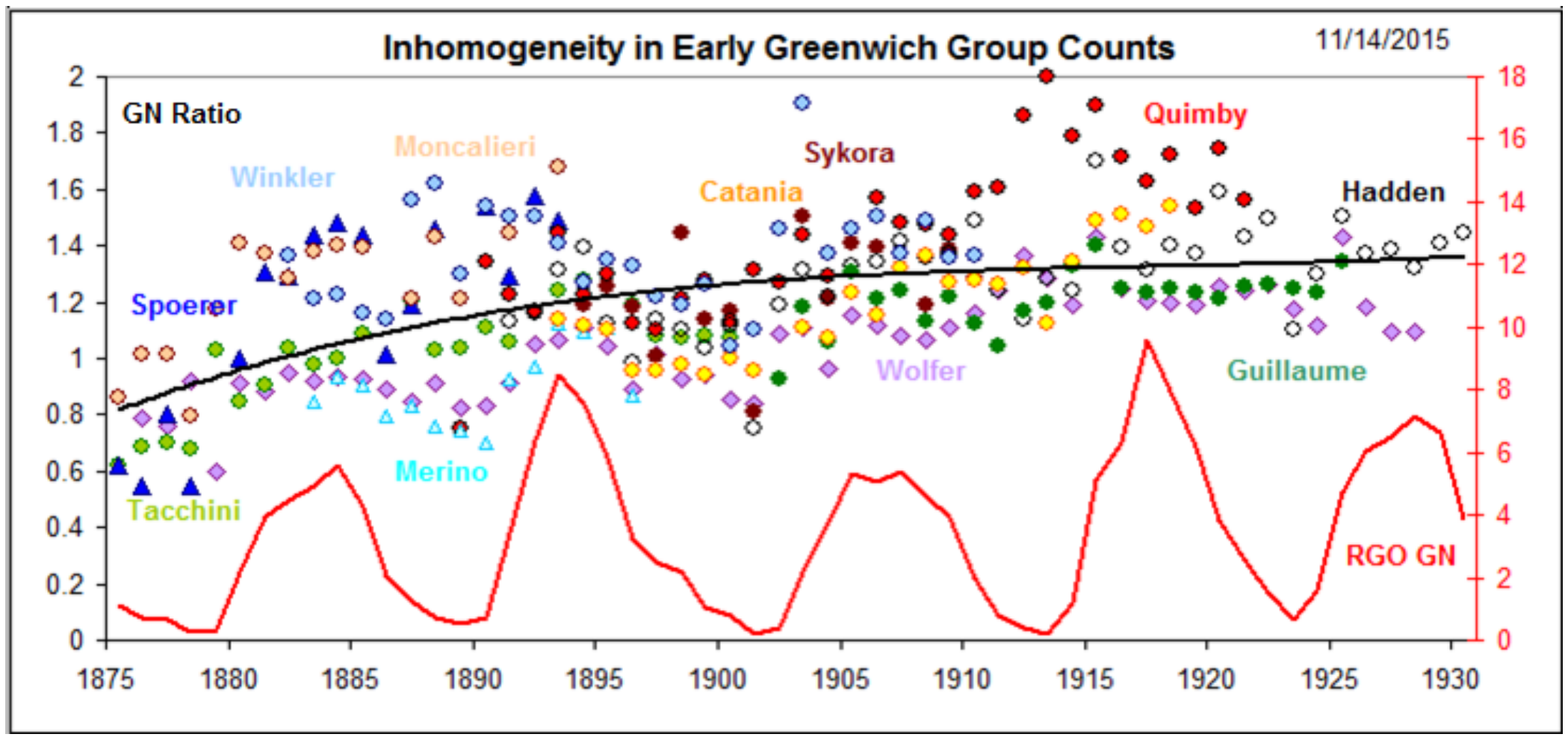

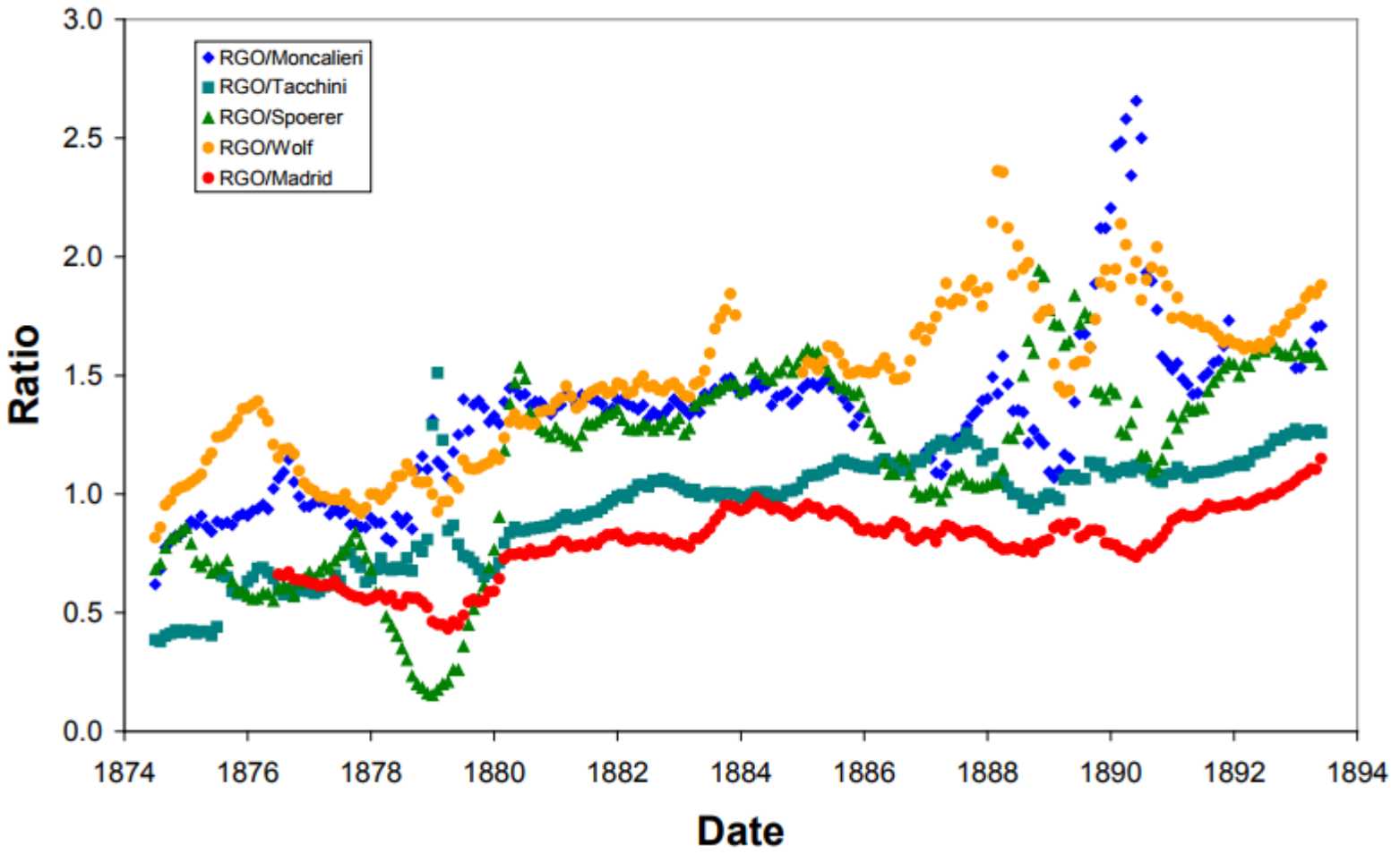

**Figure 18**. Upper panel: Ratios between annual average group counts by RGO and selected high-quality, long-serving observers. The RGO group count itself is shown by the lower (red) curve. Lower Panel: From Vaquero [2012].

## 7. Reconstruction of Open Solar Flux

Lockwood et al. [2016b] notes that "The OSF reconstruction from geomagnetic activity data is also completely independent of the sunspot data. There is one solar cycle for which this statement needs some clarification. Lockwood et al. (2013a) used the early Helsinki geomagnetic data to extend the reconstructions back to 1845, and **Svalgaard (2014) used sunspot numbers to identify a problem** with the calibration of the Helsinki data in the years 1866–1874.5 (much of solar cycle 13)."

The latter part of this claim is patently incorrect, as the problem was identified by Svalgaard comparing the purely geomagnetic indices IDV [Svalgaard & Cliver, 2005, 2010] and IHV [Svalgaard & Cliver, 2007] calculated separately from the horizontal component (*H*) and from the declination (*D*) for the Helsinki Observatory (Figure 19 from Svalgaard [2014]), and collegially communicated to Lockwood et al., prompting them to reconsider and hastily revise yet another otherwise embarrassing publication.

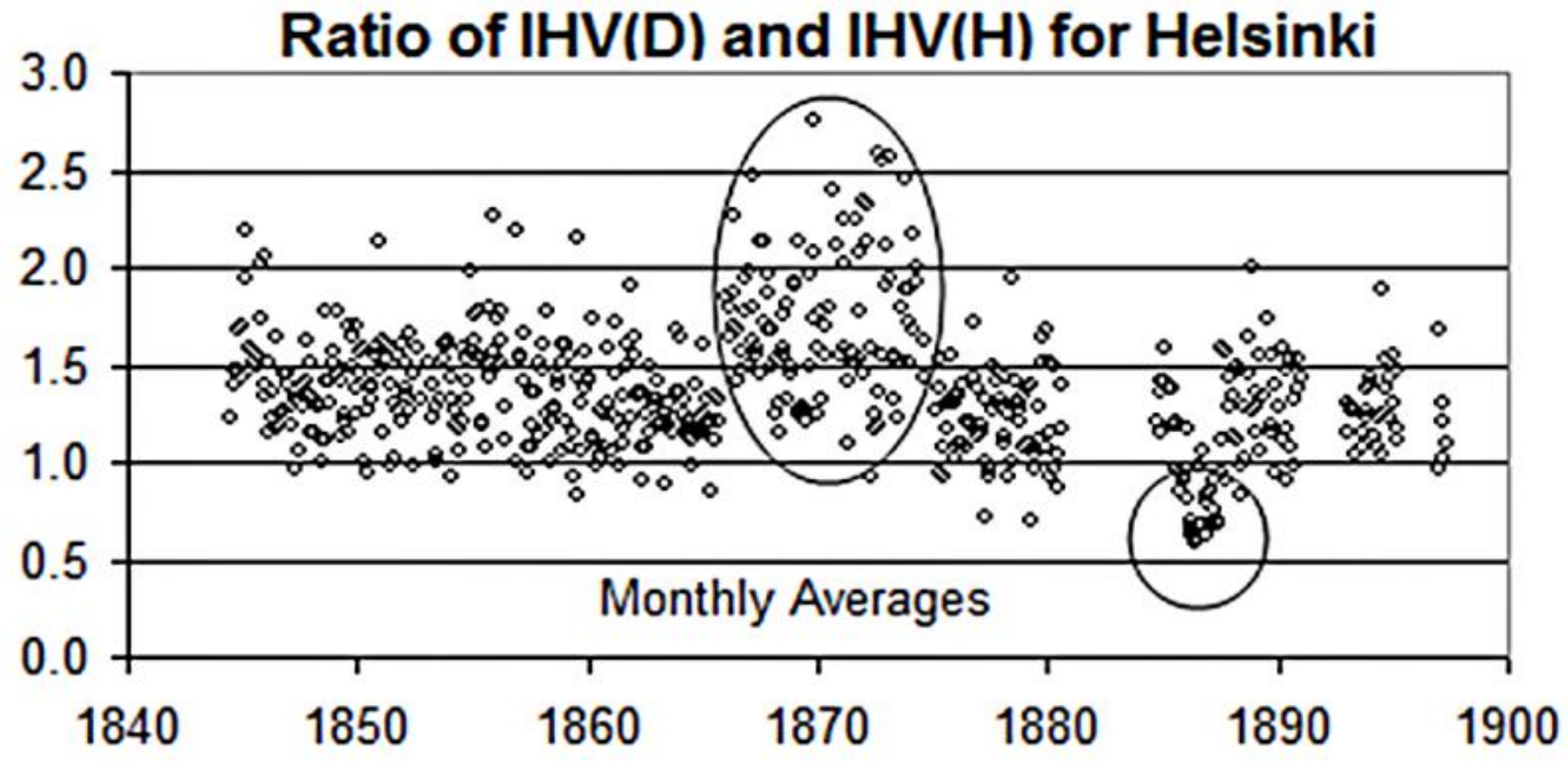

**Figure 19**. The ratio between monthly values of the IHV-index calculated using the declination, IHV(*D*), and of IHV calculated using the horizontal force, IHV(H) for Helsinki. The ovals show the effect of the scale value for *H* being too low in the interval 1866-1874.5 and of the scale value for *D* being too low for the interval 1885.8-1887.5 (From Svalgaard, 2014).

Lockwood et al. continued: "but it is important to stress that the correction of the Helsinki data for solar cycle 11 made by Lockwood et al. [2014b], and subsequently used by Lockwood et al. [2014a], was based entirely on magnetometer data and did not use sunspot numbers, thereby maintaining the independence of the two data sets." This is disingenuous because the need for correction was not discovered by Lockwood et al. [2014] but by **Svalgaard [2014] who did NOT use sunspot numbers** to identify and to quantify the discrepancy as clearly laid out in the Svalgaard [2014] article. On the other hand, the Sunspot Group Number **does** indicate precisely the same discrepancy, see Figure 20, thus actually **validating** the Group Sunspot Number for the years in question, contrary to the vacuous claim by Lockwood et al. that "The geomagnetic OSF reconstruction provides a better test of sunspot numbers than the quiet-day geomagnetic variation because the uncertainties in the long-term drift in the relationship between the two are understood" as we show in the following section.

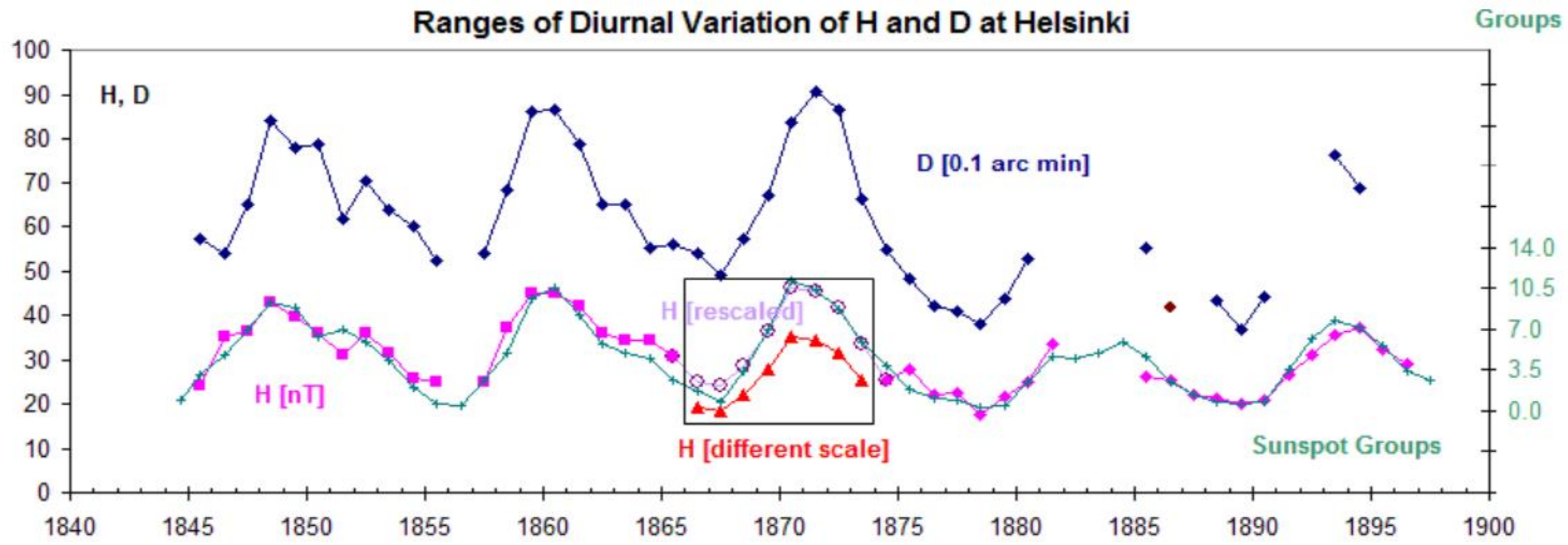

**Figure 20**. Yearly average ranges for declination, (*D* in 0.1 arc minute units), blue curve, and for horizontal force (*H* in nT units), pink curve. Because of the strong seasonal variation only years with no more than a third of the data missing are plotted.

The green curve (with "+" symbols) shows the number of active regions (sunspot groups) on the disk scaled to match the pink curve (*H*). As expected, the match is excellent, except for the interval 1866–1874 (in box), where the *H* range would have to be multiplied by 1.32 for a match as shown with purple open circles. (From Svalgaard, 2014).

## 8. Agreement with Terrestrial Proxies

The extensive analysis by Svalgaard [2016] of more than 40 million hourly values from 129 observatories covering the 176 years, 1840-2015, shows that there is a very tight and stable relationship between the annual values of the daily variation of the geomagnetic field and the Sunspot Group Number, Figure 21.

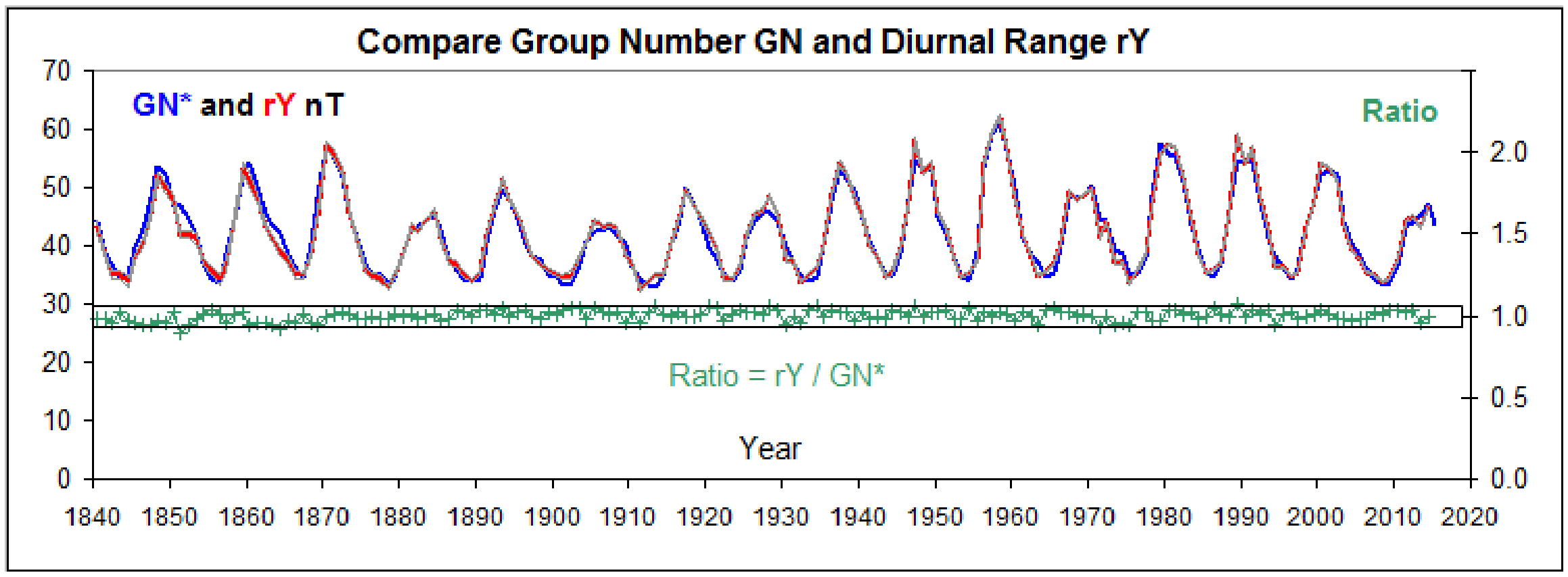

**Figure 21**. The Sunspot Group Number, GN (blue curve), scaled to match the Diurnal Range (red curve) using the regression equation GN* = 2.184 GN + 32.667, $R^2 = 0.96$. The ratio (green symbols) between the two measures is unity with a Standard Deviation of 0.03 (box). (After Svalgaard, 2016).

The OSF reconstruction is based on the geomagnetic effect of the solar wind magnetic field which indirectly does depend on the solar magnetic field and thus the sunspot number as discovered by Svalgaard at al. [2003]. The main sources of the equatorial components of the Sun's large-scale magnetic field are large active regions. If these emerge at random longitudes, their net equatorial dipole moment will scale as the square root of their number. Thus their contribution to the average Heliospheric Magnetic Field [HMF] strength will tend to increase as the square root of the sunspot number (e.g. Wang and Sheeley [2003]; Wang et al. [2005]). This is indeed what is observed [Svalgaard et al., 2003], Figure 22. We would not expect a very high correlation between HMF in the solar wind [being a point measurement] and the disk-averaged solar magnetic field, but we would expect – as observed – an approximate agreement, especially in the overall levels, see Figures 22-24.

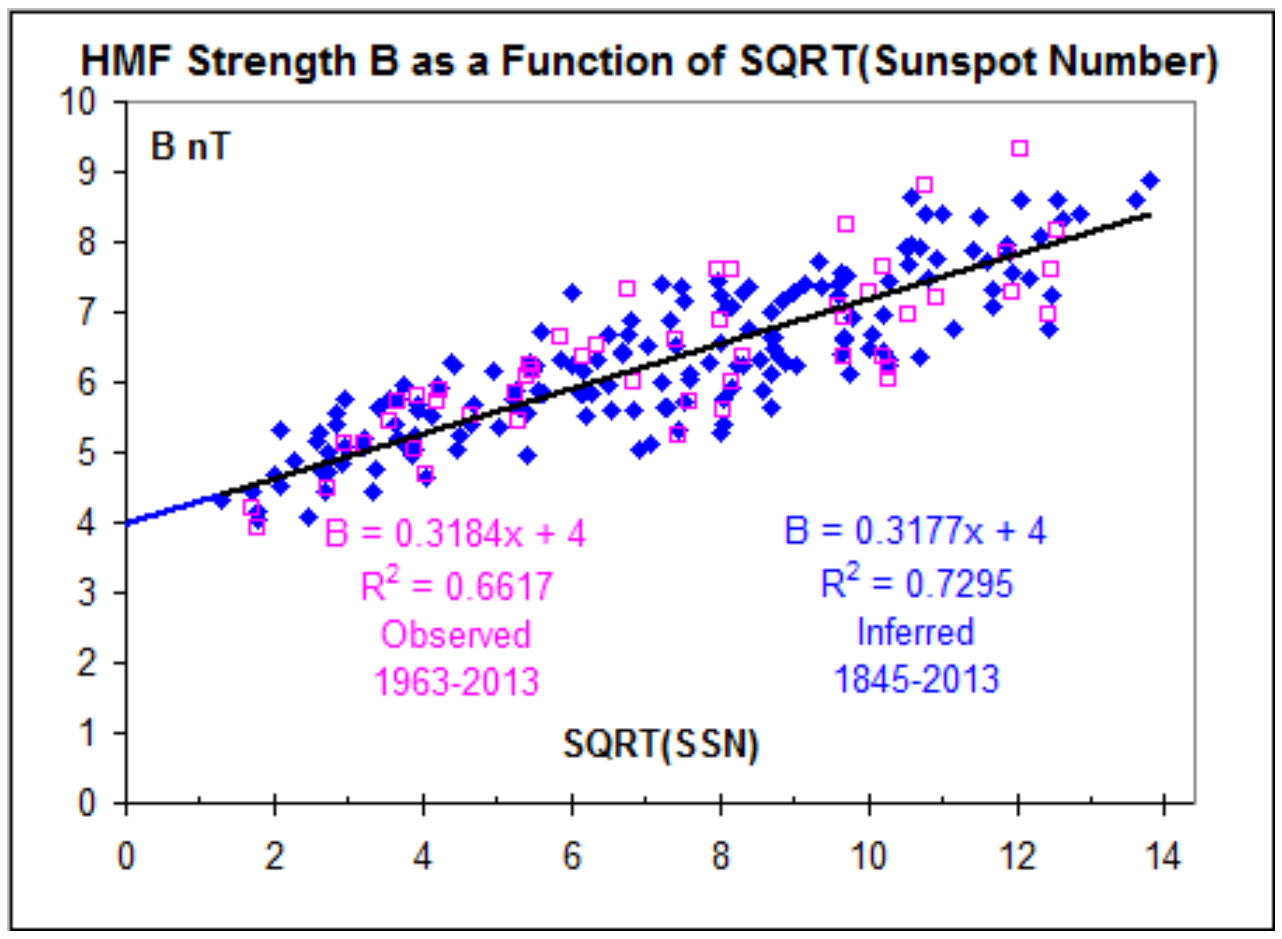

**Figure 22**. Heliospheric Magnetic Field magnitude (yearly averages) as a function of the square root of the sunspot number (the old version 1 differing mainly from the modern version 2 by a scale factor change). The observed magnitude *B* is shown as open pink squares. *B* inferred from the geomagnetic IDV-index is shown as blue diamonds. The data are consistent with a variation riding on top of a solar-activity-independent 'floor' of ≈4 nT.

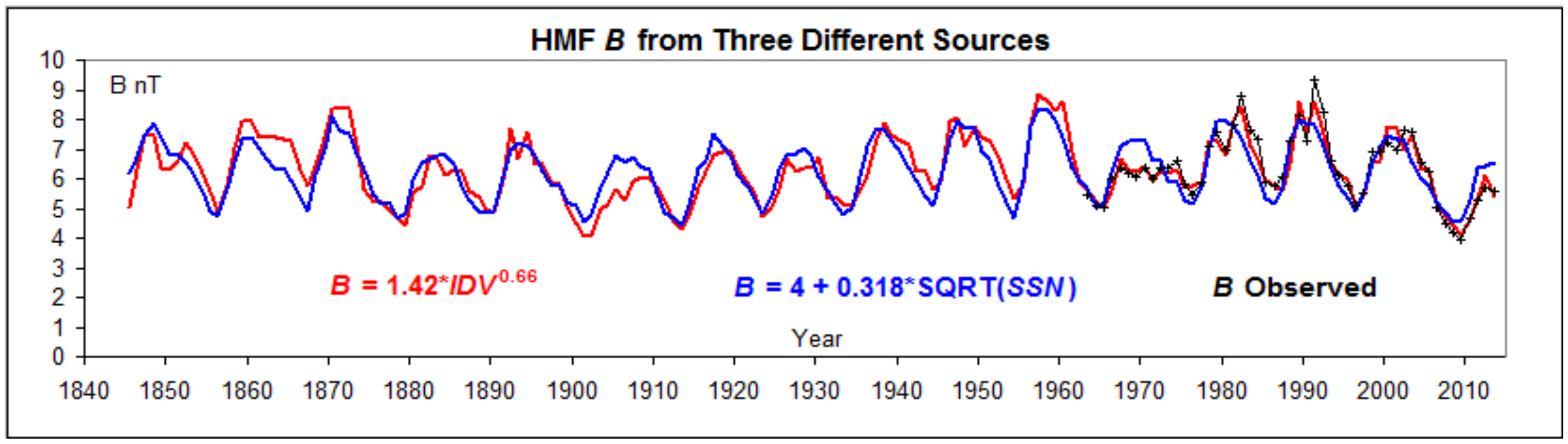

**Figure 23**. The Heliospheric Magnetic Field strength, *B*, observed (black with plus-symbols), derived from the IDV index (red) and the sunspot number (blue, version 1). There is considerable agreement (Owens et al. [2016]) about the fidelity of the IDV-derived reconstruction pioneered by Svalgaard & Cliver [2005].

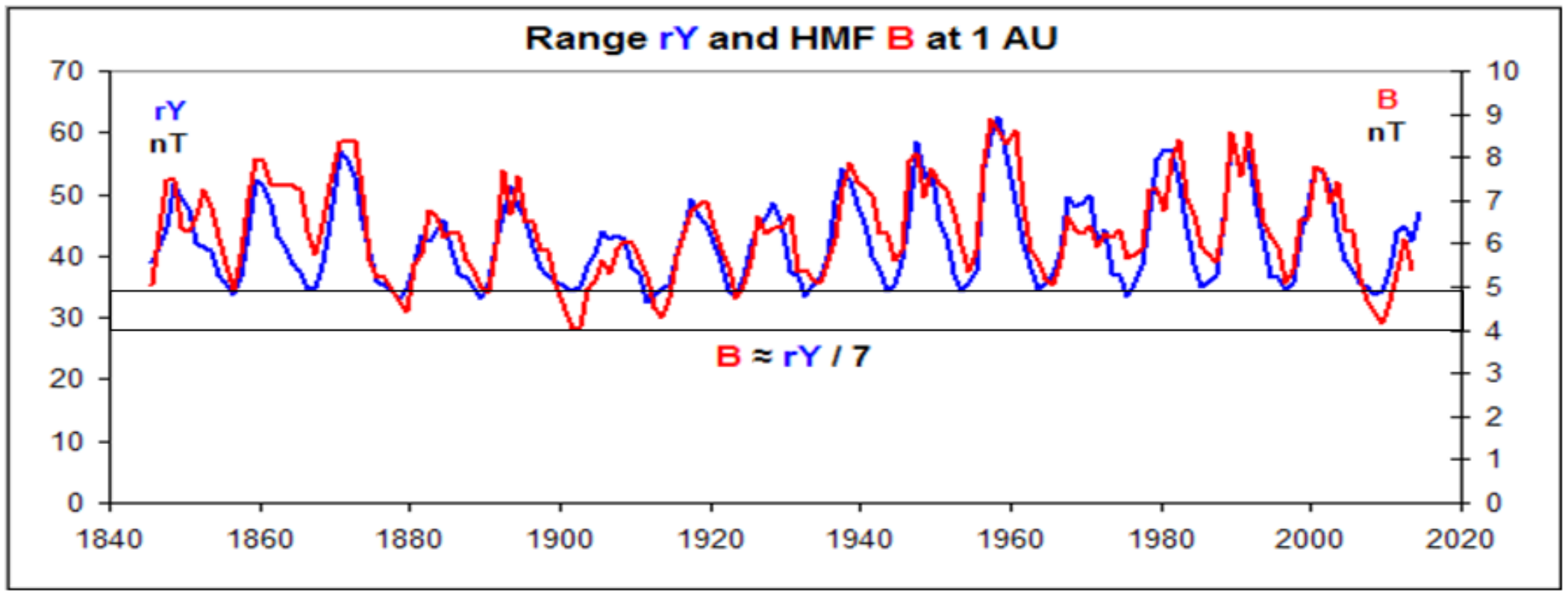

**Figure 24**. The magnetic field in the solar wind (the Heliosphere) ultimately arises from the magnetic field on the solar surface filtered through the corona, and one would expect an approximate relationship between the network field (EUV and range of the daily geomagnetic variation *rY*) and the Heliospheric field (*B*), as observed.

For both parameters (*B* and *rY*) we see that there is a constant 'floor' upon which the magnetic flux 'rides'. We see no good reason and have no good evidence that the same floor should not be present at all times, even during a Grand Minimum [see also Schrijver et al., 2011].

On the other hand, Lockwood et al. [2016b] are correct that the HMF [the base for their Open Solar Flux, OSF] is a useful indicator of the general level of solar magnetism, validating the conclusion that there is no significant trend in solar activity since at least the 1840s. It is pleasing and underscores the self-correcting nature of science to see that Lockwood now after more than a decade of struggle has finally approached and nearly matched the findings of Svalgaard & Cliver of so long ago [Svalgaard & Cliver, 2005, 2007; Owens et al., 2016], so congratulations are in order for that achievement. This is real progress. What is needed now is to build on that secure foundation laid by Svalgaard et al. [2003].

## 9. The 'Correction Matrix'

Lockwood et al. [2016b] also laments "that the practice of assuming proportionality, and sometimes even linearity, between data series (and hence using ratios of sunspot numbers) is also a cause of serious error, Usoskin et al. [2016]." As we have just shown, this not the case, as proportionality on annual time scales is not an assumption, but an observational fact. Further in Usoskin et al. [2016] they stress that "a proper comparison of the two observers is crucially important". We agree completely, but then Usoskin et al. [2016] go on to mar their paper by tendentious verbiage, such as "for a day with 10 groups reported by Wolf, the Svalgaard & Schatten, [2016] scaling would imply 16-17 groups reported by Wolfer. But Wolfer never reported more than 13 groups for [the total of four!] days with $G_{\text{Wolf}} = 10$. It is therefore clear that the results […] contradict the data". Ignoring, that for three days with $G_{\text{Wolf}} = 7$, Wolfer reported 14 groups, much more than the proportional scaling would imply.

Basing sweeping statements ("The scaling by Svalgaard & Schatten [2016] introduces very large errors at high levels of solar activity, causing a moderate [*sic*] level to appear high. This is the primary reason of high solar cycles claimed by Svalgaard & Schatten [2016] and Clette et al. [2014] in the eighteenth and nineteenth centuries") on less than a handful of cases is bad science that should have been caught during the reviewing process of the Usoskin et al. article.

Usoskin et al. [2016] advocate that "corrections must be applied to daily values […] and only after that, can the corrected values be averaged to monthly and yearly resolution". We address this issue now by first computing the 'correction matrix' for Wolf-to-Wolfer, see Table 1 and Figure 25:

| **Wolf** | **Wolfer** | N | **Wolf** | **Wolfer** | N |
|---|---|---|---|---|---|
| 0 | 0.42 | 1350 | 6 | 7.94 | 127 |
| 1 | 1.92 | 922 | 7 | 9.64 | 53 |
| 2 | 3.60 | 607 | 8 | 9.88 | 16 |
| 3 | 4.99 | 513 | 9 | 10.83 | 6 |
| 4 | 6.05 | 391 | 10 | 11.75 | 4 |
| 5 | 7.05 | 277 | 11.8 | 13.60 | 5 |
| 6 | 7.94 | 127 | 11-13 | | |

**Table 1**. Number of groups reported by Wolfer (columns 2 and 5) for each echelon of groups reported by Wolf (columns 1 and 4) for the common years 1876-1893. The number N of simultaneous observations [on same days] is given in columns 3 and 6. This is [almost] identical to the Wolf-to-Wolfer 'correction matrix' of Usoskin et al. [2016]. For ex.: they have $G_{\text{Wolfer}} = 7.12$ for $G_{\text{Wolf}} = 5$ versus our 7.05. The reason for the small (and insignificant) discrepancies is not clear, but may be related to slightly different quality-assurance procedures in the digitization and selection of the original data. The bins 11-13 have been combined into one bin.

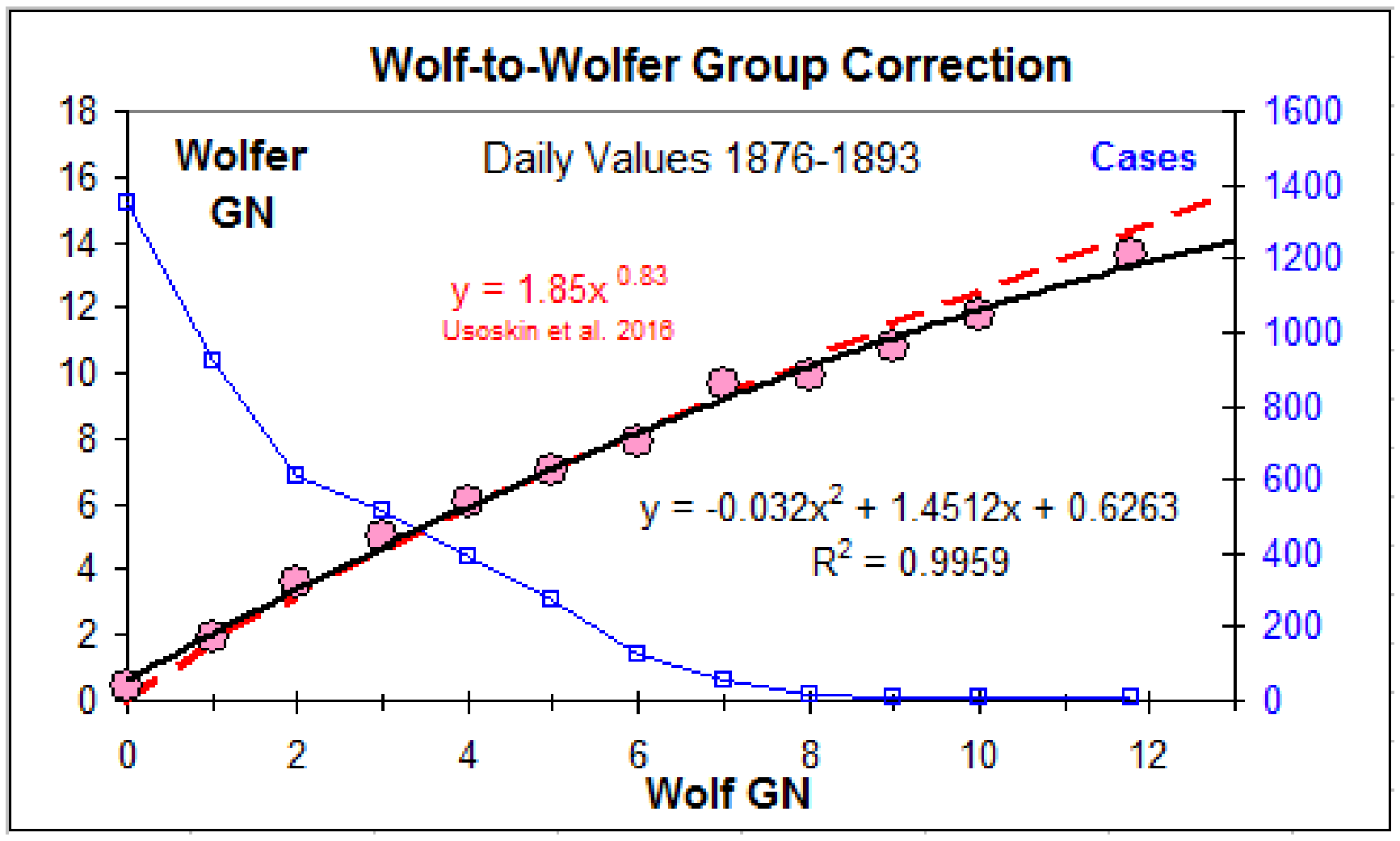

**Figure 25**. The average group counts for Wolfer as a function of the group count by Wolf (pink dots) and their 2nd degree fit (black curve). The blue curve [with open squares] shows the number of observations in each bin. The power-law through the origin (red dashed curve) is the 'correction matrix' determined by Usoskin et al. [2016].

We then follow Usoskin et al. [2016]'s admonition that "corrections must be applied to daily values and that only after that, can the corrected values be averaged to monthly and yearly resolution". Figure 26 shows the result of correcting daily values for the six months centered on the solar cycle maximum in 1884.0. We note that contrary to the baseless assertion by Usoskin et al. [2016] that "the scaling by Svalgaard & Schatten [2016] introduces very large errors at high levels of solar activity, causing a moderate level to appear high", it is the Usoskin et al. [2016] scaling that causes high levels of solar activity to appear artificially **lower** than observations indicate. This is also borne out by the data when the daily-corrected counts are averaged to monthly resolution, Figure 27. The Usoskin et al. [2016] scaling is too high for low activity (boxes (a)), and too low for high activity (boxes (c)) and only by mathematical necessity correct for medium activity (boxes (b)).

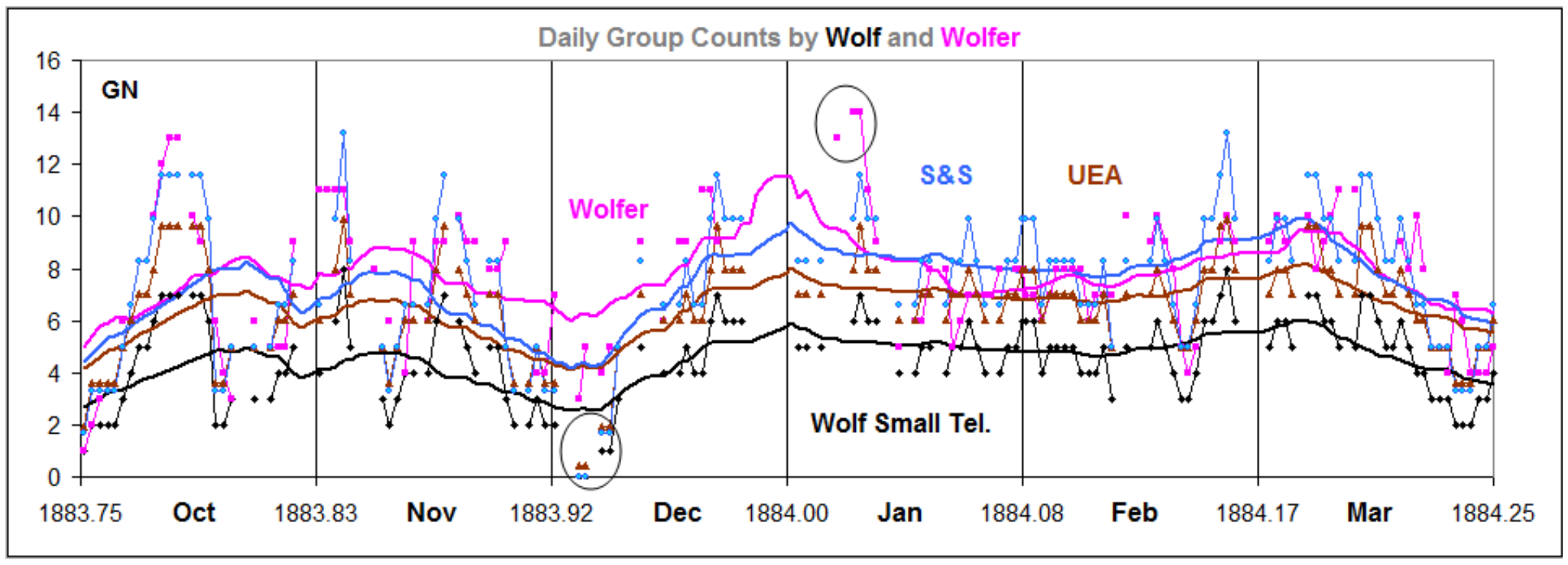

**Figure 26**. The daily group counts for Wolfer (pink curve with squares; six-month mean value = 7.70±0.25), for Wolf (black curve with diamonds, mean 4.52±0.14), and 'corrected' using the Usoskin et al. [2016] method (brown curve with triangles; too small with mean 6.50±0.17). The blue curve with dots (mean 7.46±0.24) shows the harmonized values using the Svalgaard & Schatten [2016] straightforward method, clearly matching the observational data within the errors. Heavy curves are 27-day running means. A few, sparse outlying points (in ovals) unduly influence the running means.

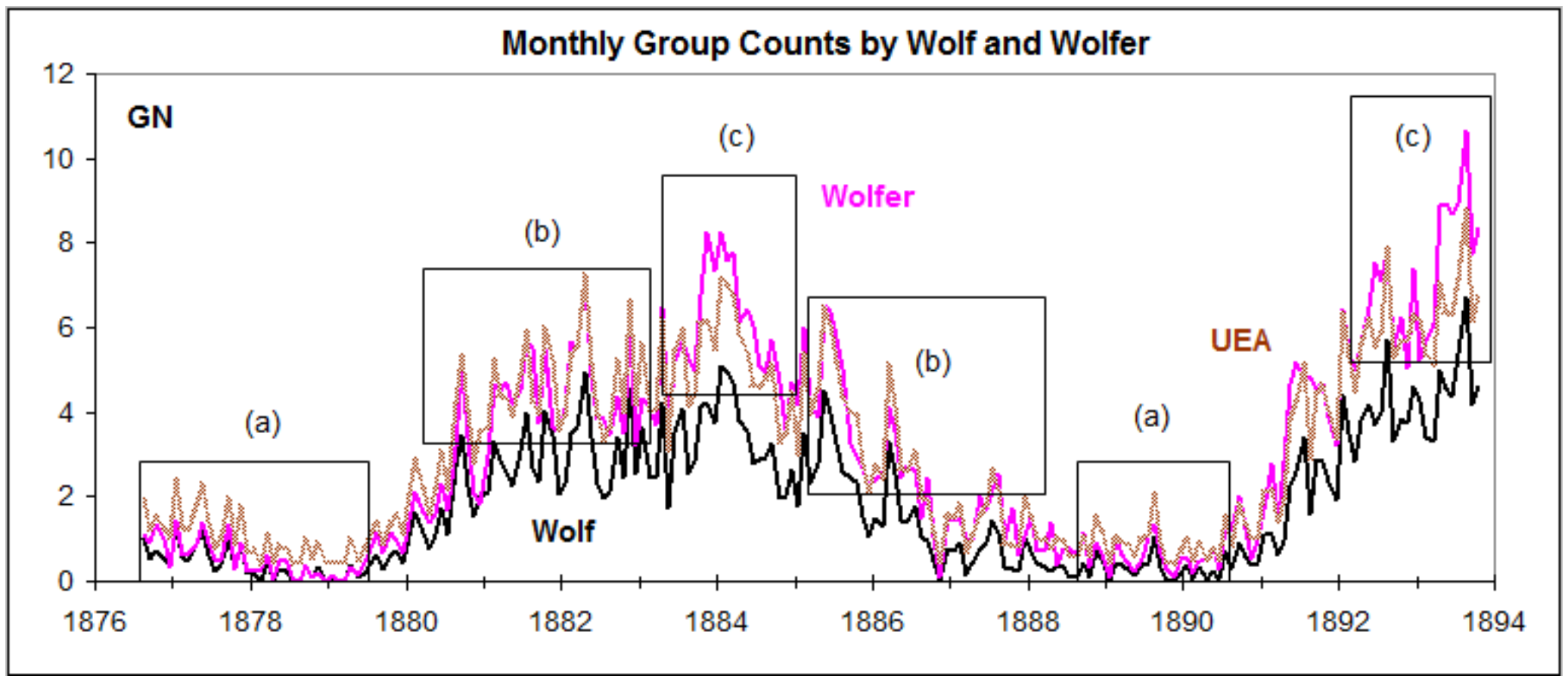

**Figure 27**. The monthly-averaged group counts for Wolfer (upper, pink), for Wolf (lower, black) and computed from the daily values 'corrected' following Usoskin et al. [2016] (UEA, brown). For low activity, the UEA values are too high (see boxes (a)). For high activity, the UEA values are too low (boxes (c)).

For people who have difficulty seeing this, we offer Figure 28 that shows for each month of simultaneous observations by Wolf and Wolfer (covering the years 1876-1893) the observed average Wolfer group counts versus the corresponding average Wolf counts (blue diamonds). A simple linear fit through the origin (blue line) is a good representation of the relationship. The pink open squares and the 2$^{nd}$-order fit to those data points show the monthly values computed using the Usoskin et al. [2016] 'correction factors' applied to daily values. It is clear that those values result in reconstructed Wolfer counts that are too small for activity higher than 3 groups (by Wolf's count) and too large for activity lower than 3 groups, contrary to the claims by Usoskin et al. [2016] and by Lockwood et al. [2016b] that the Svalgaard & Schatten's [2016] reconstructions (that so closely match

Wolfer's counts) are "seriously in error" and that the too low Usoskin et al. [2016] values are correct and preferable.

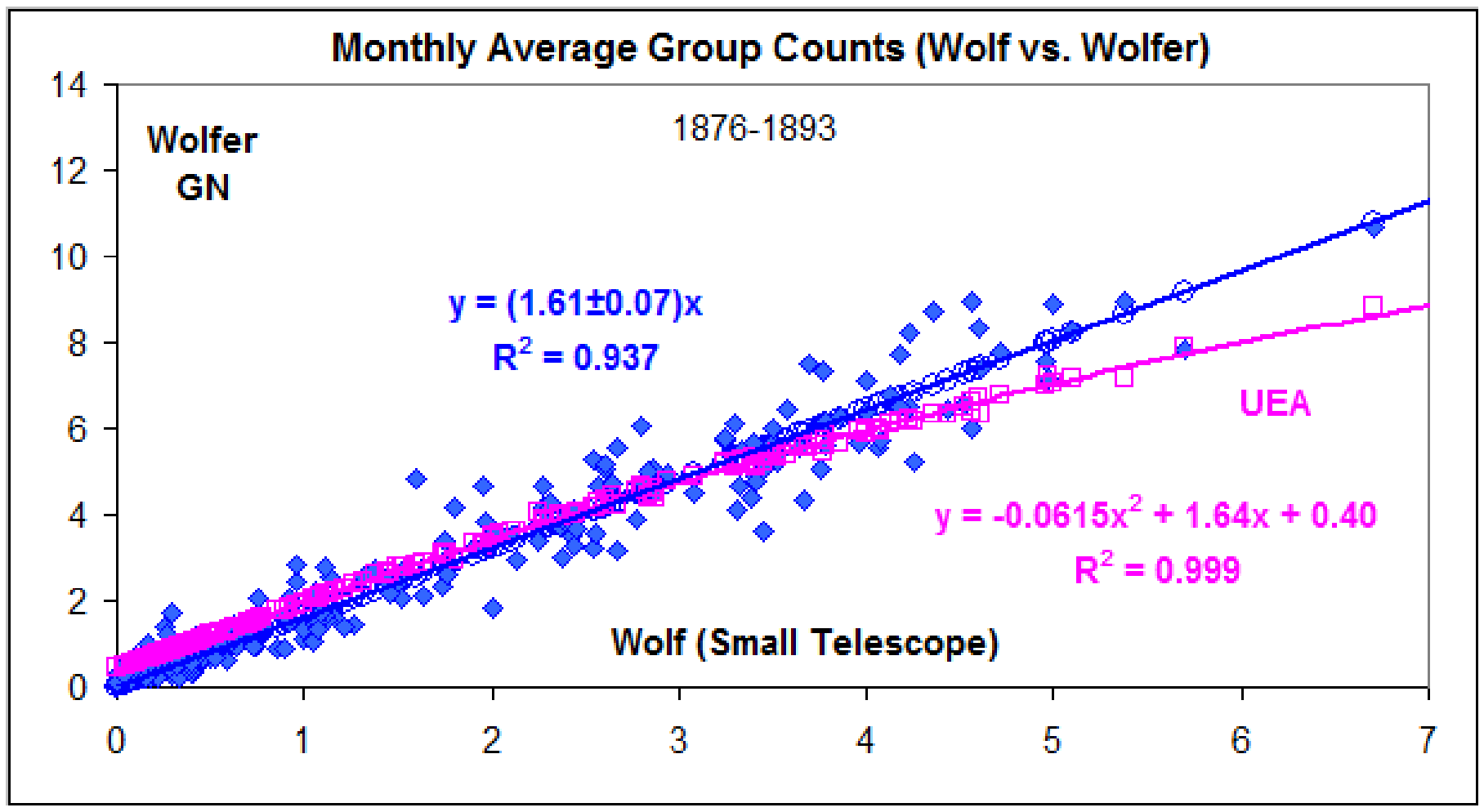

**Figure 28**. The monthly-averaged group counts for Wolfer compared to the corresponding counts by Wolf (blue diamonds) for the same months. The pink open squares (and their 2nd-order fit) show the values computed by averaging the daily counts by Wolf after applying the Usoskin et al. [2016] 'correction' method. They are clearly not a good fit to the actual data, thus invalidating the rationale for using them.

Since the Svalgaard & Schatten [2016] reconstruction is based on annual values it is critical to compare annual values. We do this in Figure 29, from which it is clear that the persistent claim that the Svalgaard & Schatten [2016] Backbone Reconstructions are "seriously in error for high solar activity" and that this is the "primary reason of high solar cycles claimed by Svalgaard & Schatten [2016] and Clette et al. [2014] in the eighteenth and nineteenth centuries" has no basis in reality and is without merit.

But why is it that the eminently reasonable procedure of constructing the monthly and annual values by averaging corrected daily values seems to fail? During minimum there really are no spots and groups for months on end, regardless of telescope used and the observer acuity, so for days with no groups reported, we should not 'correct' those zeros to 0.42 groups [as per Table 1]. For moderate activity there is no problem, but for the (rarer) high activity there must be enough differences in the distributions to make a difference in the averages or is it simply just mathematical compensation for the values that are too high for low activity. At any rate, the Backbone Reconstructions match the observations which must remain the real arbiter of success.

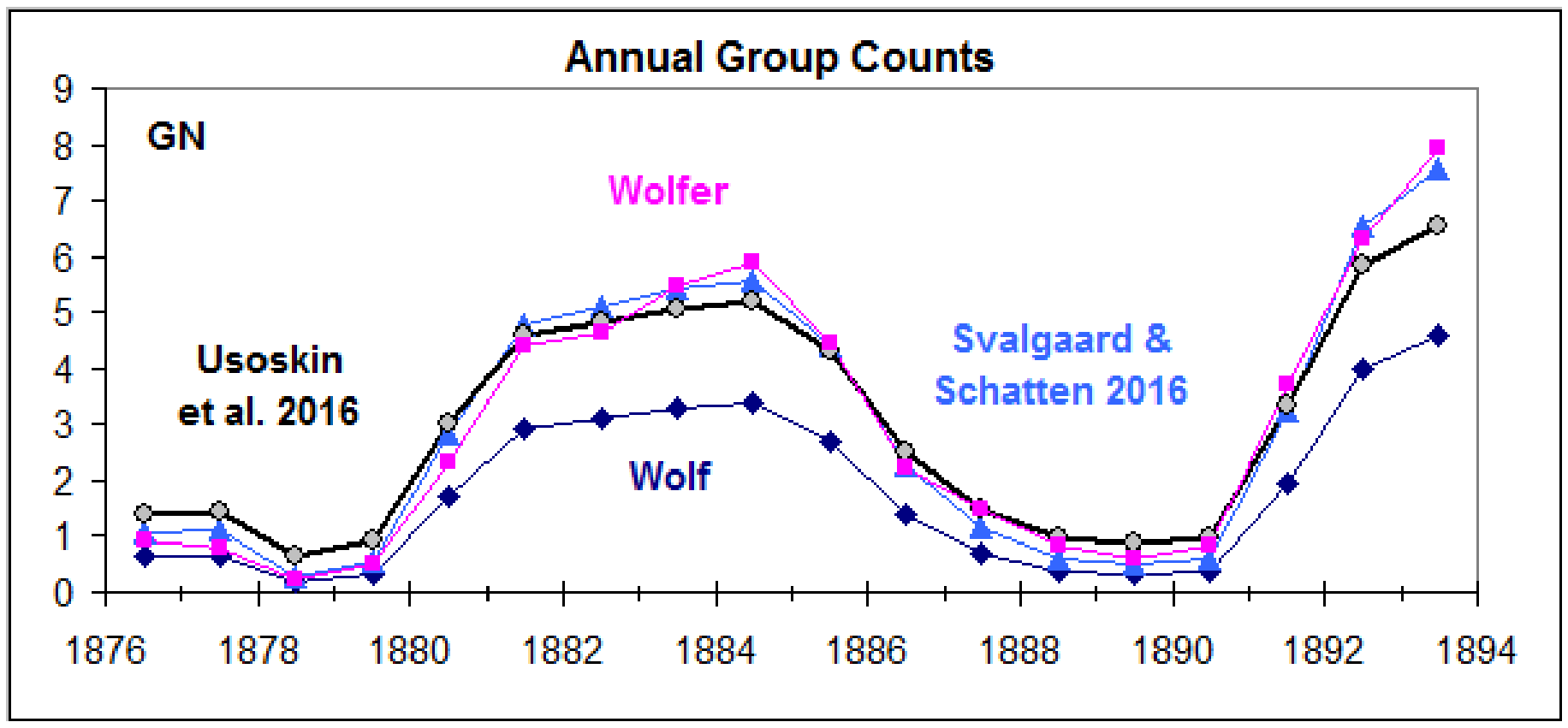

**Figure 29**. The annual group counts for Wolf (dark blue diamonds) compared to the corresponding counts by Wolfer (pink squares) for the same years. The light blue triangles show the Wolf values scaled by Svalgaard & Schatten [2016]. The gray dots on the black curve show the values computed by averaging the daily counts by Wolf after applying the Usoskin et al. [2016] 'correction' method. They are clearly not the "optimum" (used 11 times by Lockwood et al. [2016b]) fit to the actual data. In particular, they are too small for high solar activity.

The goal of normalizing or harmonizing an observer to another observer is to reduce the series by one observer to a series that is as close as possible to the other observer for the time interval of overlap. This is the principle we have followed when applying the observed proportional scaling factors. As almost all depictions of solar activity over time show annual averages, it is important to get them right. Usoskin et al. [2016] describe how their use of weighted averages is not optimal as the number of observations may vary strongly from month to month. In Svalgaard & Schatten [2016] we first compute monthly values from directly observed daily values, and only then compute the annual simple average from the available (unweighted) monthly values in order to avoid the unwanted distortions caused by an uneven distribution of observations.

## 10. On Smoothing

The Usoskin et al. [2016] article abounds with misrepresentations, perhaps designed to sow general FUD ( https://en.wikipedia.org/wiki/Fear,_uncertainty_and_doubt ) about the revisions of the sunspot series. E.g. it is claimed that Svalgaard & Schatten [2016] used "heavily smoothed data". Quantitative correlations and significance tests between heavily smoothed data are, indeed, suspect, but presumably Usoskin et al. should know that smoothing is a process that replaces each point in a series of signals with a suitable average of a number of adjacent points, which is not what computing a yearly average does. A measure of solar activity in a given year can reasonably be defined as the total number of groups (or other solar phenomena) observed during that year (taking into account the number of days with observations) and this measure was, indeed, what Schwabe [1844] used when discovering the sunspot cycle. Since tropical years have constant lengths (365.24217 days), the simple daily average (= total / number of days) over the year is then an equivalent measure of the yearly total, and does not constitute a "heavily smoothed" data point.

## 11. On Daisy-Chaining

Similarly, great importance is assigned to the deleterious effect of "daisy-chaining" as a means to discredit the Backbone Method by Svalgaard & Schatten [2016]. To wit: Usoskin et al. [2016] utter the sacred mantra "daisy-chaining" 11 times, while Lockwood et al. [2016b] use it a whopping 29 times. Lockwood et al. [2016b] usefully describe daisy-chaining as follows: "if proportionality ($k$-factors) is assumed and intercalibration of observer numbers $i$ and ($i$+1) in the data composite yields $k_i/k_{i+1} = f_i^{\,i+1}$, then daisy-chaining means that the first ($i$ = 1) and last ($i = n$) observers' $k$-factors are related by $k_1 = k_n\,\Pi^n_{i=1}(f_i^{\,i+1})$, hence daisy-chaining means that all sunspot and sunspot group numbers, relative to modern values, are influenced by all the intercalibrations between data subsets at subsequent times", as shown in panel (a) of Figure 30. We note that $n$ has to be at least 3 for true daisy-chaining to occur as there must be at least 1 intermediate observer.

But this is not how the backbones are constructed. All observers in a given backbone are only compared to exactly **one** other observer [the same primary observer], so there is no 'chain' from a first to a last observer through an number of intermediate observers and therefore no accumulation of errors along the [non-existent] chain. Figure 30 illustrates when and how daisy-chaining occurs (see caption for detail).

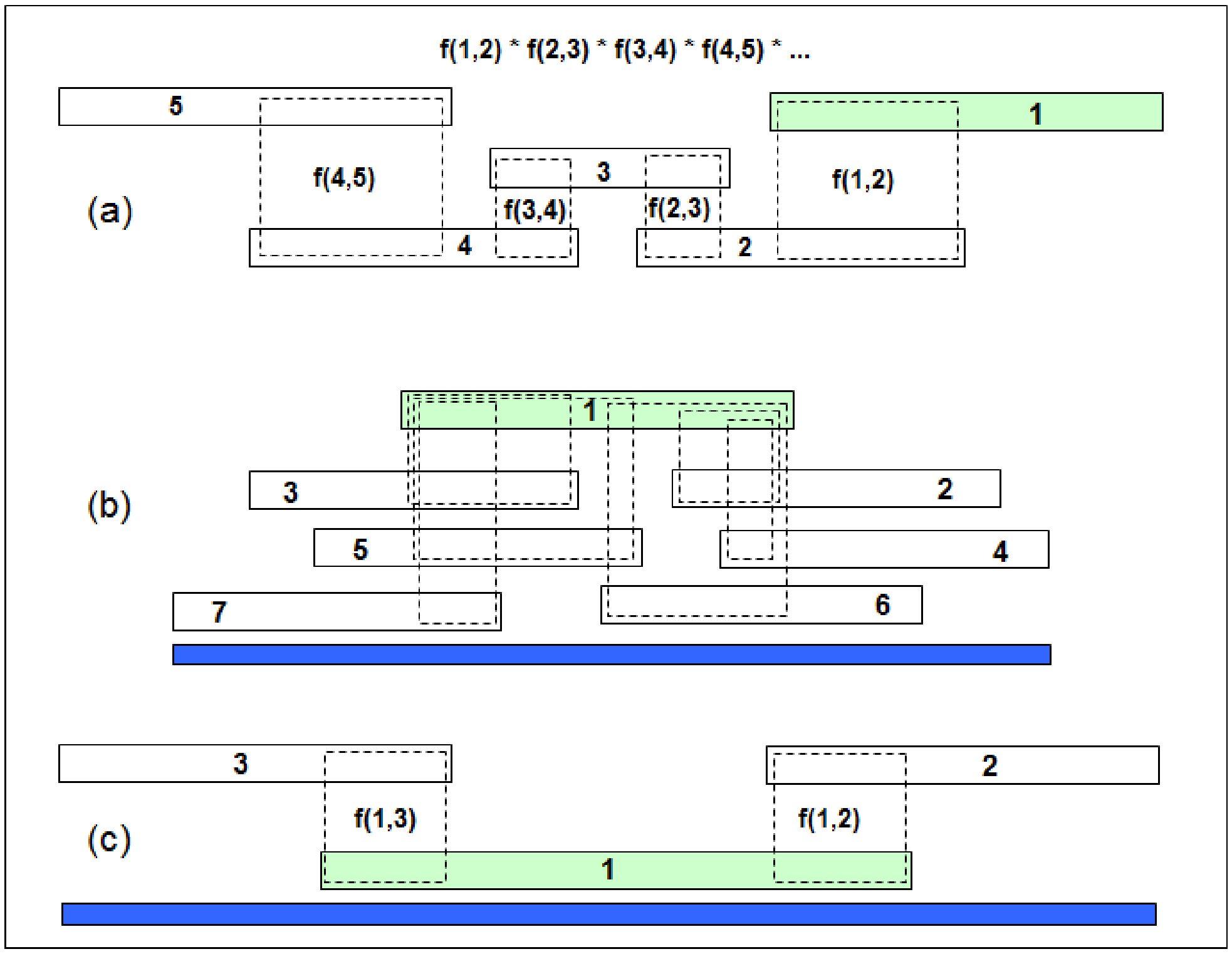

**Figure 30**. Daisy-chaining is a technique for harmonizing a series (usually over time) of observers by placing separated observations on the same scale. Panel (a) shows how to put observers (1) and (5) on the same scale (that of observer (1)) using a chain of intermediate observers (2), (3), and (4). The conversion factors (which could be functions rather than simple constants) $f$(1,2),…, $f$(4,5) transfer the scale from one

observer to the next through their product which also accumulates the uncertainty along the chain. Panel (b) shows how to construct a backbone by comparing each of the observers (2) through (7) to the 'spine' formed by the primary observer (1). Since there are no accumulating multiplications involved, there is no accumulation of errors and the entire composite backbone (shown by the blue bar) is free of the detrimental effect of daisy-chaining. Panel (c) shows how a composite (daisy-chain free) backbone can be constructed by linking surrounding and overlapping backbones (2) and (3) directly to a 'base' backbone (1) via the two independent transfer factors $f(1,2)$ and $f(1,3)$ without accumulation of uncertainty. 'Base' backbones defining the overall scale of the composites are marked as green boxes.

So the composite Wolfer Backbone extending more than one hundred years from 1841 through 1945 with Wolfer's own observations (with unchanged telescope) constituting a firm "spine" from 1876 through 1928 has no daisy-chaining whatsoever. Lockwood et al. [2016b] incorrectly claim that "until recently, **all** composites used "daisy-chaining" whereby the calibration is passed from the data from one observer to that from the previous or next observer". This seems to be based on ignorance about how the composites were constructed e.g. the relative sunspot numbers of Wolf were determined by comparing only with the Zürich observers and not by passing the calibration along a long chain of secondary observers. Similarly, the Hoyt & Schatten [1998] Group Sunspot Number after 1883 [Cliver & Ling, 2016] was based on direct comparison with the RGO observations without any daisy-chaining, and, as we have just reminded the reader, the individual Backbones were constructed also with no daisy-chaining (their primary justification).

Good examples of true daisy-chaining in action can be seen in Lockwood et al.'s [2014] use of several intermediate observers to bridge the gap between the geomagnetic observatories at Nurmijärvi and Eskdalemuir in the 20th century back to Helsinki in the 19th and to propagate the correlation with the modern observed HMF back in time, and in Usoskin et al.'s [2016] use of intermediate observers (their 'two-step' calibration) between Staudach in the 18th century and RGO in the 20th.

## 12. Comparison with H&S

Cliver & Ling [2016] have tried to reproduce the determination of the $k$-values determined by Hoyt & Schatten [1998] for observers before 1883 and have failed because the procedure was not described in enough detail for a precise replication; in particular, it is not known which secondary observers were used in calculating the $k$-factors. On the other hand, Hoyt & Schatten [1998] in their construction of the Group Sunspot Number did not use daisy-chaining (i.e. secondary observers) for data after 1883 because they had the RGO group counts as a continuous (and at the time believed to be good) reference with which to make direct comparisons. For the years after about 1900 when the RGO drift seems to have stopped or, at least abated, the Hoyt & Schatten [1998] Group Sunspot Numbers agree extremely well with the Svalgaard & Schatten [2016] Group Numbers (Figure 31), and incidentally also with various Lockwood and Usoskin reconstructions ("$R_{UEA}$ is the same as $R_G$ after 1900").

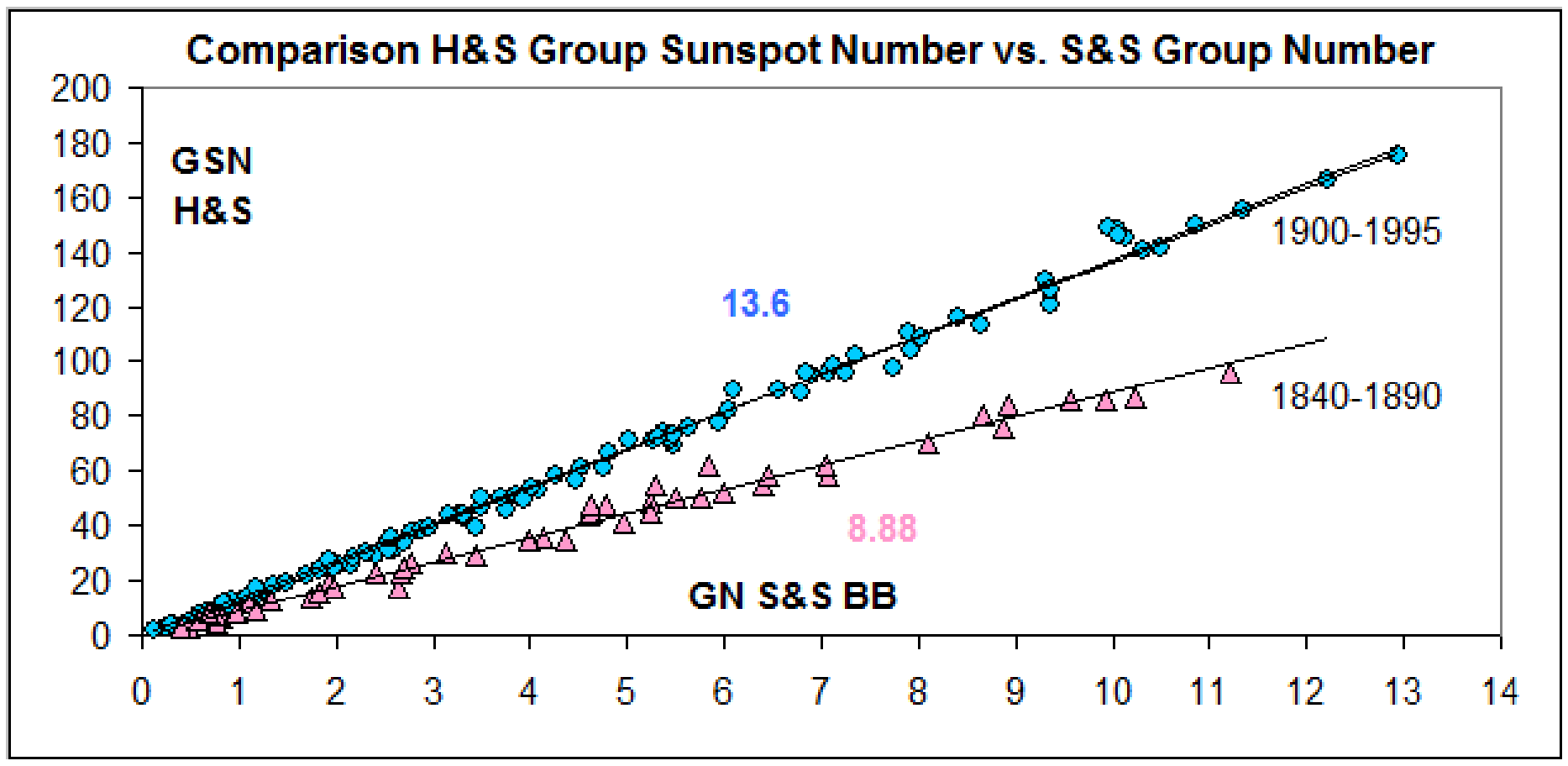

**Figure 31**. Annual averages of the Hoyt & Schatten Group Sunspot Number [GSN; often called $R_G$] compared to the Svalgaard & Schatten [2016] Group Number [GN]. For the data since 1900 (light-blue dots) there is a constant proportionality factor of 13.6 between the two series. For earlier years, the drift of the RGO counts combined with daisy-chaining the too-low values back in time lowers the factor to 8.88 (pink triangles).

For the years 1840-1890 there is also a strong linear relationship, but with a smaller slope because the drift of RGO has been daisy-chained to all earlier years (Lockwood et al. [2016b]: "Because calibrations were daisy-chained by Hoyt & Schatten (1998), such an error would influence all earlier values of $R_G$", which indeed it did). Because Wolf's data go back to the 1840s, Wolf's counts form a firm 'spine', preventing further progressive lowering of the early data resulting from the RGO problem, as observers could be scaled directly to Wolf, thus obviating daisy-chaining. The factor to 'upgrade' the early part of the series to the 'RGO-drift-free' part is 13.6/8.88 = 1.53, consistent with Figure 13. Figure 32 shows the result of 'undoing' the damage caused by the RGO drift. Hoyt & Schatten did not discover the RGO drift because their *k*-factor for Wolf to Wolfer (inexplicably) was set as low as 1.021, i.e. Wolf and Wolfer were assumed to see essentially the same number of groups relative to RGO and to each other, in spite of Wolf himself using a factor of 1.5 (albeit for the relative sunspot number of which the group number makes up about half). It is possible that this was due to not noticing that Wolf changed his instrument to a smaller telescope when he moved to Zürich.

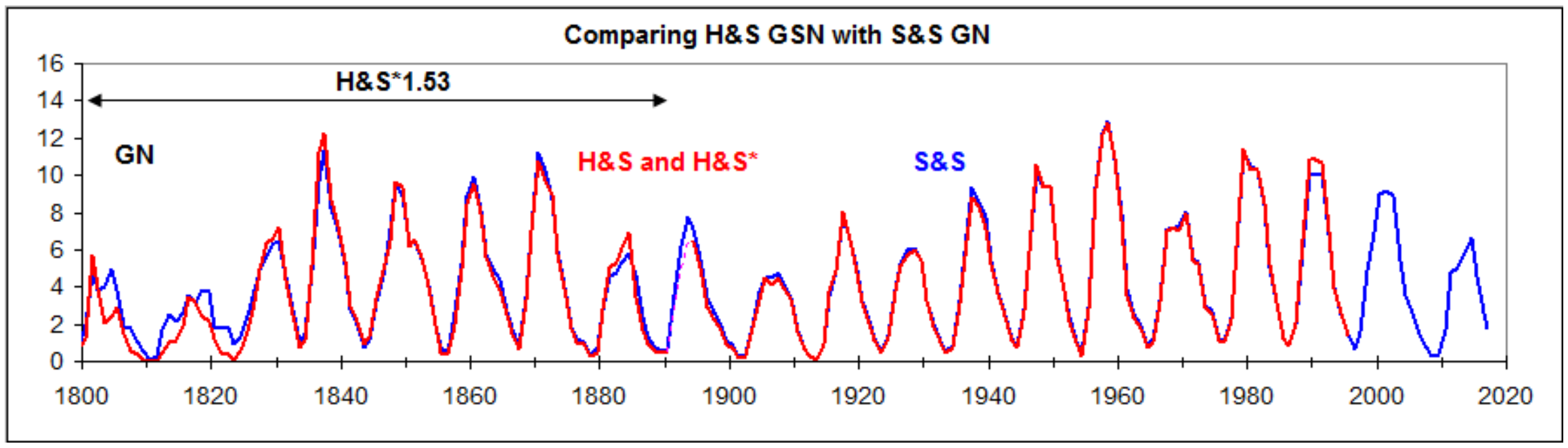

**Figure 32**. Annual averages of the Hoyt & Schatten [H&S] Group Sunspot Number divided by 13.6 (red curve) after 1900 compared to the daisy-chain free part of Svalgaard & Schatten [S&S, 2016] Group Number [GN](blue curve). For the years 1800-1890, the H&S values were then scaled up by 13.6/8.88=1.53. This brings H&S into agreement with S&S, effectively undoing the damage caused by the single daisy-chain step at the transition from the 19th to the 20th century.

## 13. Error Propagation

In addition, the 'base' for the Svalgaard & Schatten [2016] backbone reconstructions is the Wolfer Backbone directly linked to the overlapping Schwabe [1794-1883] and Koyama [1920-1996] Backbones, with no need for intermediate observers, and thus there is a daisy-chain free composite backbone covering the more than two hundred years from 1794 to 1996. The backbone method was conceived to make this possible. As the Wolfer 'reference backbone' is in the middle of that two-hundred year stretch, there is no accumulation of errors as we go back in time from the modern period. Any errors would rather propagate ***forward*** in time from Wolfer until today as well as backwards from Wolfer until the 18th century, thus minimizing total error-accumulation. Before 1800, the errors are hard to estimate, let alone the run of solar activity. Our best chance for tracing solar activity that far back and beyond may come from non-solar proxies, such as the cosmic ray record.

## 14. The Cosmic Ray Record

Cosmogenic radionuclides offer the possibility of obtaining an alternative and completely independent record of solar variability. However, the records are also influenced by processes independent of solar activity (e.g. by climate). Regardless of these uncertainties, the recent work by Muscheler et al. [2016] and Herbst et al. [2017] show very good agreement between the revised sunspot records and the $^{10}$Be records from Antarctica and the $^{14}$C-based activity reconstructions, see Figure 33, lending strong support for the revisions, at least after 1750.

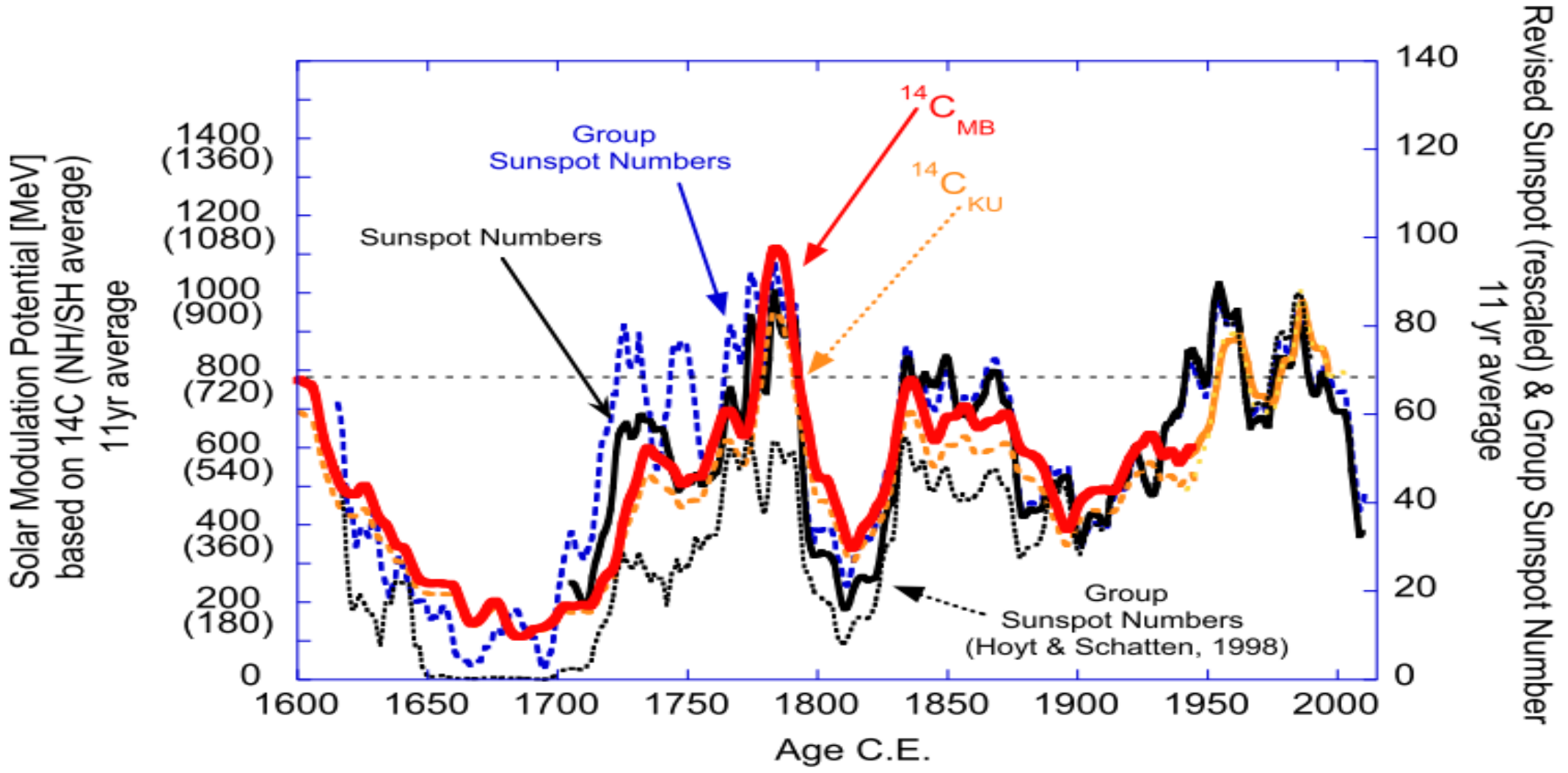

**Figure 33**. Comparison of the $^{14}$C based solar-modulation function with the revised sunspot (black) and (scaled) group sunspot (dashed-dark blue) numbers. All records are shown as running 11-year averages. The red (orange) curve shows the $^{14}$C (neutron monitor)-based results using the production calculations of Masarik and Beer (labeled

$^{14}C_{MB}$). The dashed-orange curves show the results based on Kovaltsov, Mishev, and Usoskin (labeled $^{14}C_{KU}$, with the left y-axes numbers in brackets). The sunspot data have been rescaled to allow for a direct comparison to the Group Sunspot Number data. The old group sunspot record from Hoyt and Schatten is shown as the black dotted curve (From Muscheler et al. [2016]).

Asvestari et al. [2017] attempt to assess the accuracy of reconstructions of historical solar activity by comparing model calculations of the OSF with the record of the cosmogenic radionuclide $^{44}Ti$ measured in meteorites for which the date of fall is accurately known. The technique has promise although the earliest data are sparse, and as the authors note: "The exact level of solar activity after 1750 cannot be distinguished with this method".

## 15. The Active Day Fraction

Usoskin et al. [2016] suggest using the ratio between the number of days per month when at least one group was observed and the total number of days with observations. This Active Day Fraction, ADF, is assumed to be a measure of the acuity of an observer and thus might be useful for calibrating the number of groups seen by the observer by comparing her ADF with a reference observer. For an example, see Figure 34.

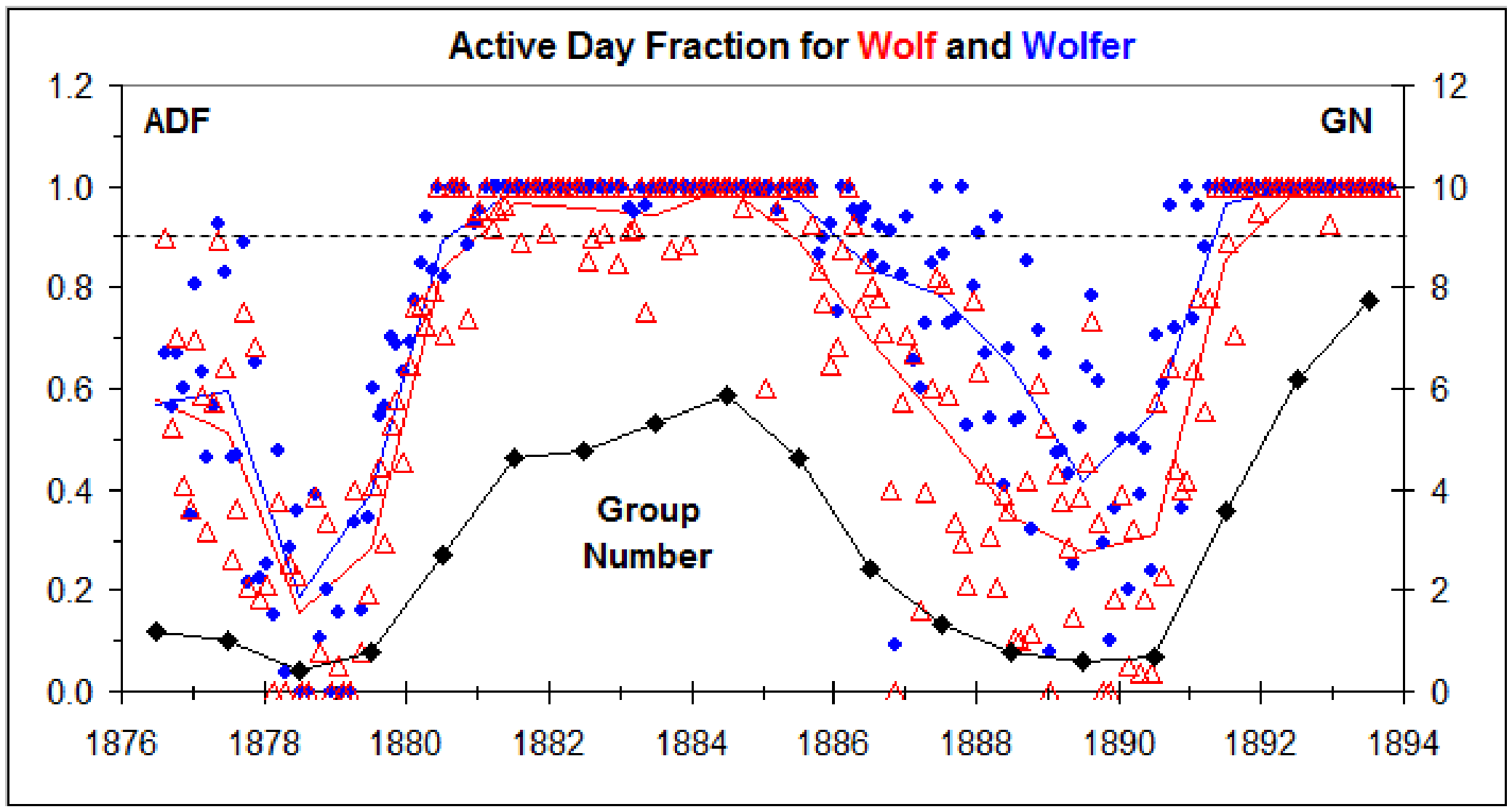

**Figure 34**. The Active Day Fraction, ADF (the ratio between the number of days per month when at least one group was observed and the total number of days with observations) for Wolf (red triangles) and for Wolfer (blue diamonds). Thin lines show the annual mean values. The annual Group Numbers indicating solar cycle maxima and minima are shown (black symbols) at the bottom of the graph with the right-hand scale.

A problem with the ADF is that at sunspot maximum every day is an 'active day' so ADF is nearly always unity and thus does not carry information about the statistics of high solar activity. This 'information shadow' occurs for even moderate group numbers. Information gleaned from low-activity times must be extrapolated to cover solar maxima under the assumption that such extrapolation is valid regardless of activity. Usoskin et al. [2016] applied the ADF-technique to 19th century observers, and the technique was not validated with well-observed modern data. As they admit: "We stopped the calibration in 1900 since the reference data set of RGO data is used after 1900". As it does not make much sense to attempt the use the ADF when it is unity, Usoskin et al. [2016] limit their analysis to times when ADF < 0.9 (dashed line in Figure 34). It is interesting to note that for the low-activity years 1886-1890 the average ADF for Wolfer was 1.50 times higher than for Wolf, close to the *k*-value Wolfer had established for Wolf.

## 16. What Happened to Their Views From 2015?

In a 2015 paper by 16 illustrious luminaries in our field [Usoskin et al., 2015], reconstructions of the OSF and the Solar Modulation Potential were presented. The authors assume that the open solar magnetic flux (OSF) is one of the main heliospheric parameters defining the heliospheric modulation of cosmic rays. It is produced from surface magnetic fields expanding into the corona from where they are dragged out into the heliosphere by the solar wind. The authors use what they call a simple, "but very successful model" to calculate the OSF from the sunspot number series and an assumed tilt of the heliospheric current sheet. Using an updated semi-empirical model the authors have computed the modulation potential for the period since 1610.

Figure 35 shows how their OSF and the modulation potential compare with the Svalgaard & Schatten [2016] Group Number series. With the possible exception of the Maunder Minimum (which is subject to active research), the agreement between the three series is remarkable, considering the simplifications inherent in the models. All three series do away with the notion of an exceptionally active sun in the 20$^{th}$ century, consistent with the findings of Berggren et al. [2009] that "Recent $^{10}$Be values are low; however, they do not indicate unusually high recent solar activity compared to the last 600 years."

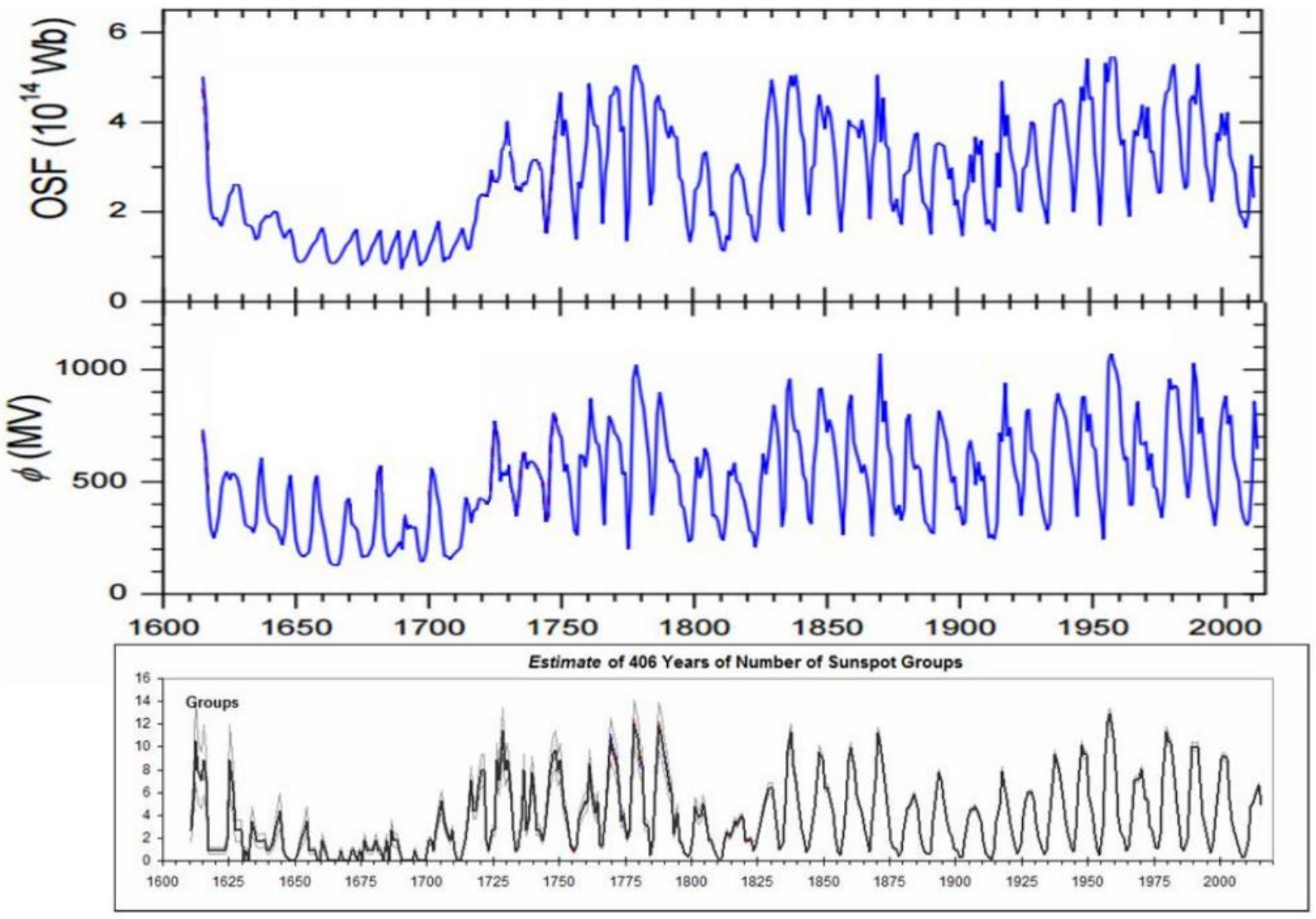

**Figure 35**. (Top) Reconstruction of the Open Solar Flux (adapted with permission from Usoskin et al., 2015). (Middle) Reconstruction (ibid) of the cosmic ray modulation

potential since 1600. (Bottom) The sunspot group number from Svalgaard & Schatten [2016].

Since Lockwood et al. [2016b] and Usoskin et al. [2016] severely criticize (they use the word 'error' 63 times) the Svalgaard & Schatten [2016] backbone-based Sunspot Group Number series, does this mean that they now disavow and repudiate the 2015 paper that they claimed was so "very successful"? It would seem so. The community is ill served with such a moving target.

## 17. Comparing With Simple Averages

A spreadsheet with the raw, observed annual group counts and their values normalized to Spörer's count can be found here http://www.leif.org/research/Sporer-GN-Backbone.xls. As we have found for the other backbones, the simple, straightforward averages of all observers for each year are surprisingly close to the normalized values [see Figure 36], thus apparently making heated discussions about how to normalize seem less important. In our [2016] discussion of Hoyt & Schatten [1998] we noted that "it is remarkable that the raw data with no normalization at all closely match (coefficient of determination for linear regression $R^2$ = 0.97) the number of groups calculated by dividing their GSN by an appropriate scale factor (14.0), demonstrating that the elaborate, and somewhat obscure and, in places, incorrect, normalization procedures employed by Hoyt & Schatten [1998] have almost no effect on the result".

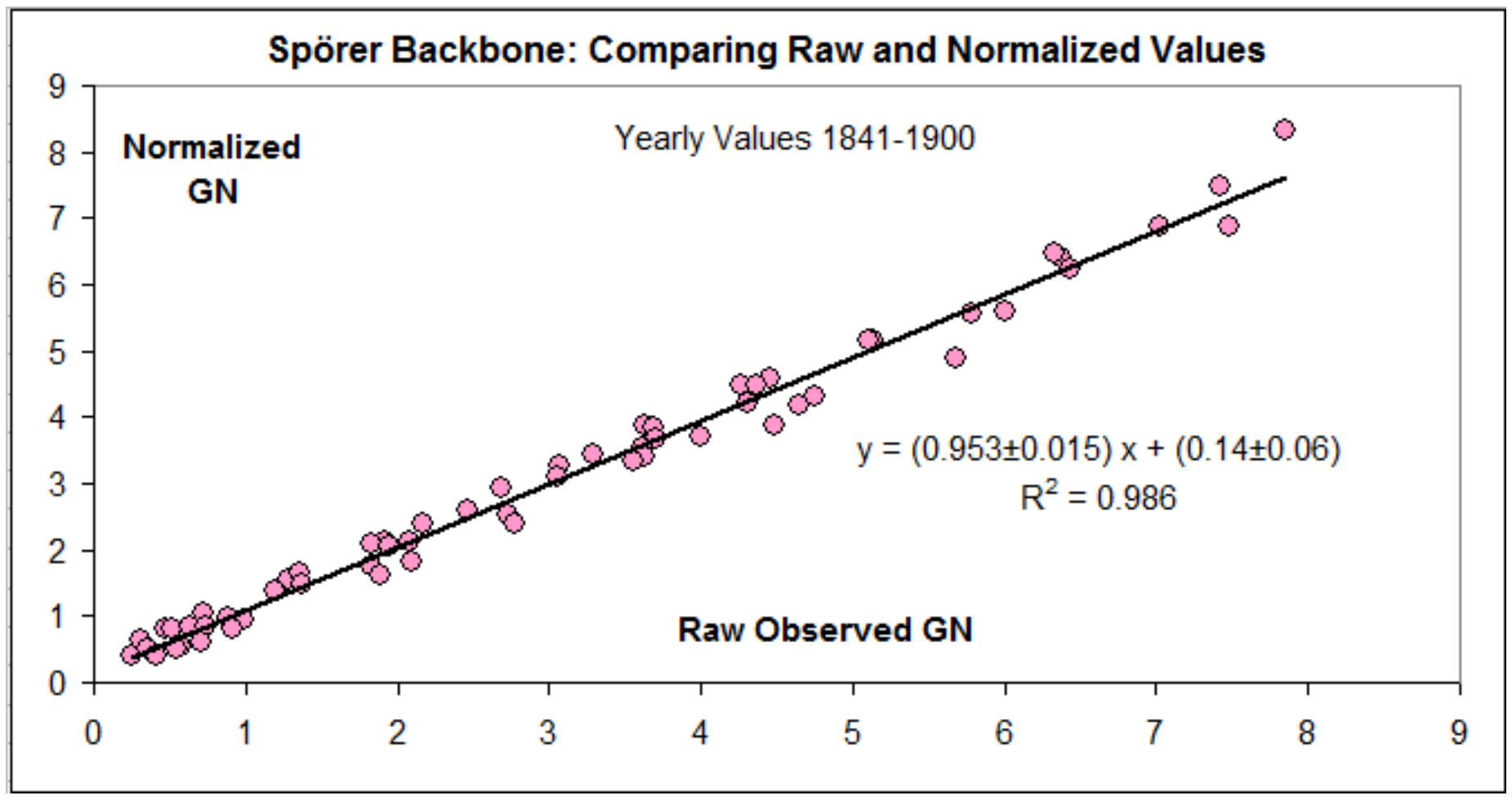

**Figure 36.** Comparison of the Normalized Group Numbers and the Raw, Observed Group Numbers for the Spörer Backbone 1841-1900.

This remarkable result might simply indicate that a sufficient number of observers span the typical values that could be obtained by telescopes and counting methods of the time so that the averages span the true values corresponding to the technology and science of the day, which then becomes the determining factors rather than the acuity and ability of observers.

## 18. The ‘Correction Matrix’ Method

An application of the ‘correction matrix’ method has recently been published by Chatzistergos et al. [2017]. Unfortunately, the article is marred by the usual misrepresentations. E.g.:“The homogenization and cross-calibration of the data recorded by earlier observers was **always** performed through a daisy-chaining sequence of linear scaling normalization of the various observers, using the *k*−factors. This means that starting with a reference observer, the *k*−factors are derived for overlapping observers. The latter data are in turn used as the reference for the next overlapping observers, etc.”

This is simply not correct. For the Wolf sunspot series, observers were directly normalized to the Zürich observers for the interval ~1850-1980 without any intermediate observers. And the secondary observers were only used to fill-in gaps in the Zürich data. For the Hoyt & Schatten [1998] Sunspot Group Number series there was no daisy-chaining used after 1883, and for the Svalgaard & Schatten [2016] Group Number series there was no daisy-chaining used for the two-hundred year long series from 1798-1996.

Further: “Firstly, such methods assume that counts by two observers are proportional to each other, which is generally not correct.” … “All of these sunspot number series used calibration methods based on the linear scaling regression to derive constant *k*−factors. However, this linear *k*−factor method has been demonstrated to be unsuitable for such studies (Lockwood et al. 2016a; Usoskin et al. 2016), leading to errors in the reconstructions that employ them.” On the contrary, as we have shown, proportionality is generally directly observed and only in rare cases is there weak non-linearity which in any case is handled suitably.

And: “Svalgaard & Schatten (2016) also used the method of daisy-chaining *k*−factors. But these authors introduced five key observers (called ‘backbones’, BB hereafter) to calibrate each overlapping secondary observer to these BBs. Thus, they seemingly reduced the number of daisy-chain steps because some daisy-chain links are moved into the BB compilation rather than being eliminated. The problem with this method is that most of the BB observers did not overlap with each other. Thus their inter-calibration was performed via series extended using secondary observers with lower quality and poorer statistics.” Again, this is incorrect. The secondary observers are compared directly to the primary observer with no intermediate steps. This is not a ‘problem’ but a virtue that prevents the bad effects of daisy-chaining.

On the other hand, when their reference observer (RGO) was good (since 1900) the Chatzistergos et al. [2017] reconstruction shows a remarkable linear agreement with Svalgaard & Schatten [2016], Figure 37.

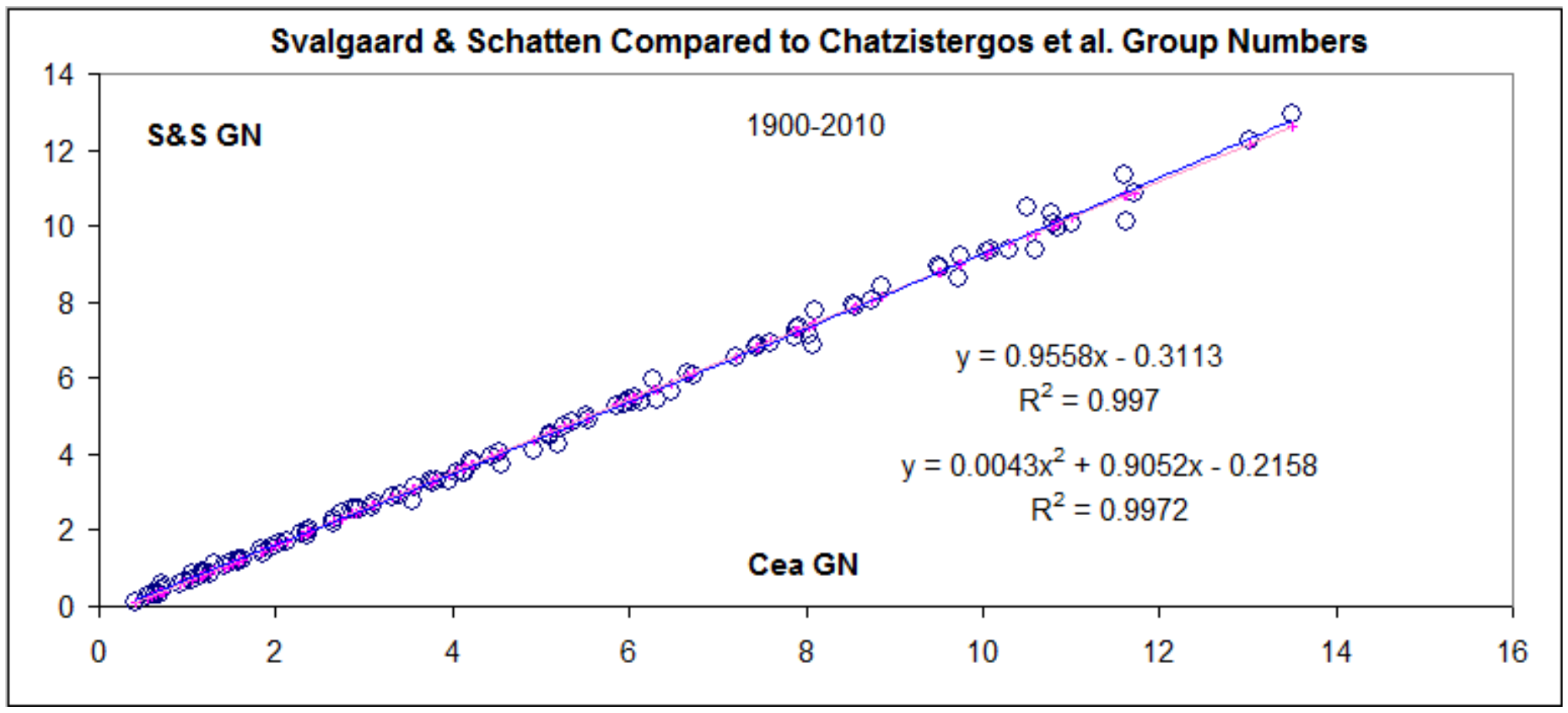

**Figure 37.** Comparison of the Chatzistergos et al. [2017] reconstruction of the Sunspot Group Number and the Svalgaard & Schatten [2016] Backbone method since 1900 (annual values).

As we noted in Section 9, one should not invent group numbers when there is no activity. The Chatzistergos et al. [2017] reconstruction has the usual problem shared with Usoskin et al. [2016] of being too high by ~0.3 groups at sunspot minimum, otherwise the relationship with the Svalgaard & Schatten [2016] reconstruction shows close to perfect proportionality ($R^2$ = 0.997), belying their claim that "such methods assume that counts by two observers are proportional to each other, which is generally not correct". Down-scaling the annual Chatzistergos et al. [2017] values by the linear fit y = 0.956 x – 0.311 to put them on the Wolfer Backbone scale established by Svalgaard & Schatten [2016] removes the solar minimum anomaly and shows that the two methods (when the data are good) agree extremely well, Figure 38, regardless of the persistent claim that the Svalgaard & Schatten [2016] backbone method is generally invalid and unsound compared to the "modern and non-parametric" methods advocated by Chatzistergos et al. [2017] and Usoskin et al. [2016].

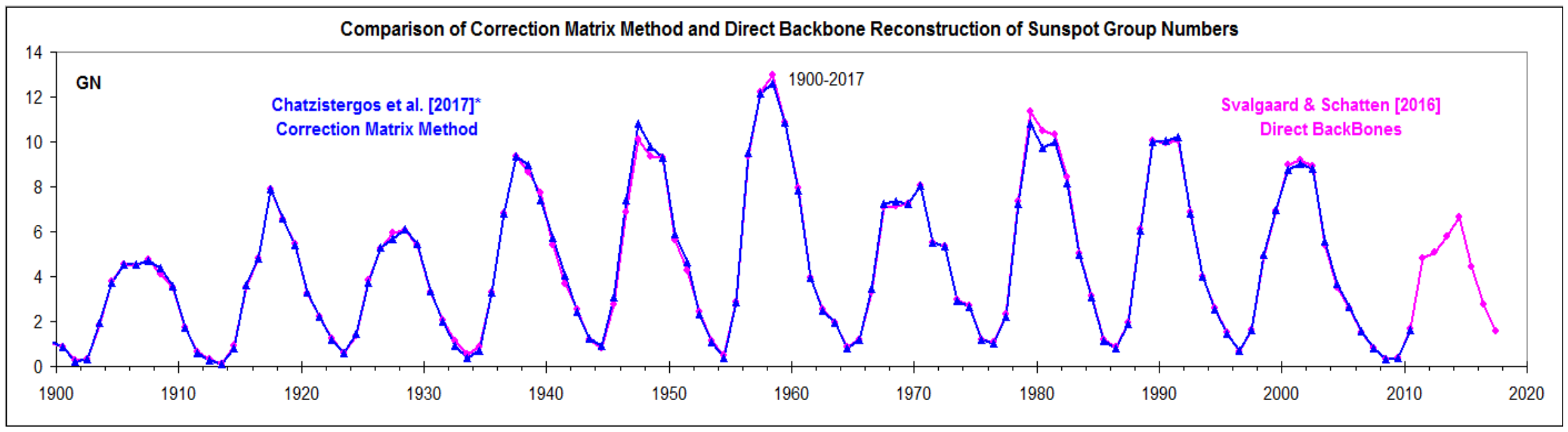

**Figure 38.** Comparison of the down-scaled Chatzistergos et al. [2017] Correction Matrix-based reconstruction of the Sunspot Group Number (blue triangles) and the Svalgaard & Schatten [2016] Backbone method (pink dots) since 1900.

So, it is clear that those 'concerns' about methods are unfounded. As the major objective of our detractors seems to be to maintain their notion that the Modern Maximum was a *Grand* Maximum, possibly unique in the past several thousand years, we should now look at the Chatzistergos et al. [2017] reconstruction for times before 1900, Figure 39.

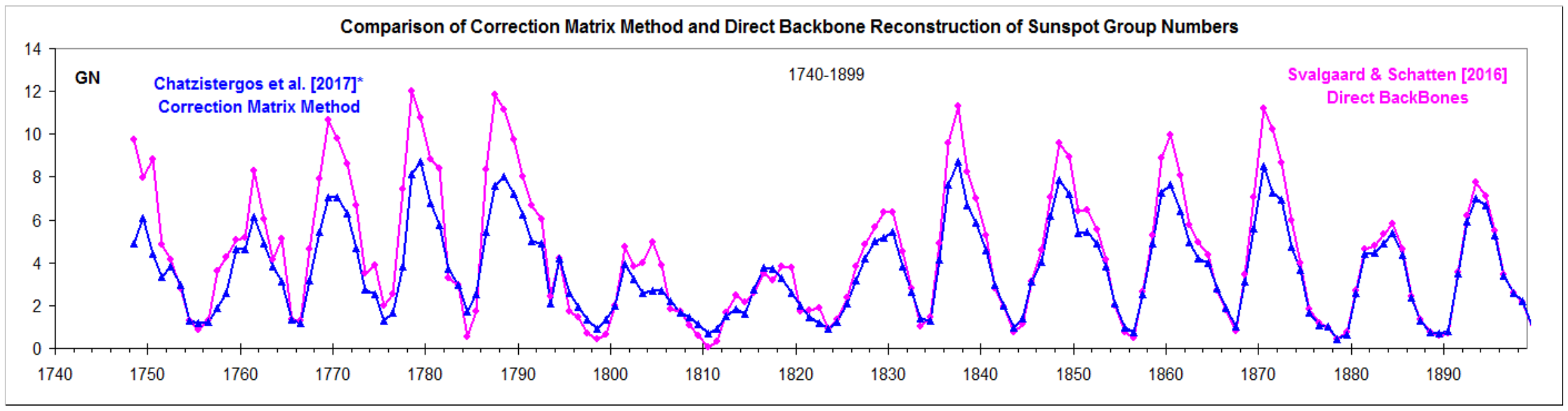

**Figure 39.** Comparison of the scaled Chatzistergos et al. [2017] Correction Matrix-based reconstruction of the Sunspot Group Number (blue triangles) and the Svalgaard & Schatten [2016] Backbone method (pink dots) before 1900.

From this comparison it appears that the Chatzistergos et al. [2017] reconstruction for times before 1900 is seriously too low (or as they would put it: the Svalgaard & Schatten [2016] Backbones are seriously in error, being too high for medium or high solar activity).

How can we resolve this discrepancy? The first (in going towards earlier times) major differences occur for the cycles peaking in 1870 and 1860. Just prior to that time, Wolf was moving from Berne to Zürich and even though a Fraunhofer-Merz telescope was installed in 1864 in the newly built observatory, Wolf never used it after that (but his assistants, in particular Wolfer later on, did). Instead Wolf used smaller telescopes until his death in late 1893; see Figure 40.

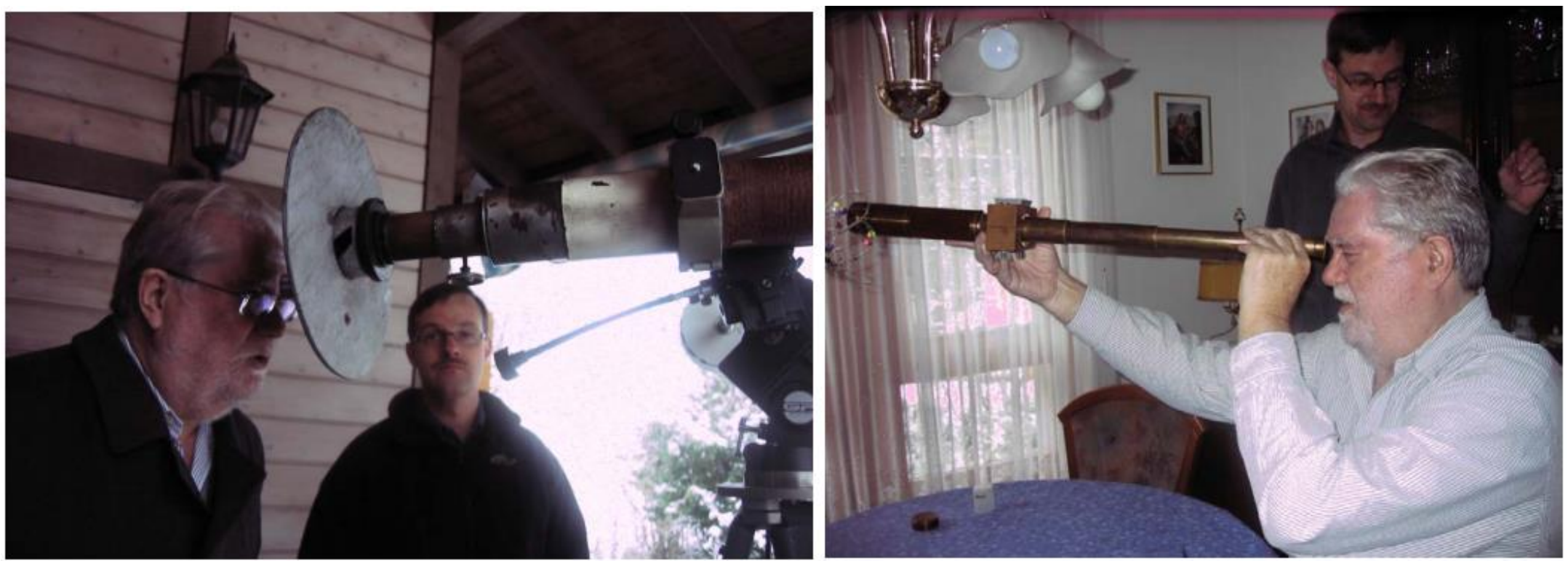

**Figure 40.** (Left) The 82 mm aperture (magnification X64) refractor used mostly by Wolf's assistants at the Zürich Observatory since 1864, designed by Joseph Fraunhofer and manufactured in 1822 at the Fraunhofer factory by his assistant Georg Merz. The telescope still exists and is being used daily by Thomas Friedli (person at center). (Right) One of several small, portable, handheld telescopes (~40 mm aperture, magnification X20) used by Wolf almost exclusively from 1860 on, and still in occasional use today. More on the telescopes can be found at Friedli [2016]. (Photos: Vera De Geest).

We have 18 years of (very nearly) simultaneous observations by Wolf and Wolfer, and as we determined in Section 5, Wolfer on an annual basis naturally saw 1.65 times as many groups with the larger telescope as Wolf saw with the smaller telescopes; in addition, Wolf did not count the smallest groups that would only be visible at moments of very good seeing, nor the umbral cores in extended active regions.

So, we can compare Chatzistergos et al. [2017] with the Wolfer Backbone of Svalgaard & Schatten [2016] and with the Wolf counts, Figure 41. Needless to say, Wolf scaled with the 1.65 factor agrees very well with the Wolfer Backbone. The progressive difference between the reconstructions becomes evident going back from ~1895, strongly suggesting that the daisy-chaining used by Chatzistergos et al. [2017] to connect the earlier data to their post-1900 RGO reference observer is skewing their reconstruction towards lower values, aptly illustrating the danger of daisy-chaining. In particular, the cycles peaking in 1860 and 1870 are clearly too low compared to both Wolf's and Wolfer's counts. The deleterious effect is even greater for the 18th century (Figure 39).

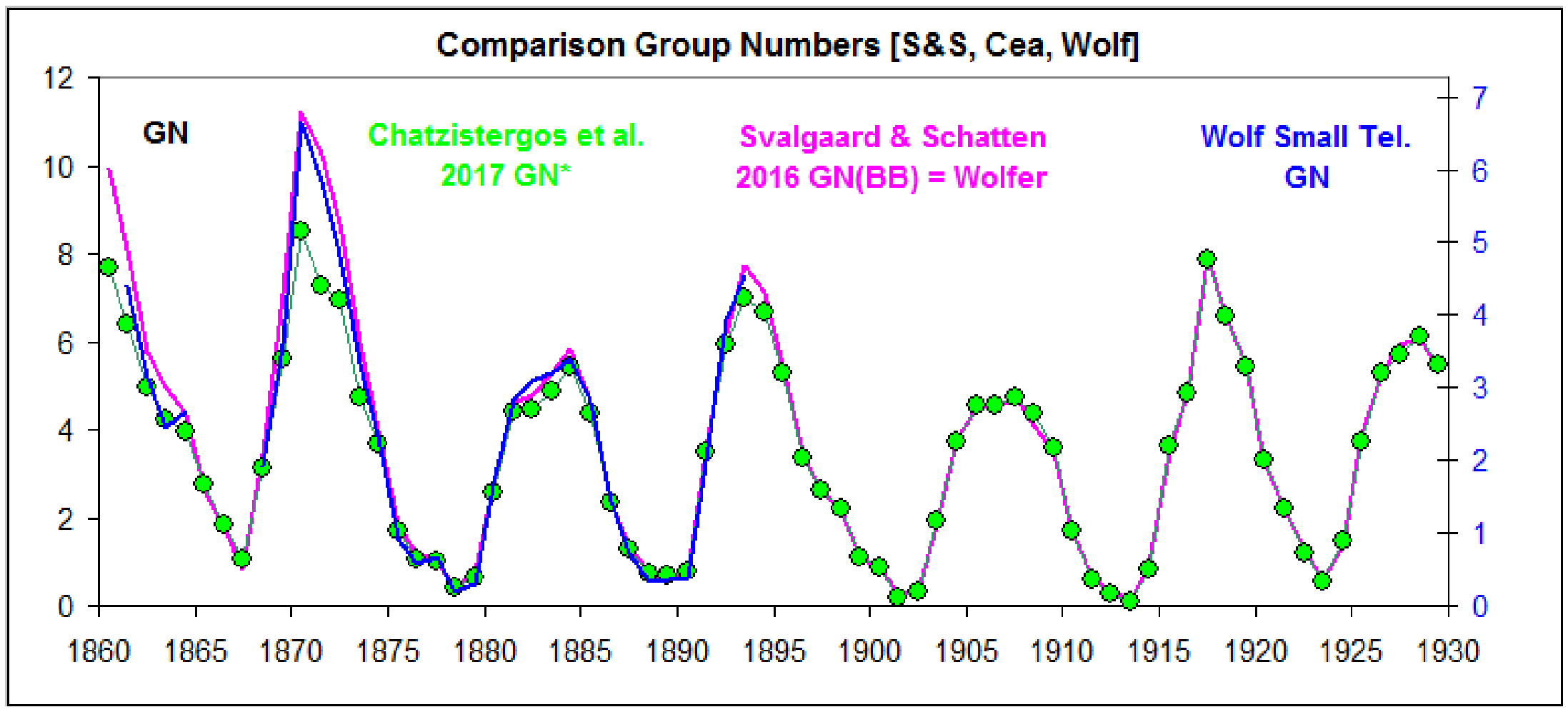

**Figure 41.** Comparison of the annual scaled Chatzistergos et al. [2017] sunspot group numbers (green dots), to the Group Number for the Wolfer Backbone by Svalgaard & Schatten [2016] (pink curve) and the Wolf counts with the 'small telescopes' (blue curve matching the Wolfer Backbone) using the right-hand scale (1.65 times smaller than the left-hand Wolfer scale).

As the derivation of the daisy-chain from RGO to Wolfer by Chatzistergos et al. [2017] is not transparent enough for closer analysis and cannot be replicated, it is not clear exactly how the lower values before ~1895 come about.

## 19. More On the Active Day Fraction Method

Yet another article extolling the virtues of the Active Day Fraction Method [Willamo et al., 2017] have just been published. When the observers' counts are compared to the reference observer (RGO) after 1900, the result is very similar to the Svalgaard & Schatten [2016] group number series, scaled to the same mean: Figure 42, including a strong linear relationship, Figure 43.

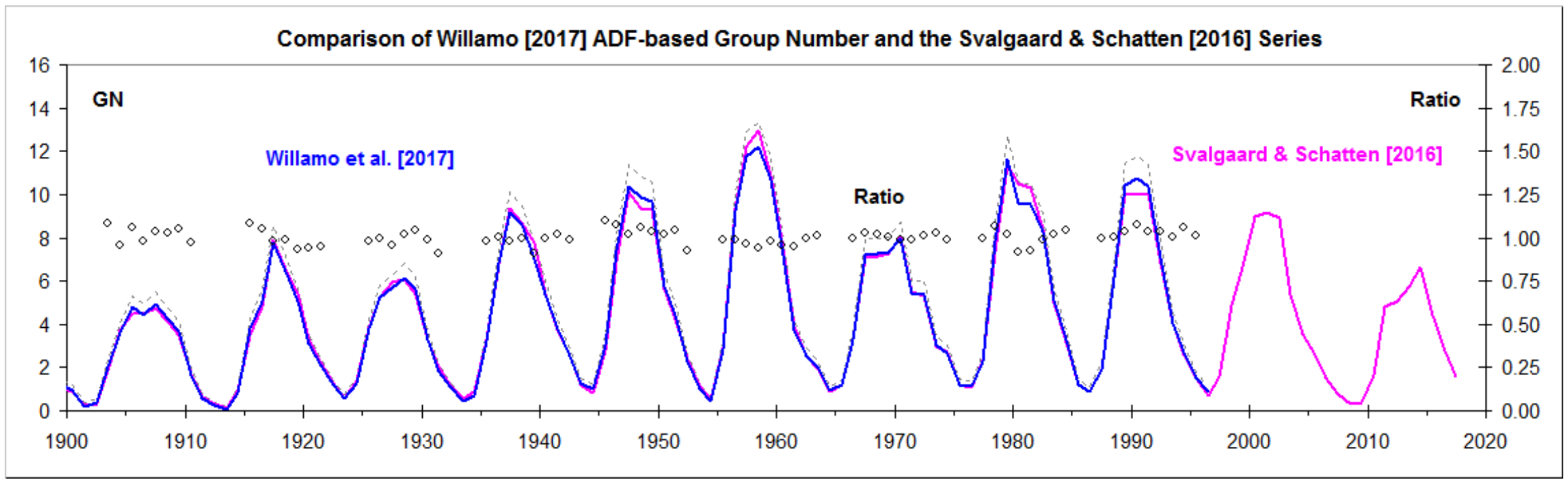

**Figure 42.** Comparison of the annual scaled Willamo et al. [2017] sunspot group numbers (blue curve), to the Group Number by Svalgaard & Schatten [2016] (pink curve). The scaling function (see Figure 43) is y = 0.925 x – 0.139 ($R^2$ = 0.994). The ratio between the two series (for years with group numbers greater than 1.5) is shown by small open circles and is not significantly different from unity.

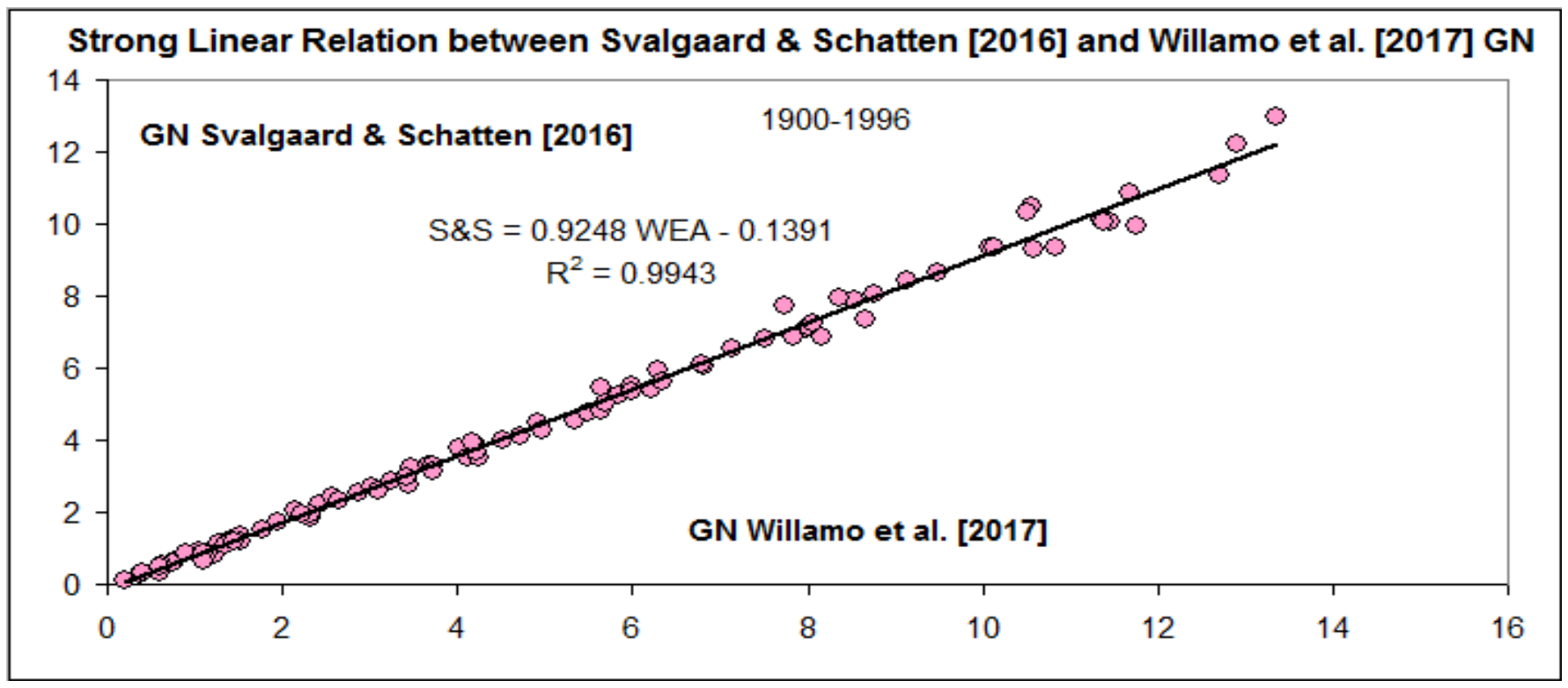

**Figure 43.** Because the Group Numbers are normalized to different observers (RGO and Wolfer) their values are not necessarily identical, This Figure gives the linear scaling function y = 0.925 x – 0.139 ($R^2$ = 0.994) to bring the RGO-based values onto the Wolfer scale.

As with the 'correction matrix' method, an artificial non-zero offset must first be removed. After that, the agreement is extraordinary, showing that the ADF-based method works well for observers overlapping directly with the RGO reference observer and presumably sharing the modern conception of what constitutes a sunspot group as well as conforming to the same PDF. This is, however, not the case for observers before 1900, Figure 44, where the bad effects of the assumption that the PDF for RGO can be transferred unchanged to earlier times become apparent.

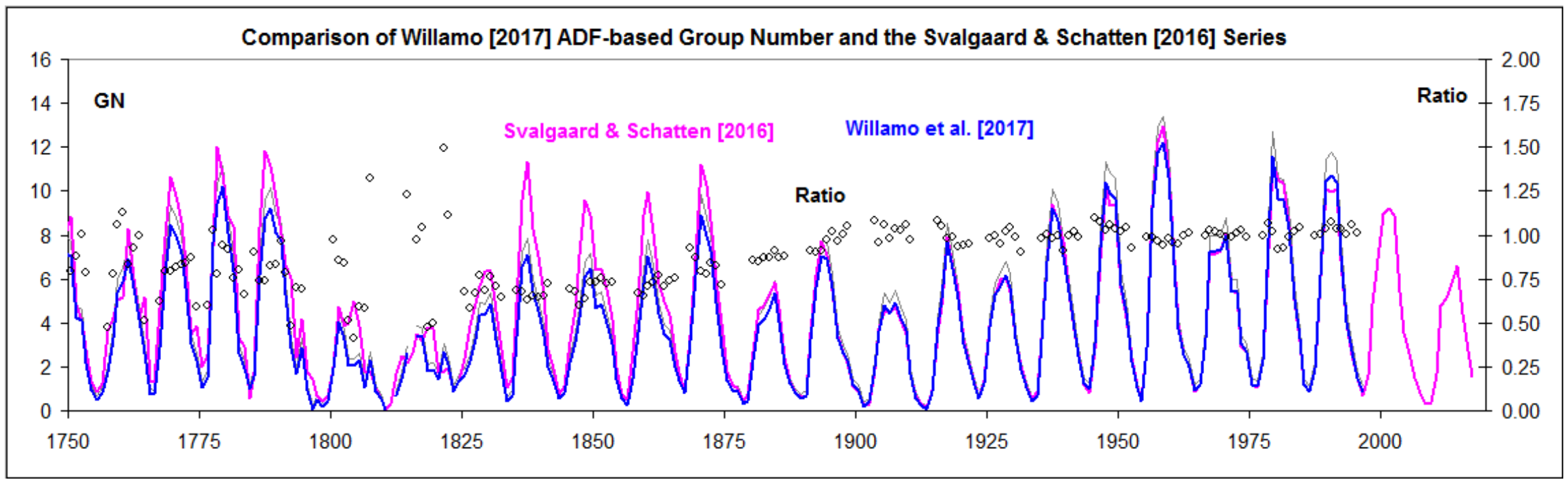

Figure 44. Comparison of the scaled Willamo et al. [2017] group number series (blue) to that of Svalgaard & Schatten [2016] (pink) for the entire interval 1750-1996. Their ratio (for years with group number greater than 1.5) is not different from unity after 1900, but shows a steady decline going back most of the century before that.

It is clear that the ratio is falling steadily gong back from ~1900 to ~1825 and that the noise in the 18th century data [e.g. too few days with no spots were reported] is too large to place much trust in the ADF-method for those years. So, we have the curious situation that when the data is good, the vilified Svalgaard & Schatten [2016] methodology using "unsound procedures and assumptions" yields an astounding agreement with a 'modern and non-parametric' method. We take this as verification of both methods when applied to modern data with common understanding of the nature of solar activity, and as failure of the ADF-method when older data based on inferior technology and, in particular, outdated understanding are used.

## 21. The ADF Methods Fails for 'Equivalent Observers'

We identify several pairs of 'equivalent' observers defined as observers with equal or nearly equal 'observational threshold' areas of sunspots on the solar disk as determined by the 'Active Day Fraction' method [e.g. Willamo et al., 2017]. For such pairs of observers, the ADF-method would be expected to map the actually observed sunspot group numbers for the individual observers to two reconstructed series that are very nearly equal and (it is claimed) represent 'real' solar activity without arbitrary choices and deleterious, error-accumulating 'daisy-chaining'. We show that this goal has not been achieved (for the critical period at the end of the 19th century and the beginning of the 20th), rendering the ADF-methodology suspect and not reliable nor useful for studying the long-term variation of solar activity.

The Active Day Fraction is assumed to be a measure of the acuity of the observer and of the quality of the telescope and counting technique, and thus might be useful for calibrating the number of groups seen by the observer by comparing her ADF with a modern reference observer.

A problem with ADF is that near sunspot maximum, every day is an 'active day' so ADF at such times is nearly always unity and thus does not carry information about the statistics of high solar activity. This 'information shadow' occurs for even moderate group numbers greater than three. Information gleaned from low-activity times must then be extrapolated to cover solar maxima under the hard-to-verify assumption that such extrapolation is valid regardless of activity, secularly varying observing technique and counting rules, and instrumental technology.

In this section we test the validity of the assumptions using pairs of high-quality observers where within each pair the observers every year reported very nearly identical group counts distributed the same way for several decades. The expectation on which our assessment rests is that the ADF method shall duly reflect this similarity and yield very similar reconstructions, for both observers within each pair. If not, we shall posit that the ADF method has failed (at least for the observers under test) and that the method therefore cannot without qualification be relied upon for general use.

The original Hoyt & Schatten catalog has been amended and in places corrected and the updated and current version [Vaquero et al., 2016] is now curated by the World Data Center for the production, preservation and dissemination of the international sunspot number in Brussels: http://www.sidc.be/silso/groupnumberv3[4]. Ilya Usoskin has kindly communicated the data extracted from the above that were used for the calculation [Willamo et al., 2017] of the ADF-based reconstruction of the Group Number. We have used that selection (taking into account the correct Winkler 1892 data[3]) for our assessment (can be freely downloaded from http://www.leif.org/research/gn-data.htm). We compute monthly averages from the daily data, and yearly averages from months with at least 10 days of observations during the year. It is very rare that this deviates above the noise level from the straight yearly average of all observations during that year.

## 23. Winkler and Quimby are Equivalent Observers

Winkler and Quimby form the first pair. Wilhelm Winkler (1842-1910) - a German private astronomer and maecenas [Weise et al., 1998] observed sunspots with a Steinheil refractor of 4-inch aperture at magnification 80 using a polarizing helioscope from 1878 until his death in 1910 and reported his observations to the Zürich observers Wolf and Wolfer who published them in full in the 'Mittheilungen' whence Hoyt & Schatten [1998] extracted the group counts for inclusion in their celebrated catalog of sunspot group observations[5]. The Reverend Alden Walker Quimby of Berwyn, Pennsylvania observed from 1892-1921 with a 4.5-inch aperture telescope with a superb Bardou lens (1889-1891 with a smaller 3-inch aperture). The observations were also published in full in 'Mittheilungen' and included in the Hoyt & Schatten catalog. As we shall see below, Winkler and Quimby have identical group *k'*-values with respect to Wolfer and thus saw and reported comparable number of sunspot groups.

Figure 45 shows that Winkler and Quimby have (within the errors) the same *k'*-factors (1.295±0.035 and 1.279±0.034) with respect to Wolfer, based on yearly values. For monthly values, the factors are also equal (1.25±0.02 and 1.27±0.02) so it must be accepted that Winkler and Quimby are very nearly equivalent observers.

---

[4] Also available at http://haso.unex.es/?q=content/data

[5] Unfortunately, the data in the original Hoyt & Schatten data files for Winkler in 1892 are not correct. The data for Winkler in the data file are really those for Konkoly at O-Gyalla for that year. L.S. has extracted the correct data from the original source [Wolf, 1893].

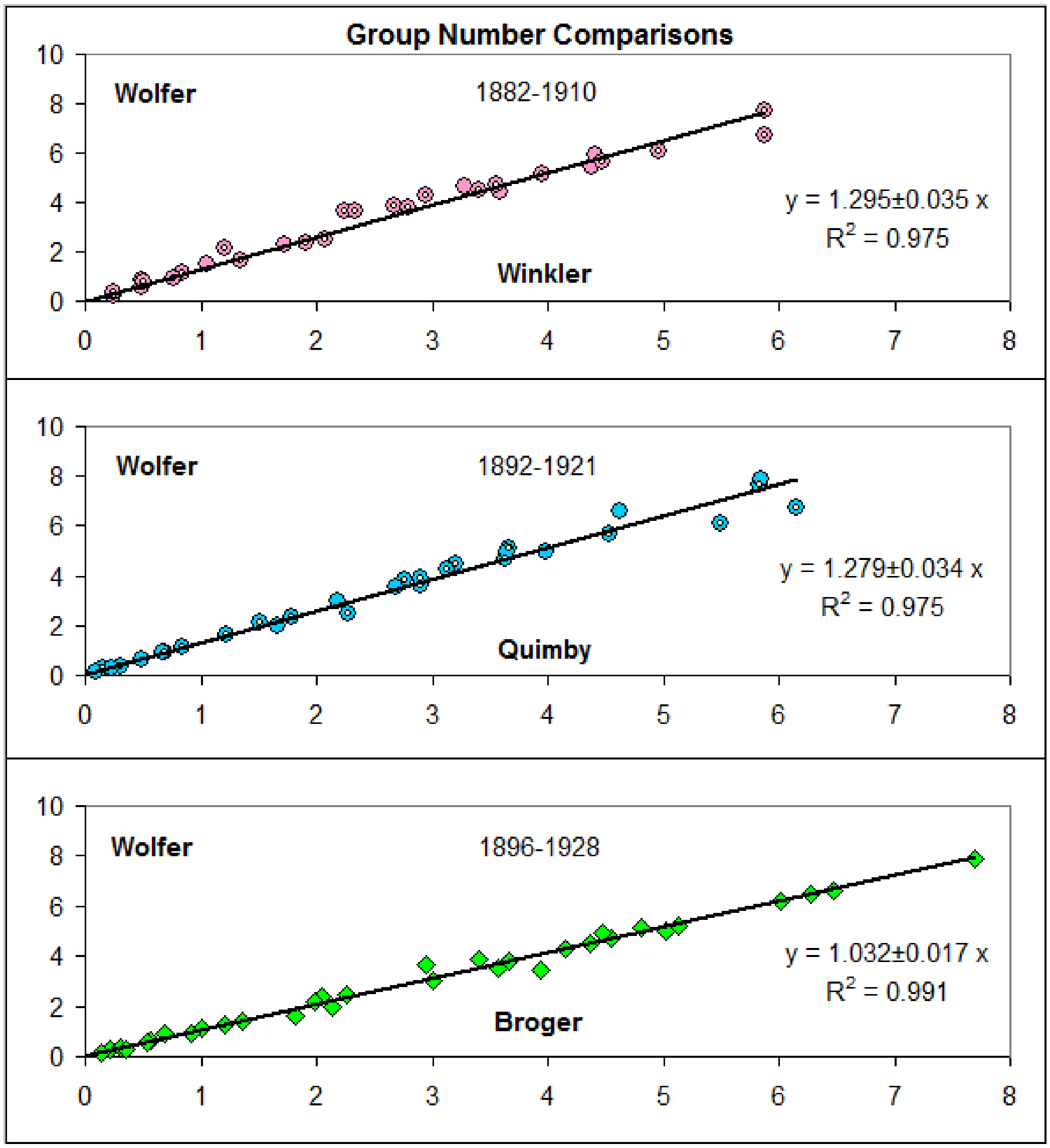

**Figure 45.** (Top) The average number of groups per day for each year 1882-1910 for observer Winkler compared to the number of groups reported by Wolfer. (Middle) The average number of groups per day for each year 1892-1921 for observer Quimby compared to the number of groups reported by Wolfer. Symbols with a small central dot mark common years between Winkler and Quimby. (Bottom) The average number of groups per day for each year 1896-1928 for the Zürich observer Broger compared to the number of groups reported by Wolfer. The slope of the regression line and the coefficient of determination $R^2$ are indicated on each panel. The offsets for zero groups are not statistically significant.

For days when two observers have both made an observation, we can construct a 2D-map of the frequency distribution of the simultaneous daily observations of the group counts *occurrence(groups(Observer1), groups(Observer2))*, i.e. showing on how many days Observer1 reports G1 groups while Observer2 reports G2 groups, varying G1 and G2 from 0 to a suitable maximum. Figure 46 (Upper Panels) shows such maps for Winkler and Quimby (Observers1) versus Wolfer (Observer2). It is clear that the maps are very similar and 'well-behaved', with narrow ridges stretching along the regression lines.

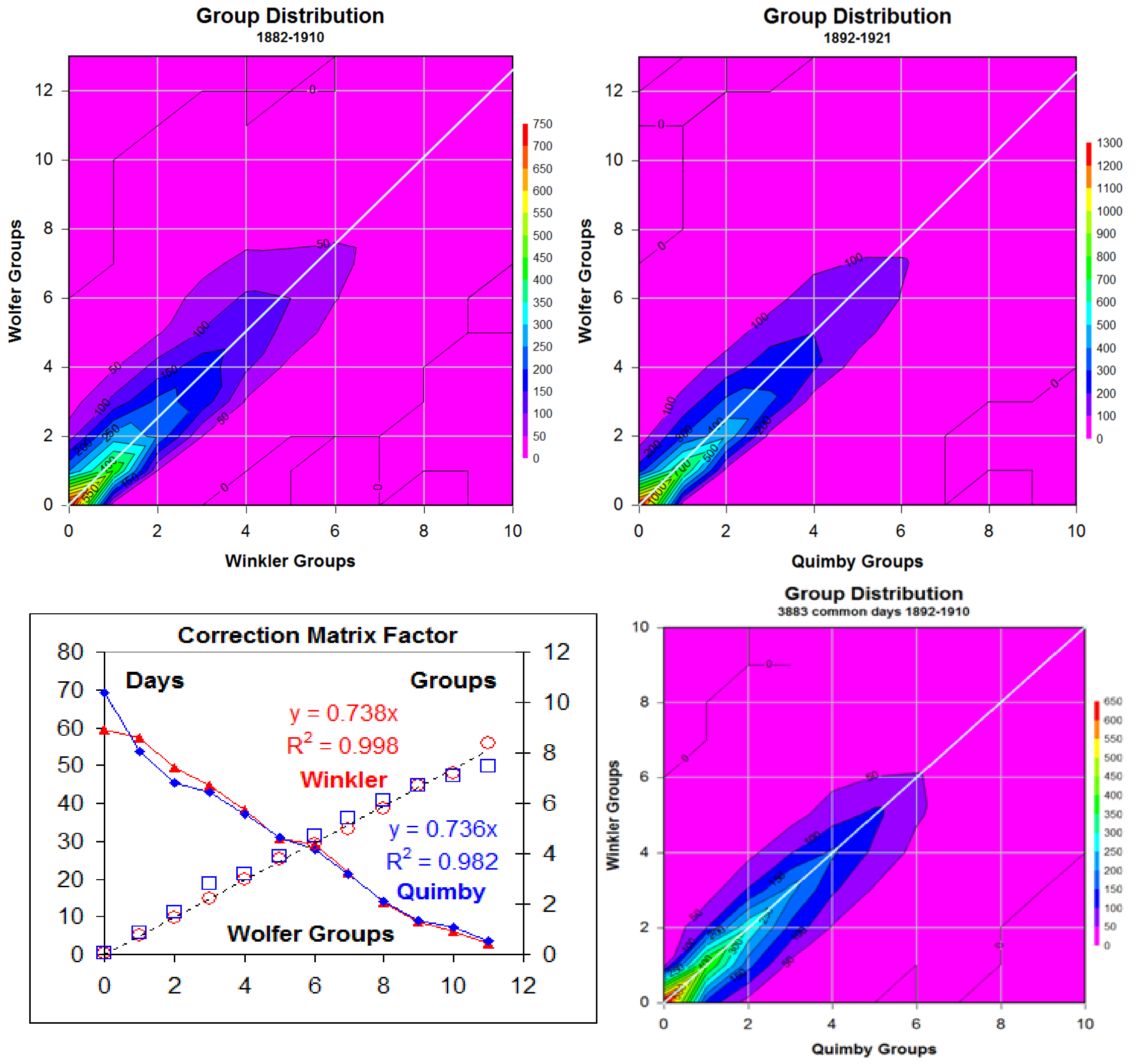

**Figure 46.** (Upper Panel Left) Distribution of simultaneous daily observations of group counts showing on how many days Winkler reported the groups on the abscissa while Wolfer reported the groups on the ordinate axis, e.g. when Winkler reported 5 groups, Wolfer reported 6 groups on 100 days during 1882-1910. (Upper Panel Right) Same, but for Quimby and Wolfer. The diagonal lines lie along corresponding group values determined by the daily *k'*-factors (≈1.25). (Lower Panel Left) The number of groups reported by Winkler (red circles) and by Quimby (blue squares) as a function of the number of groups reported by Wolfer on the same days. Also shown are the average number of days per year (left-hand scale) when those groups were observed (Winkler red triangles; Quimby blue diamonds). The factors are based on the 99% of the days where the group count is less than 12. Above that, the small-number noise is too large. (Lower Panel Right) Distribution of simultaneous daily observations of group counts showing on how many days Quimby reported the groups on the abscissa while Winkler reported the groups on the ordinate axis, e.g. on days when Quimby reported 4 groups, Winkler also reported 4 groups on about 150 days during 1892-1910.

In Figure 46 (Lower Panels) we plot the number of groups reported by Winkler against the number of groups reported by Quimby on the same day, to show that Winkler and Quimby are equivalent observers. The diagonal line marks equal frequency of groups reported by both observers.

The 'Correction Factor' is the average factor to convert a daily group count by one observer to another. Figure 46 (Lower Panel) showed that Winkler and Quimby have almost identical factors for conversion from Wolfer with almost identical distributions in time. This is again an indication that Winkler and Quimby are equivalent observers. If so, the yearly group numbers reported by the two observers should be nearly equal, which Figure 47 shows that they, as expected, are.

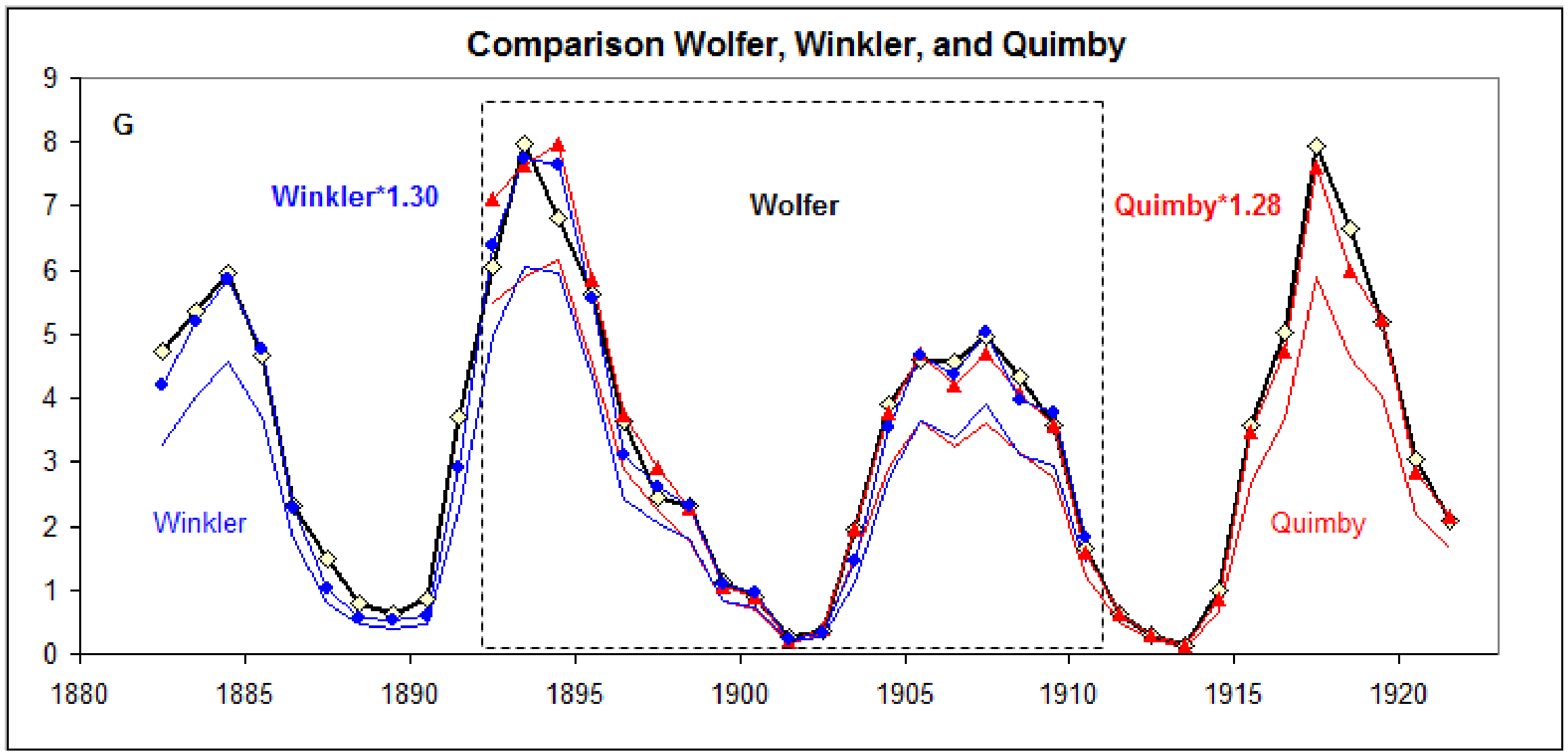

**Figure 47.** Yearly average reported group counts by Winkler (thin blue line without symbols) and Quimby (thin red line without symbols). The dashed line box outlines the years with common data. If we multiply the raw data by the *k'*-factors we get curves for Winkler (blue line with diamonds) and Quimby (red line with triangles) that should (and do) reasonably match the raw data for Wolfer (black line with light-yellow diamonds).

We have shown that Winkler and Quimby are equivalent observers and that their data multiplied by identical (within the errors) *k'*-factors reproduce the Wolfer observations.

## 24. Broger and Wolfer are Equivalent Observers

Broger and Wolfer form a second pair. Max Broger (18XX-19ZZ) was hired as an assistant at the Zürich Observatory and observed 1896–1936 using the same (still existing) Fraunhofer-Merz 82mm 'Norm telescope' at magnification 64 as director Wolfer. Alfred Wolfer (1854-1931) started as an assistant to Wolf in 1876 and observed until 1928. Broger had a *k'*-value of unity with respect to Wolfer and thus saw and reported comparable number of sunspot groups. In addition, there probably was institutional consensus as to what would constitute a sunspot group. The observations were direct at the eyepiece and all were published in the 'Mitteilungen' and from 1880 on in the Hoyt & Schatten catalog.

In Figure 45 we showed the average number of groups per day for each year 1896-1928 for Broger compared to the number of groups reported by Wolfer. The *k'*-factor for Broger is unity within 2-σ, indicating that Broger and Wolfer are equivalent observers. For days when two observers have both made an observation, we can construct a 2D-map of the occurrence distribution of the 6778 simultaneous daily observations of counts during 1896-1928 similar to Figure 46. Figure 48 (right) shows the map for Broger versus Wolfer.

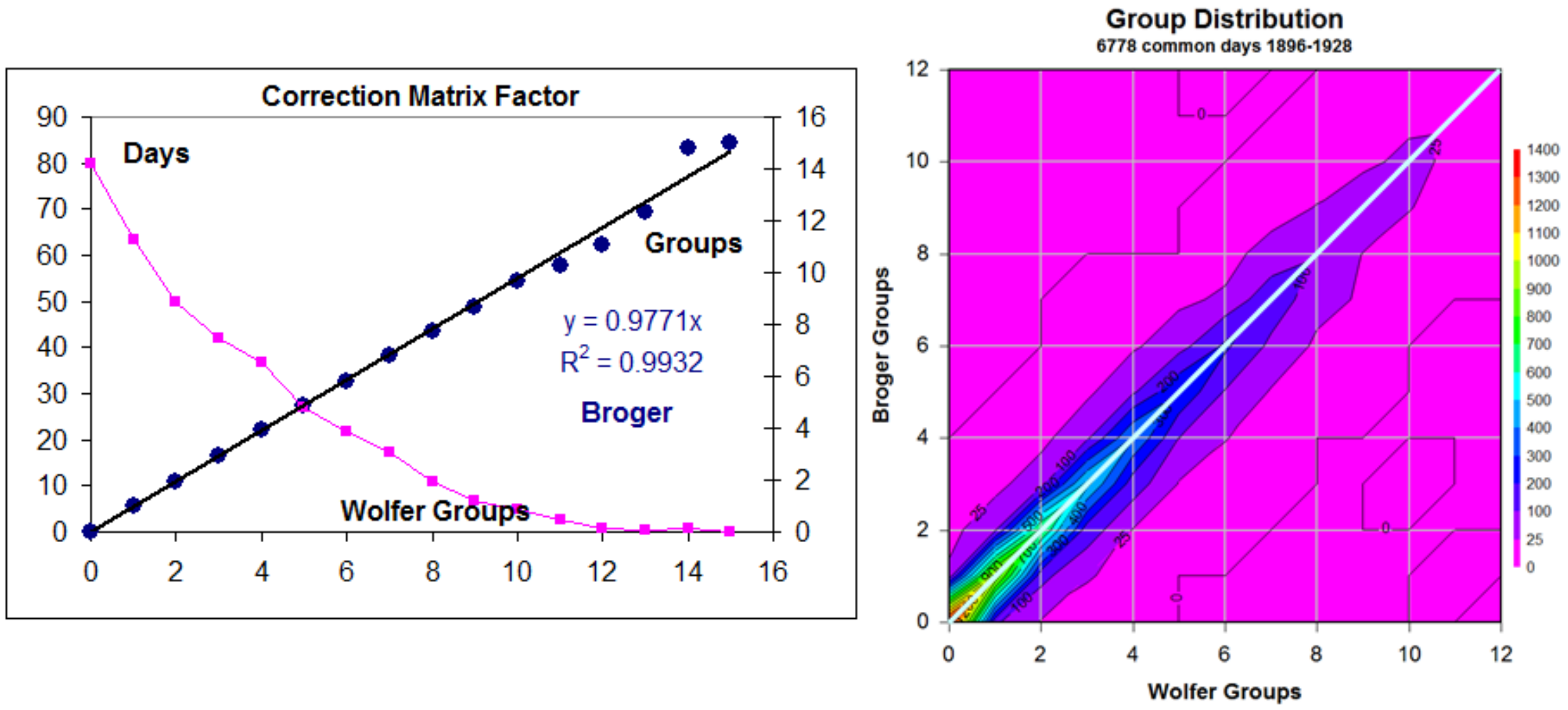

**Figure 48.** (Right) Distribution of simultaneous daily observations of group counts showing on how many days Wolfer reported the groups on the abscissa while Broger reported the groups on the ordinate axis, e.g. on days when Wolfer reported 4 groups, Broger also reported 4 groups on about 400 days during 1896-1928. (Left) The number of groups reported by Broger (dark-blue dots) as a function of the number of groups reported by Wolfer on the same days. Also shown are the average number of days per year (left-hand scale) when those groups were observed (pink squares).

Figure 48 shows that Broger and Wolfer have almost identical distributions in time. This is again an indication that Broger and Wolfer are equivalent observers. If so, the group numbers reported by the two observers should be nearly equal, which Figures 49 and 50 show that they, as expected, are.

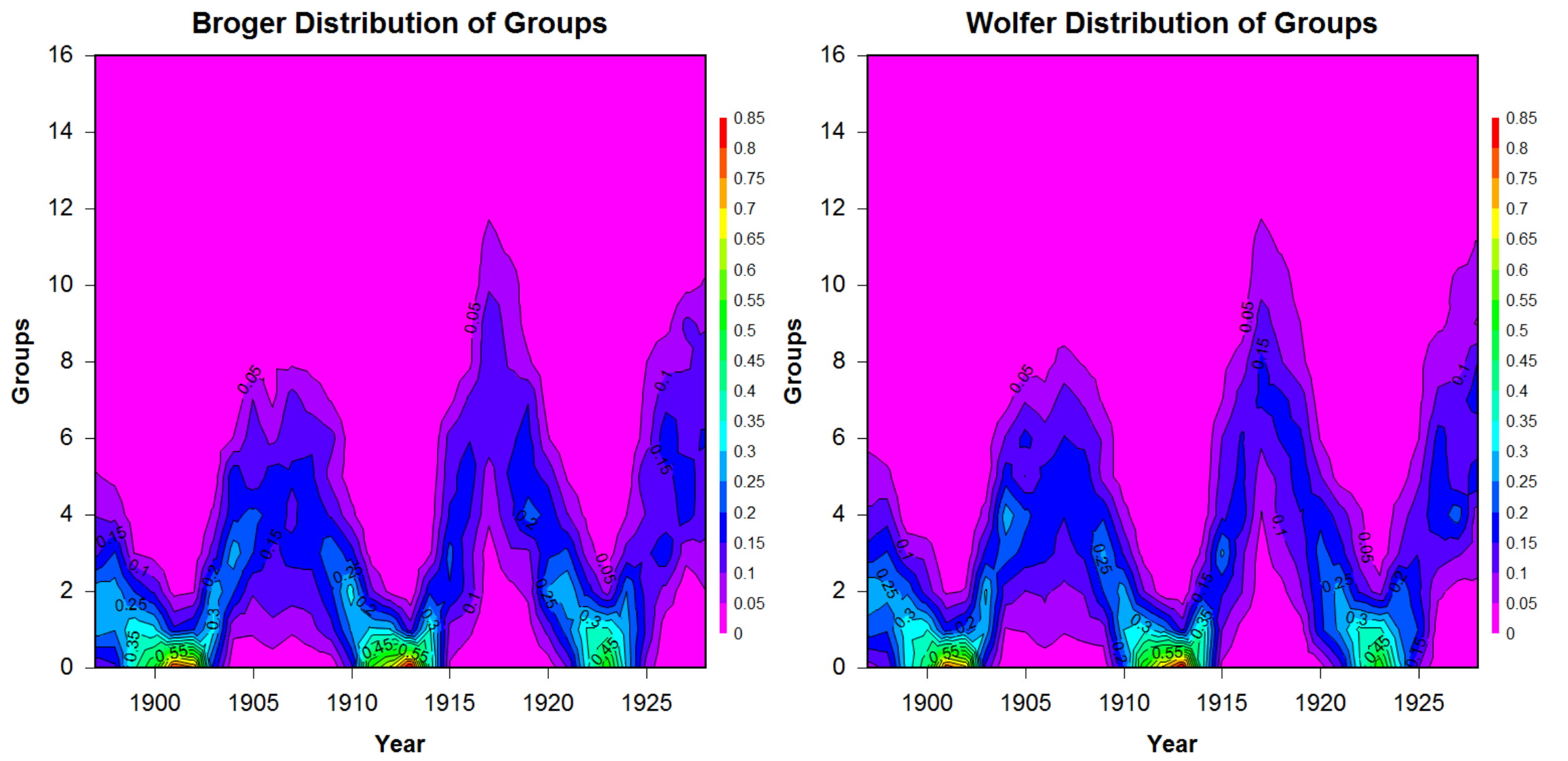

**Figure 49.** Distribution in time of daily observations of group counts showing the fraction of days per year Broger (left) and Wolfer (right) reported the groups on the ordinate axis).

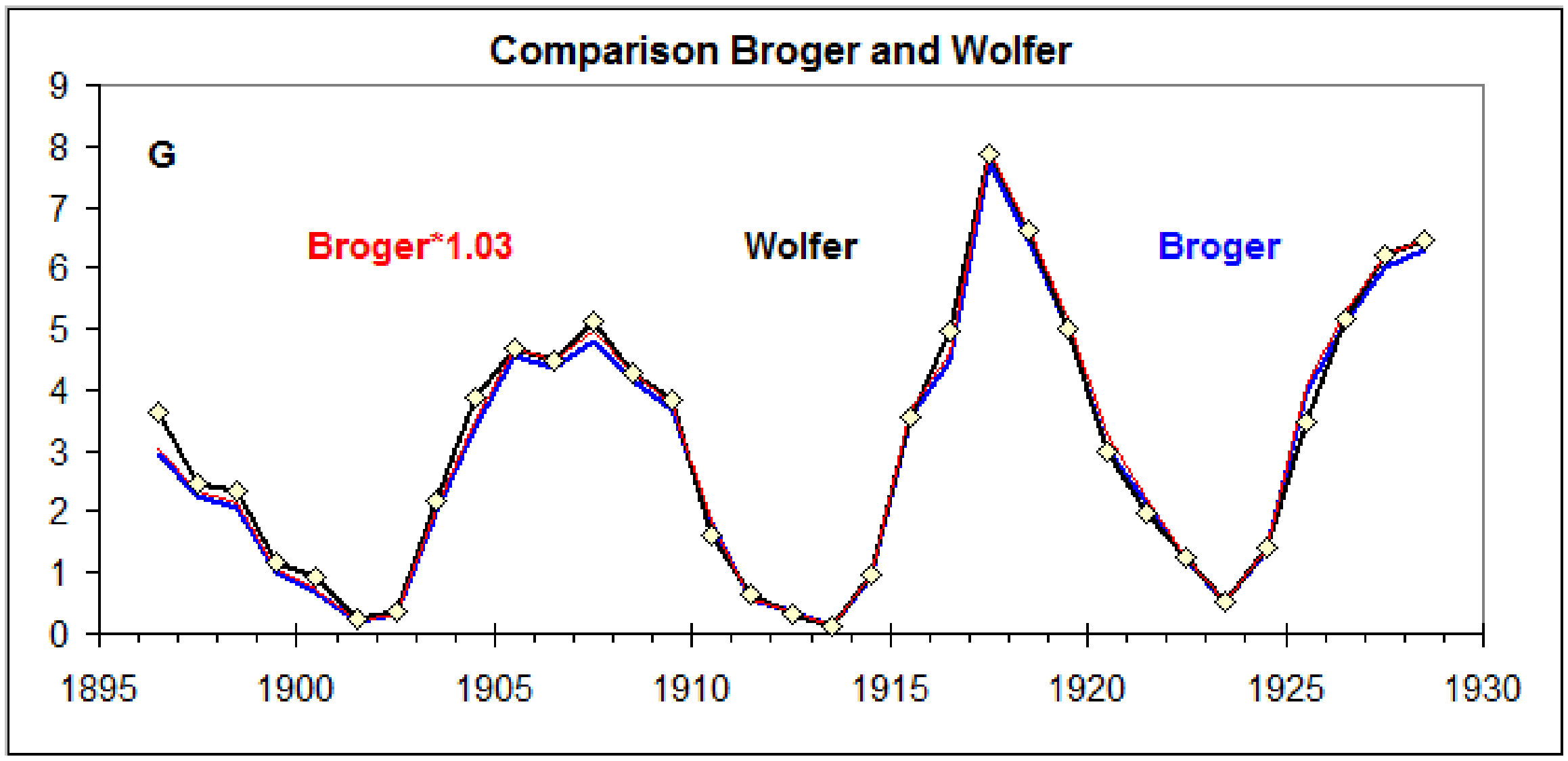

**Figure 50.** Yearly average reported group counts by Broger (blue line) and Wolfer (black line with light-yellow diamonds). If we multiply Broger's raw data by his *k'*-factor with respect to Wolfer we get the thin red line curve. There might be a hint of a slight learning curve for Broger for the earliest years.

We have shown that Broger and Wolfer are equivalent observers and that Broger's data reproduce the Wolfer observations. Combining the data in Figures 47 and 50 provides us with a firm and robust composite reconstruction of solar activity during the important transition from the 19$^{th}$ to the 20$^{th}$ centuries, Figure 51:

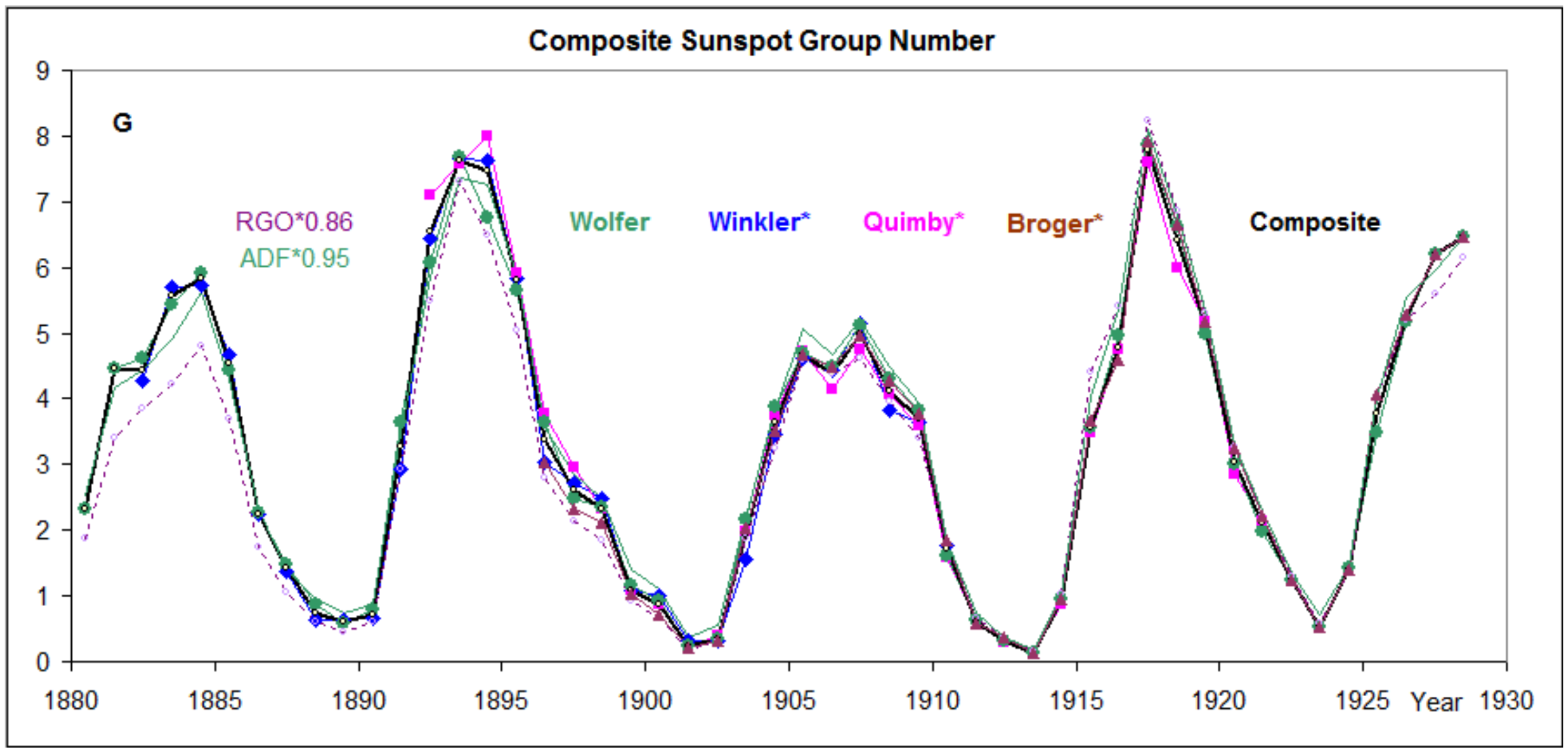

**Figure 51.** Composite Group Number series from Wolfer (green dots), Winkler (blue diamonds), Quimby (pink squares), and Broger (purple triangles). The dashed line shows the RGO (Royal Greenwich Observatory) group number scaled by a factor 0.86 derived from a fit with Wolfer spanning 1901-1928. The thin green line without symbols shows the ADF-based values from Willamo et al. [2017] scaled to fit Wolfer.

The consistency between Wolfer, Broger*, Quimby*, and Winkler*[6] throughout the years 1880-1928 suggests that there have been no systematic long-term drifts in the Composite. On the other hand, the well-known deficit for RGO before about 1890 is clearly evident. The ADF-based values seem at first blush to match the Composite reasonably well. Unfortunately, the agreement is spurious as we shall show in the following sections.

## 25. The ADF Observational Threshold

The ADF-method [Willamo et al., 2017] is based on the assumption that the 'quality' of each observer is characterized by his/her acuity given by an observational threshold area *S*[7], on the solar disk of all the spots in a group. The threshold (all sunspot groups with an area smaller than that were considered as not observed) defines a calibration curve derived from the cumulative distribution function (CDF) of the occurrence in the reference dataset (RGO) of months with the given ADF. A family of such curves is produced for different values of *S*. The observational threshold for each observer is defined by fitting the actual CDF curve of the observer to that family of calibration curves. The best-fit value of *S* and its 68% (±1σ) confidence interval were defined by the $\chi^2$ method with its minimum value corresponding to the best-fit estimate of the observational threshold. Table 2 gives the thresholds for the observers considered in this article.

**Table 2.** The columns are: the name of the observer, the Fraction of Active Days, the lower limit of *S* for the 68% confidence interval, the observational threshold area *S* in

[6] The asterisks denote the raw values multiplied by the *k'*-factor.
[7] Simplified form of the $S_S$ used by Willamo et al. [2017].

millionth of the solar disk, the upper limit of *S*, and the observer's code number in the Vaquero et al. [2016] database. (From Willamo et al., [2017]).

| Observer | ADF % | S low | S µsd | S high | Code |
|---|---|---|---|---|---|
| RGO | 86 | - | **0** | - | 332 |
| Spörer | 86 | 0 | **0** | 2 | 318 |
| Wolfer | 77 | 1 | **6** | 11 | 338 |
| Broger | 78 | 5 | **8** | 11 | 370 |
| Weber | 81 | 20 | **25** | 31 | 311 |
| Shea | 80 | 20 | **25** | 31 | 295 |
| Quimby | 73 | 17 | **23** | 31 | 352 |
| Winkler | 75 | 51 | **60** | 71 | 341 |

## 26. Does the ADF-method Work for Equivalent Observers?

We have shown above (Section 23 and 24) that pairs of Equivalent Observers (same observational thresholds or same k'-factors) saw and reported the same number of groups. As a minimum, one must demand that the group numbers determined using the ADF-method also match the factually observed equality of a pair of equivalent observers. If the ADF-method yields significant difference between what two equivalent observers actually reported, we cannot expect the method to give correctly calibrated results for those two observers and, by extension, for any observers. We assert that this is true regardless of the inner workings and irreproducible computational details of the ADF-method (or any method for that matter).

## 27. ADF Fails for Quimby and Winkler

Figure 52 shows the ADF-based group numbers (from Willamo et al. [2017]) for the Equivalent Observers Quimby and Winkler.

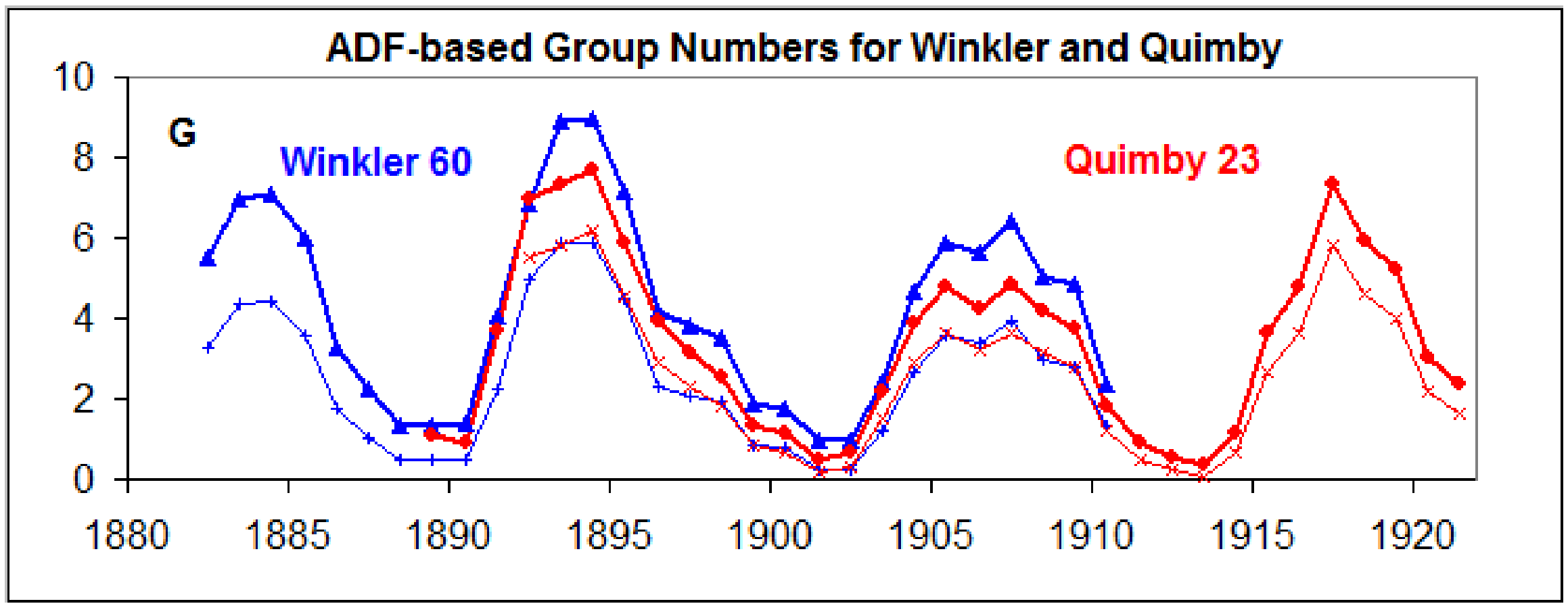

**Figure 52.** ADF-based group numbers for Winkler (*S* = 60, blue triangles) and Quimby (*S* = 23, red dots). The raw, actually observed group numbers for Winkler (*k'* = 1.3, blue plusses) and Quimby (*k'* = 1.3, red crosses) are shown below the ADF-based curves.

It should be evident that ADF-method fails to produce the expected nearly identical counts observed by these two equivalent observers, not to speak about the large discrepancy (60 vs. 23) in the *S* threshold areas.

## 28. ADF Fails for Broger and Wolfer

Figure 53 shows the ADF-based group numbers (from Willamo et al. [2017]) for the Equivalent Observers Broger and Wolfer.

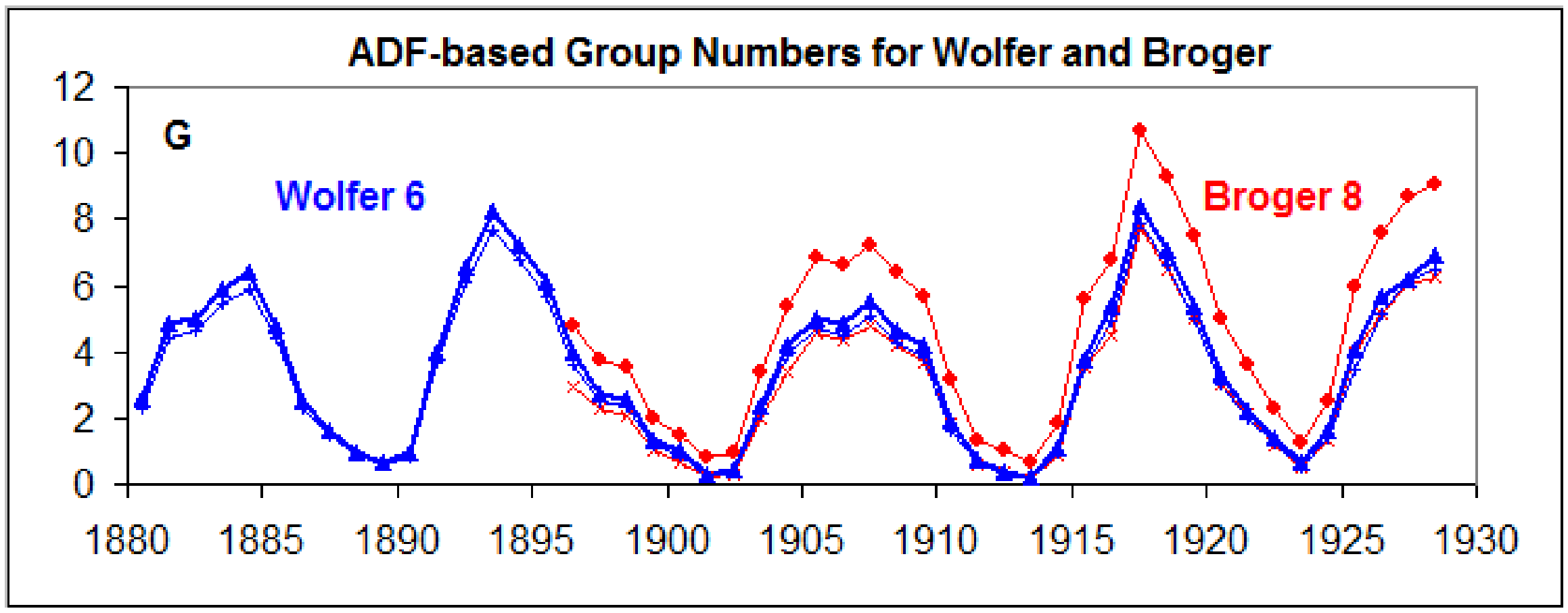

**Figure 53.** ADF-based group numbers for Wolfer (*S* = 6, blue triangles) and Broger (*S* = 8, red dots). The raw, actually observed group numbers for Wolfer (*k'* = 1.0, blue plusses) and Broger (*k'* = 1.0, red crosses) are shown below the ADF-based curves.

It should be evident that the ADF-method fails to produce the expected nearly identical counts observed by these two equivalent observers, in spite of the nearly identical *S* threshold areas.

## 29. ADF Fails for Weber and Shea

Heinrich Weber (observed 1859-1883) and Charles Shea (observed 1847-1866, 5538 drawings reduced by Hoyt & Schatten) should also be equivalent observers because they have identical *S* values of 25. Figure 54 shows the ADF-based group numbers (from Willamo et al. [2017]) and the actual observed group numbers for Weber and Shea.

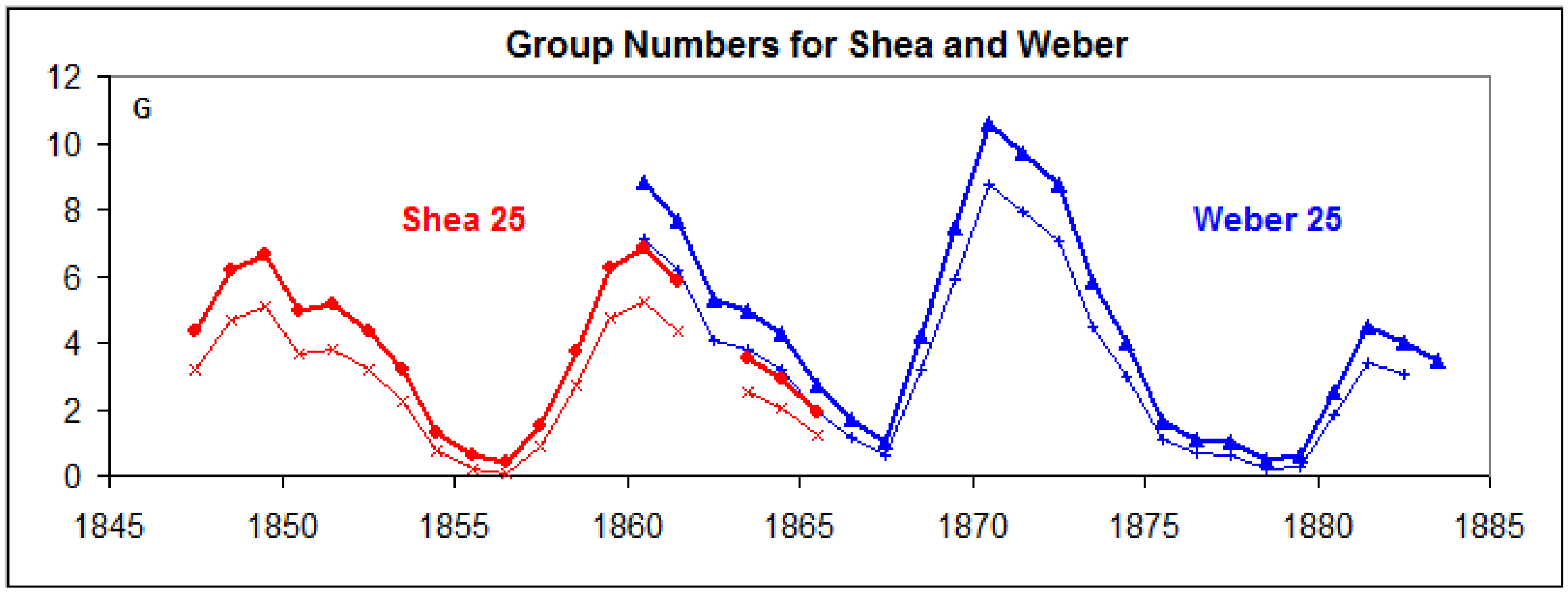

**Figure 54.** ADF-based group numbers for Weber (*S* = 25, blue triangles) and Shea (*S* = 25, red dots). The raw, actually observed group numbers for Weber (blue plusses) and Shea (red crosses) are shown below the ADF-based curves.

It should be evident that the ADF-method fails to produce the expected nearly identical counts observed by these two observers with identical *S* threshold areas. In addition, the actual observations are not consistent with equal *S* values since Weber reported 40% more groups than Shea. Data for 1862 are missing from the database. The observations by Shea are preserved in the Library of the Royal Astronomical Society (London) and bear re-examination.

## 30. ADF Fails for Spörer and RGO

Spörer was labeled a 'perfect observer' on account of his 'observational threshold $S_S$ area' being determined to be equal to zero, based on the assumption that the observer can see and report all the groups with the area larger than $S_S$, while missing all smaller groups. So, Spörer could apparently, according to the ADF calibration method, see and report all groups, regardless of size and should never miss any. This suggests a very direct test: compute the yearly average group count for both Spörer and the 'perfect observer' exemplar, the Royal Greenwich Observatory (RGO), and compare them. They should be identical within a reasonable (very small) error margin. We find that they are not and that RGO generally reported 45% more groups than Spörer, and that therefore, the ADF-method is not generally applicable

We concentrate on the interval 1880-1893 where sufficient and unambiguous data are available from the following observers: Gustav Spörer (at Anclam), Royal Greenwich Observatory (RGO), and Alfred Wolfer (Zürich), as provided by Usoskin (Personal Communication, 2017 to Laure Lefèvre) in this format:

| Year | M | D | G | G(ADF) | $G_{Lo}$ | $G_{Hi}$ |
|---|---|---|---|---|---|---|
| 1880 | 1 | 4 | 1 | 1.04806 | 1 | 1 |
| 1880 | 1 | 7 | 2 | 2.07032 | 2 | 2 |
| 1880 | 1 | 8 | 3 | 3.09613 | 3 | 3 |

Year, M=Month, D=Day, G=Observed group count
G(ADF)= ADF-based reconstruction
$G_{Lo}$=Low Limit of G(ADF)
$G_{Hi}$=High Limit of G(ADF)

It is not clear from the data if the limits $G_{Lo}$ and $G_{Hi}$ (determining the confidence interval) are truncated or rounded to the nearest integer or if they are the actual true values. In any case, they are always identical for Spörer.

Table 1 of Willamo et al. [2017] specifies that Spörer is a 'perfect observer' with 'observational threshold $S_S$ (in millionths of the solar disk)' equal to zero, based on the assumption that the 'quality' of each observer is characterized by his/her observational acuity, measured by a threshold area $S_S$. The threshold implies that the observer can see and report all the groups with the area larger than $S_S$, while missing all smaller groups. So, Spörer could apparently, according to the ADF calibration method, see and report all groups, regardless of size and should never miss any, except for a few that evolved and died without Spörer seeing them. In fact, the $G_{Lo}$ and $G_{Hi}$ given by Usoskin are identical as they should be for perfect data without errors. If so, it suggests a very direct test: compute the yearly average group count for both Spörer and RGO and compare them. They should be identical within a reasonable (very small) error margin.

The following table gives the annual values for Spörer (calculated by Willamo et al. [2017]), Spörer (observed and reported), RGO, Wolfer, and the Svalgaard & Schatten [2016] Group Number Backbone:

| Year | Spörer(W) | Spörer(O) | RGO | Wolfer | S&S BB |
|---|---|---|---|---|---|
| 1880.5 | *2.18* | 2.11 | 2.19 | 2.69 | 2.70 |
| 1881.5 | 3.11 | 3.03 | 3.96 | 4.69 | 4.62 |
| 1882.5 | 3.56 | 3.46 | 4.48 | 4.59 | 4.78 |
| 1883.5 | 3.57 | 3.47 | 4.92 | 5.90 | 5.31 |
| 1884.5 | 3.87 | 3.78 | 5.58 | 5.53 | 5.84 |
| 1885.5 | 2.89 | 2.81 | 4.28 | 4.32 | 4.64 |
| 1886.5 | 1.93 | 1.87 | 2.04 | 2.17 | 2.41 |
| 1887.5 | 1.17 | 1.12 | 1.25 | 1.44 | 1.35 |
| 1888.5 | 0.61 | 0.57 | 0.72 | 0.73 | 0.78 |
| 1889.5 | 0.32 | 0.29 | 0.52 | 0.60 | 0.60 |
| 1890.5 | 0.59 | 0.55 | 0.71 | 1.15 | 0.69 |
| 1891.5 | 2.58 | 2.51 | 3.41 | 4.17 | 3.56 |
| 1892.5 | 4.08 | 3.98 | 6.39 | 5.98 | 6.18 |
| 1893.5 | 5.62 | 5.50 | 8.51 | 8.31 | 7.73 |
| Average | **2.577** | **2.504** | **3.497** | **3.733** | **3.656** |
| Ratio | **1.029** | **1.000** | **1.397** | **1.491** | **1.460** |

Table 2 shows that Gustav Spörer (1822-1895, observed 1861-1893) and the Greenwich observers (1884-1976) are both 'perfect observers' [Willamo et al., 2017] since their *S* value is zero[8]. We should therefore expect that they should observe and report nearly identical yearly values of the sunspot group numbers, as they have the same observational threshold and no groups should be missed.

We here posit that what Spörer actually reported (column three) is what must be compared to the reconstructions. It is thus evident that RGO is 40%, Wolfer 49%, and S&S BB 46% higher than what Spörer 'the perfect observer' saw and reported. And that therefore the test has failed. The ADF-method of calibration does not give the correct result in this simple, straightforward, and transparent example. Figure 55 shows the results in graphical form.

[8] The data for 1879 for Spörer are anomalously high because all days with zero groups were entered as missing in the Hoyt & Schatten catalog. This may have influenced slightly the determination of *S*.

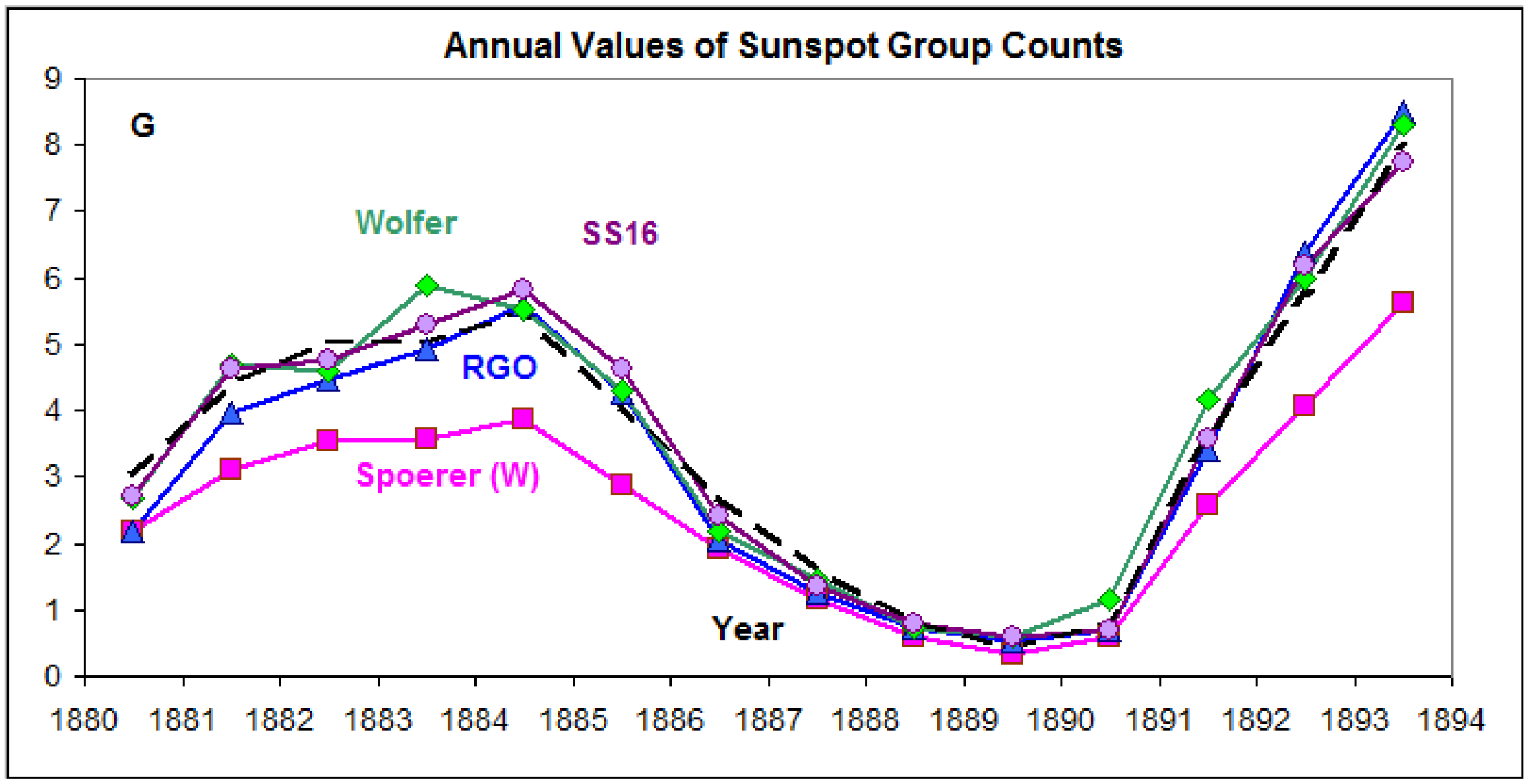

**Figure 55.** Annual values of the Sunspot Group Number for Spörer (pink squares; calculated by Willamo et al. [2017]), RGO (blue triangles), Wolfer (green diamonds), Svalgaard & Schatten [2016] (purple dots). Scaling Spörer up by a factor 1.45 yields the black dashed curve.

The difference between Spörer and the real 'perfect observer' RGO is vividly evident in Figure 46 that shows the fraction of the time where a given number of groups was observed as a function of the phase within the sunspot cycle. At high solar activity Spörer saw significantly fewer spots than RGO. It is also at such times that the ADF is close to unity (as at such times almost every day is an 'active day' in every cycle) and therefore does not carry information about the size of the cycle. The ADF-method does not yield a correct 'observational threshold $S_S$' for G. Spörer and thus does not form a reliable basis for reconstruction of past solar activity valid for all times and observers, and as such must be discarded for general use if applied blindly to less than perfect data.

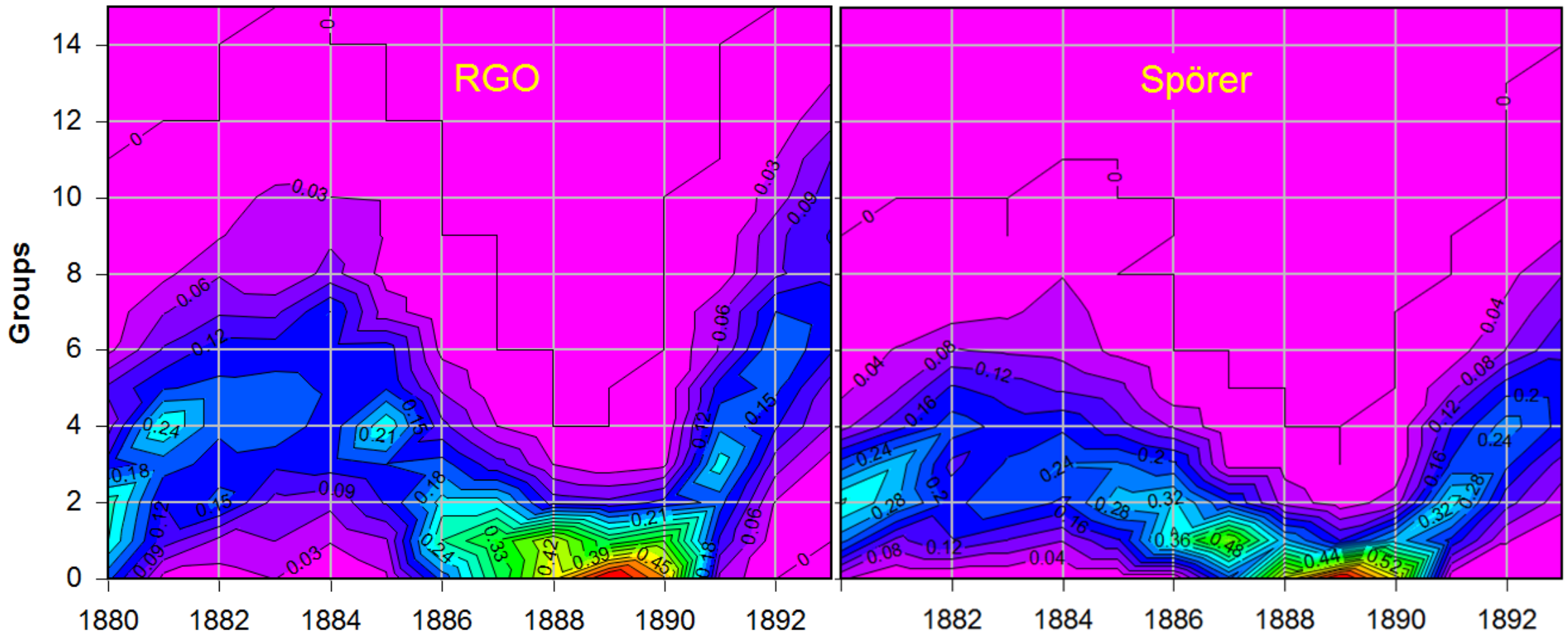

**Figure 56.** Frequency of occurrence of counts of groups on the solar disk as a function of time during 1880-1893 for RGO (left) and Spörer (right) determined for each year by the number of days where a given number of groups was observed on the disk divided by the number of days with an observation.

Spörer needs to be scaled up by a factor 1.45 to match RGO, so can hardly be deemed to be a 'perfect observer' as determined by the ADF-method.

## 31. The Problem with Zero Groups

Even if we compare two equivalent observers there will be a spread in the values. If one observer sees, say, four groups on a given day, the other observer will often observe a different number, because of variable seeing and of small groups emerging, merging, splitting, or disappearing at different times for the two observers. So there is a 'point-spread function' with a round hill of width typically one to two groups, centered on the chosen group number value, Figure 57:

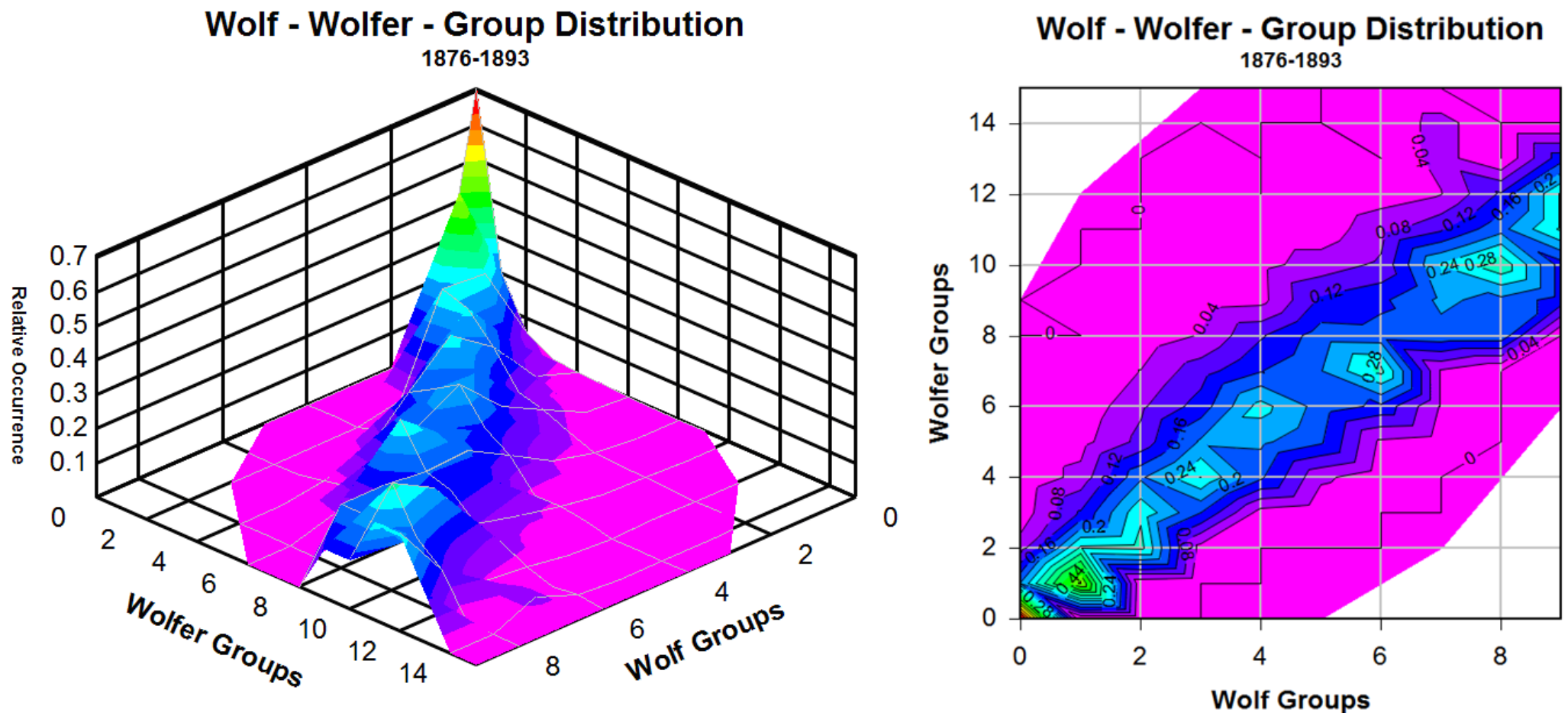

**Figure 57.** The distribution of daily values of the observed Sunspot Group Numbers for Wolfer for each bin of Wolf's group number, normalized to the sum of all groups in that bin. (Left) A 3D view of the 'hills' for each bin. (Right) A contour plot of the distribution.

So, in general, there will be a neighborhood in the distribution around a given group number 'hill' where some group numbers are a bit larger and some are a bit smaller than the top-of-the-hill number. This holds for all bins *except* for the zero bin, because there are no negative group numbers. As a result, the other observer's average group number for the first observer's zero bin will be artificially too high. This fundamental flaw can be seen in the ADF-series for all observers, rendering the ADF-values generally too high for low activity. The purpose of the ADF-method is to bring all observers considered onto the same scale. As Figure 58 shows this goal is not realized for low solar activity.

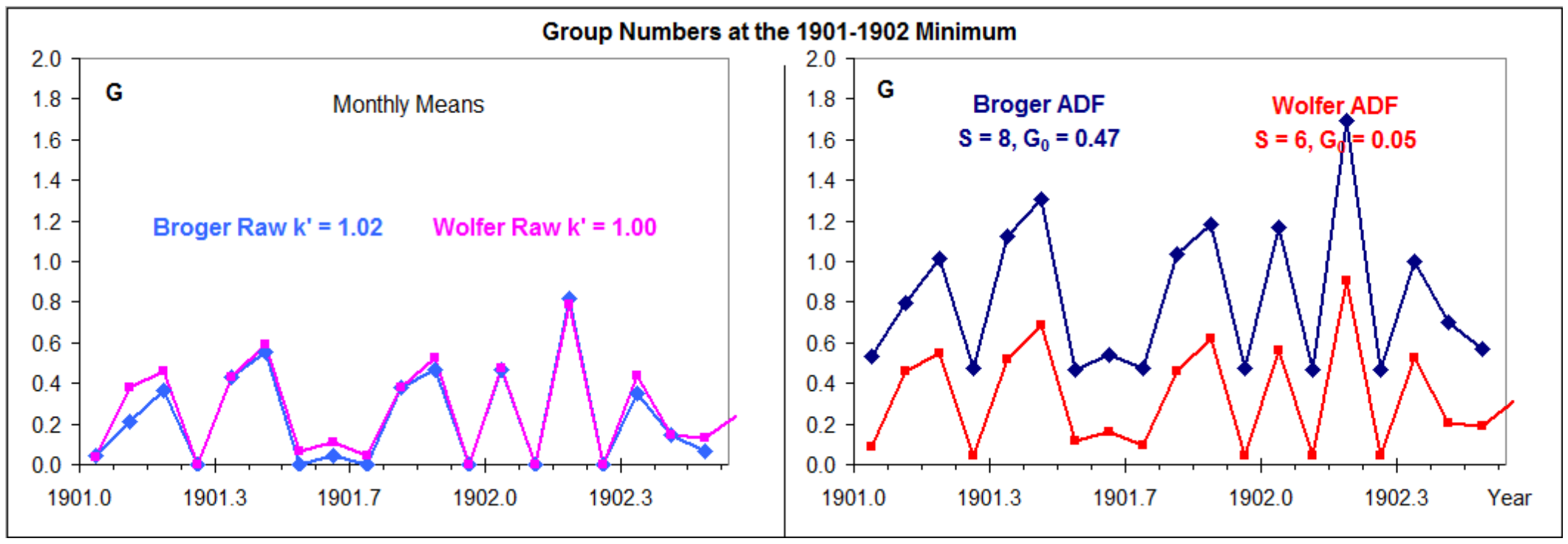

**Figure 58.** (Left) The monthly mean Group Numbers observed by the equivalent observers Broger (light-blue diamonds) and Wolfer (pink squares) during the deep solar minimum 1901.0-1902.6. (Right) The Group Numbers for Broger (dark-blue diamonds) and Wolfer (red squares) computed by Willamo et al. [2017] using the ADF-method. The artificial offset for Broger (0.47) is particularly egregious for $G_{\text{Wolfer}} = 0$.

From modern observations we know that during solar minima there are many days (e.g. for years 2008: 265, and 2009: 262, and 1913: 311) when there are no spots or groups on the disk, regardless of how strong the telescope is and how good the eyesight of the observer is. A good reconstruction method should thus not invent groups when there are none.

We have identified several pairs of 'equivalent' observers and shown that the group numbers computed using the ADF-method do not reproduce the equality of the group numbers expected for equivalent observers, rendering the vaunted[9] ADF-methodology suspect and not reliable nor useful for studying the long-term variation of solar activity. We suggest that the claim [Willamo et al., 2017] that their "new series of the sunspot group numbers with monthly and annual resolution, […] is forming a basis for new studies of the solar variability and solar dynamo for the last 250 years" is self-aggrandizing, and, if their series is used, will hinder such research. It is incumbent on the community to resolve this issue [Cliver, 2016] so progress can be made, not just in solar physics, but in the several diverse fields using solar activity as input.

## 32. ADF Calibration is No Better then Straight Average

Figure 59 shows that the ADF-derived group number for the time interval 1840-1930 is simply equal (within a constant factor of 1.2) to the average group number computed from the raw data in the Vaquero et al. [2016] database with no normalization at all, but differ before ~1885 from the Svalgaard & Schatten [2016] backbone-derived group number, while agreeing well since 1885. As already pointed out [Svalgaard & Schatten, 2017] this agreement continues up to the present time. Such wholesale agreement since 1840 is not expected because of the change in group recognition and definition since the time of Wolfer following Wolf's death in 1893. A simple explanation may be that the ADF-method just adds noise to the observational raw data with the noise washing out in the average, so that what we see is just a reflection of the changed definition of a group

[9] frequentative of Latin vanare: "to utter empty words"

combined with changes in technology and observing modes rather than a change in solar activity.

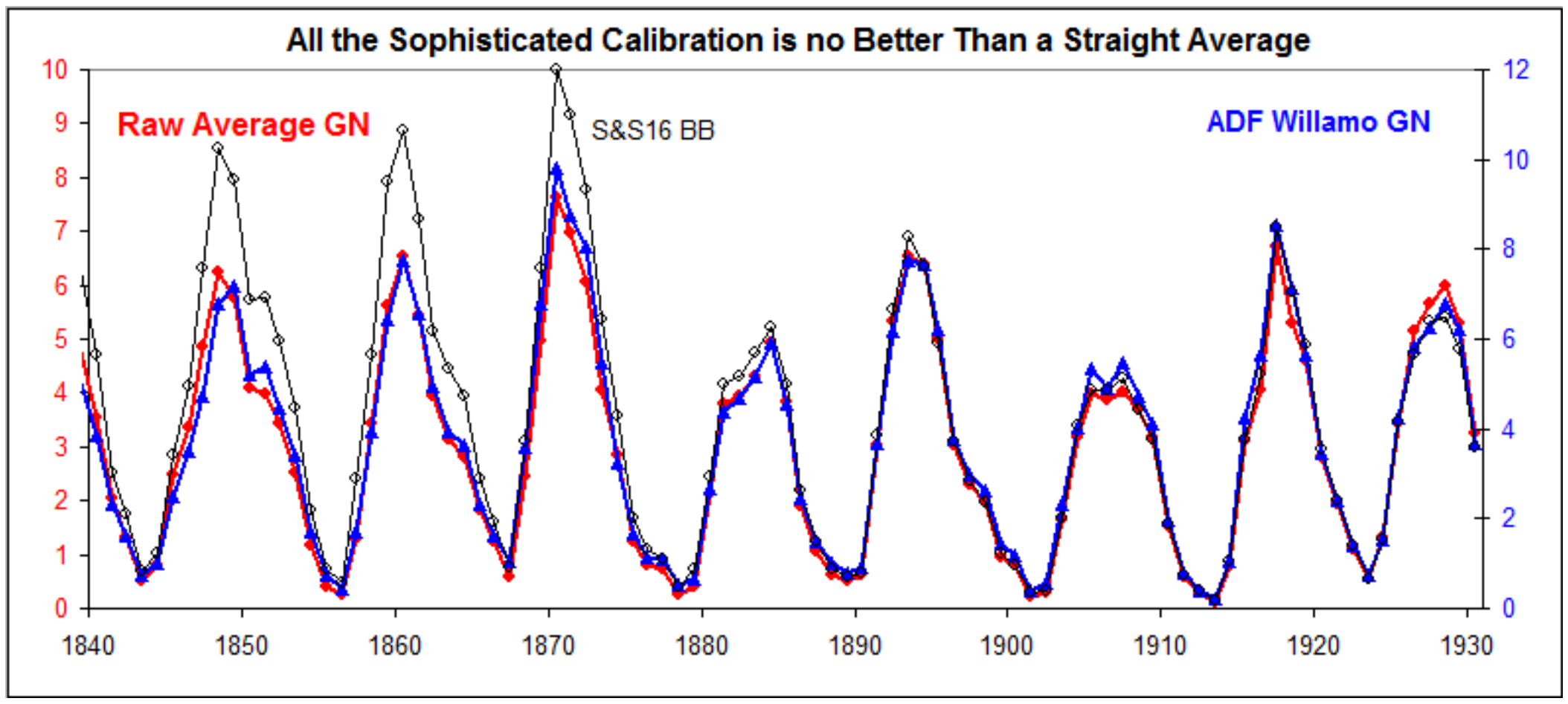

**Figure 59.** Yearly average values of the Sunspot Group Number computed using the ADF-method [Willamo et al., 2017] (blue triangles and curve; right-hand scale) and computed as a simple average of the raw, un-normalized group numbers (red diamonds and curve; left-hand scale); both scaled by a constant factor (1.2) to match each other. The Svalgaard & Schatten [2016] backbone is shown by the black open circles and curve, scaled to match after 1890.

## 33. Conclusion

We have shown that the criticism by Lockwood et al. [2016b] and by Usoskin et al. [2016] expressed by the statement that "our concerns about the backbone reconstruction are because it uses unsound procedures and assumptions in its construction, that it fails to match other solar data series or terrestrial indicators of solar activity, that it requires unlikely drifts in the average of the calibration *k*-factors for historic observers, and that it does not agree with the statistics of observers' active-day fractions" is unfounded, baseless, and without merit. Let us recapitulate our responses to each of those concerns in sequence:

1) "it uses unsound procedures and assumptions in its construction". This is primarily about whether it is correct to use a constant proportionality factor when calibrating observers to the primary observer. We showed in Section 2 that proportionality is an observational fact within the error of the regression. In addition, we clarify in Section 11 some confusion about daisy-chaining and show that no daisy-chaining was used for the period 1794-1996 in the construction of the backbones.

2) "it fails to match other solar data series or terrestrial indicators of solar activity". We showed in Section 8 that our group numbers match the variation of the diurnal amplitude of the geomagnetic field and the HMF derived from the geomagnetic IDV index and in Sections 14 and 16 that they match the (modeled) cosmogenic radionuclide record.

3) “it requires unlikely drifts in the average of the calibration *k*-factors for historic observers “ We showed in Section 6 that the RGO group counts were drifting during the first twenty years of observation and that other observers agree during that period that the RGO group count drift is real.
4) “it does not agree with the statistics of observers’ active-day fractions”. We show that the ADF-method fails for observers that the method itself classifies as equivalent observers and that the method thus is not generally applicable and that it therefore is not surprising that it fails to agree with the backbone group number series.
5) We identified several misrepresentations and (perhaps) misunderstandings.

We are nevertheless pleased that the subject of revising the records of solar activity has become an active area of research, but it should be done right.

## Acknowledgements

L.S. thanks Stanford University for support. We appreciate comments by Frédéric Clette, Laure Lefèvre, and Ed Cliver. We specifically thank Laure Lefèvre for providing Ilya Usoskin’s data files pertaining to the ADF method.

## References

Asvestari, E., Usoskin, I. G., Kovaltsov, G. A., Owens, M. J., Krivova, N. A., Rubinetti, S., Taricco, C.: Assessment of different sunspot number series using the cosmogenic isotope $^{44}$Ti in meteorites, Monthly *Notices Royal Astron. Soc.* **467**(2), 1608, doi:10.1093/mnras/stx190, 2017.

Balmaceda, L., Solanki, S. K., Krivova, N. A, and Foster, S.: A homogeneous database of sunspot areas covering more than 130 years, *J. Geophys. Res.* **114**, A07104, doi:10.1029/2009JA014299, 2009.

Berggren, A.-M., Beer, J., Possnert, G., Aldahan, A., Kubik, P., Christl, M., Johnsen, S. J., Abreu, J., and Vinther, B. M.: A 600-year annual $^{10}$Be record from the NGRIP ice core, Greenland, *Geophys. Res. Lett.* **36**, L11801, doi:10.1029/2009GL038004, 2009.

Chatzistergos, T., Usoskin I. G., Kovaltsov, G. A., Krivova, N. A., Solanki, S. K.: New reconstruction of the sunspot group number since 1739 using direct calibration and ‘backbone’ methods, *Astron. & Astrophys.*. 602, A69, doi:10.1051/0004-6361/201630045, 2017.

Clette, F., Cliver, E. W.,·Lefèvre, L., Svalgaard, L.,·Vaquero, J. M., and Leibacher, J. W.: Preface to Topical Issue: Recalibration of the Sunspot Number, *Solar Phys.* **291**(9), 2479, doi:10.1007/s11207-016-1017-8, 2016.

Clette, F., Svalgaard, L., Vaquero, J. M., and Cliver, E. W.: Revisiting the Sunspot Number - A 400-Year Perspective on the Solar Cycle, *Space Sci. Rev.* **186**, 35, doi:10.1007/s11214-014-0074-2, 2014.

Cliver, E. W.: Comparison of New and Old Sunspot Number Time Series, *Solar Phys.* **291**(9-10), 2891, doi:10.1007/s11207-016-0929-7, 2016.

Cliver, E.W. and Ling, A.G.: The Discontinuity *Circa* 1885 in the Group Sunspot Number, *Solar Phys.* **291**, 2763, doi:10.1007/s11207-015-0841-6, 2016.

Friedli, T. K.: Sunspot Observations of Rudolf Wolf from 1849-1893, *Solar Phys.* **291**(9), 2505, doi:10.1007/s11207-016-0907-0, 2016.

Herbst, K., Muscheler, R., Heber, B.: The new local interstellar spectra and their influence on the production rates of the cosmogenic radionuclides $^{10}$Be and $^{14}$C, *J. Geophys. Res. (Space Physics)*, **122**(1), 23, doi:10.1002/2016JA023207, 2017.

Hoyt D. V., Schatten K. H., and Nesmes-Ribes, E.: The one hundredth year of Rudolf Wolf's death: Do we have the correct reconstruction of solar activity? *Geophys. Res. Lett.* **21**(18), 2067, doi:10.1029/94GL01698, 1994.

Hoyt D. V., Schatten K. H.: Group Sunspot Numbers: A New Solar Activity Reconstruction, *Solar Phys.* **181**(2), 491, doi:10.1023/A:1005056326158, 1998.

Muscheler, R., Adolphi, F.,·Herbst, K., Nilsson, A.: The Revised Sunspot Record in Comparison to Cosmogenic Radionuclide-Based Solar Activity Reconstructions, *Solar Phys.* **291**(9), 3025, doi:10.1007/s11207-016-0969-z, 2016.

Lockwood, M., Nevanlinna, H., Barnard, L., Owens, M. J., Harrison, R. G., Rouillard, A. P., Scott, C. J.: Reconstruction of geomagnetic activity and near-Earth interplanetary conditions over the past 167 yr – Part 4, *Annal. Geophys.* **32**(4), 383, doi: 10.5194/angeo-32-383-2014, 2014.

Lockwood, M., Owens, M. J., Barnard, L., and Usoskin, I. G.: Tests of Sunspot Number Sequences: 3. Effects of Regression Procedures on the Calibration of Historic Sunspot Data, *Solar Phys.* **291**(9), 2829, doi:10.1007/s11207-015-0829.2, 2016a.

Lockwood, M., Owens, M. J., Barnard, L., and Usoskin, I. G.: An Assessment of Sunspot Number Data Composites over 1845–2014, *Astrophys. Journal* **824**(1), doi:10.3847/0004-637X/824/1/54, 2016b.

Owens, M. J., Cliver, E. W., McCracken, K. G., Beer, J., Barnard, L., Lockwood, M., Rouillard, A., Passos, D., Riley, P., Usoskin, I. G., and Wang, Y.-M.: Near-Earth heliospheric magnetic field intensity since 1750: 1. Sunspot and geomagnetic reconstructions. *J. Geophys. Res. (Space Physics)* **121**, 6048, doi: 10.1002/2016JA022550, 2016.

Schrijver, C. J., Livingston, W. C., Woods, T. N., and Mewaldt, R. A.: The minimal solar activity in 2008‐2009 and its implications for long‐term climate modeling, *Geophys. Res. Lett.* **38**, L06701, doi:10.1029/2011GL046658, 2011.

Schwabe, S. H.: Sonnenbeobachtungen im Jahre 1843, *Astron. Nachrichten* **21**, 233, 1844.

Svalgaard, L.: Correction of errors in scale values for magnetic elements for Helsinki, *Ann. Geophys.* **32**, 633, doi:10.5194/angeo-32-2014, 2014.

Svalgaard, L.: Reconstruction of Solar Extreme Ultraviolet Flux 1740-2015, *Solar Phys.* **291**(10), 2981, doi:10.1007/s11207-016-0921-2, 2016.

Svalgaard, L., Cliver, E. W., and Le Sager, P.: Determination of Interplanetary Magnetic Field Strength, Solar Wind Speed, and EUV Irradiance, 1890-Present. International Solar Cycle Studies Symposium, June 23-28, 2003, Tatranska Lomnica, Slovak Republic, *Proceedings (ESA SP-535)*, 15, ed. A. Wilson, Noordwijk, ISBN:92-9092-845-X, 2003.

Svalgaard, L. and Cliver, E. W.: The IDV index: Its derivation and use in inferring long-term variations of the interplanetary magnetic field strength, *J. Geophys. Res.* **110**, A12103, doi:10.1029/2005JA011203, 2005.

Svalgaard, L. and Cliver, E. W.: Interhourly variability index of geomagnetic activity and its use in deriving the long-term variation of solar wind speed, *J. Geophys. Res.* **112**, A10111, doi:10.1029/2007JA012437, 2007.

Svalgaard, L. and Cliver, E. W.: Heliospheric magnetic field 1835–2009, *J. Geophys. Res.* **115**, A09111, doi:10.1029/2009JA015069, 2010.

Svalgaard, L., and Schatten, K. H.: Reconstruction of the Sunspot Group Number: the Backbone Method, *Solar Phys.* **291**(10), 291(9), 2653, doi:10.1007/s11207-015-0815-8, 2016.

Usoskin, I. G., Arlt, R., Asvestari, E., Hawkins, E., Käpylä, M., Kovaltsov, G. A., Krivova, N., Lockwood, M., Mursula, K., O'Reilly, J., Owens, M. J., Scott, C. J., Sokoloff, D. D., Solanki, S. K., Soon, W., Vaquero, J. M.: The Maunder minimum (1645–1715) was indeed a Grand minimum: A reassessment of multiple datasets, *Astron.&Astrophys.* **581**, A95, doi:10.1051/0004-6361/201526652, 2015.

Usoskin, I. G., Kovaltsov, G. A., Lockwood, M., Mursula, K., Owens, M. J., Solanki, S. K.: A New Calibrated Sunspot Group Series Since 1749: Statistics of Active Day Fractions, *Solar Phys.* **291**(9), 2685, doi:10.1007/s11207-015-0838-1, 2016.

Vaquero, J. M.: Presentation at 2nd- Sunspot Number Workshop, Brussels, Belgium, http://www.leif.org/research/SSN/Vaquero2.pdf, 2012.

Vaquero, J. M., Trigo, R. M., and Gallego, M. C.: A Simple Method to Check the Reliability of Annual Sunspot Number in the Historical Period 1610 – 1847, *Solar Phys.* **277**(2), 389, doi:10.1007/s11207-011-9901-8, 2012.

Vaquero, J. M., Svalgaard, L., Carrasco, V. M. S., Clette, F., Lefèvre, L., Gallego, M. C., Arlt, R., Aparicio, A. J. P., Richard, J-G., Howe, R.: A Revised Collection of Sunspot Group Numbers, *Solar Phys.* **291**(10), 3061, doi:10.1007/s11207-016-0982-2, 2016.

Wang, Y.-M. and Sheeley Jr, N. R.: On the fluctuating component of the Sun's large‐scale magnetic field, *Astrophys. J.* **590**, 1111, doi:10.1086/375026, 2003.

Wang, Y.-M., Lean, J. L., and Sheeley Jr, N. R.: Modeling the Sun's magnetic field and irradiance since 1713, *Astrophys. J.* **625**, 522,doi:10.1086/429689, 2005.

Weise, W., Dorschner, J., and Schielicke, R. E.: Wilhelm Winkler (1842-1910) - a Thuringian private astronomer and maecenas, *Acta Historica Astron.* **3**, 95, 1998.

Willamo, T., Usoskin, I. G., and Kovaltsov, G. A.: Updated sunspot group number reconstruction for 1749–1996 using the active day fraction method, *Astron. & Astrophys*. **601**, A109, doi:10.1051/0004-6361/201629839, 2017.

Wolf, J. R.: *Astronomische Mittheilungen*, **LXXXII**, 58, 1893.